\documentclass[letter, 11pt]{article}
\usepackage{amssymb, amsmath, amsthm, mathtools, pdflscape, subfig, comment, lscape}
\usepackage[margin=1in]{geometry}
\allowdisplaybreaks[1]
\usepackage[longnamesfirst, round]{natbib}
\usepackage[colorlinks=true, linkcolor=blue, filecolor=blue, urlcolor=blue, citecolor=blue, driverfallback=dvipdfmx]{hyperref}
\usepackage[dvipdfmx]{graphicx}

\newtheorem{assumption}{Assumption}

\newtheorem{theorem}{Theorem}
\newtheorem{lemma}{Lemma}
\theoremstyle{definition}

\newtheorem{remark}{Remark}

\usepackage{setspace}
\allowdisplaybreaks

\title{Kernel Estimation for Panel Data with Heterogeneous Dynamics\footnote{First Version: February, 2018.}}
\author{Ryo Okui\thanks{Department of Economics and the Institute of Economic Research, Seoul National University, Building 16, 1 Gwanak-ro, Gwanak-gu, Seoul, 08826, South Korea. Email: \href{mailto:okui.ryo.3@gmail.com}{okui.ryo.3@gmail.com}} \and Takahide Yanagi\footnote{Graduate School of Economics, Kyoto University, Yoshida Honmachi, Sakyo, Kyoto, 606-8501, Japan.  Email: \href{mailto:yanagi@econ.kyoto-u.ac.jp}{yanagi@econ.kyoto-u.ac.jp}}}

\date{May, 2019}

\begin{document}

\maketitle

\begin{abstract}
This paper proposes nonparametric kernel-smoothing estimation for panel data to examine the degree of heterogeneity across cross-sectional units.
We first estimate the sample mean, autocovariances, and autocorrelations for each unit and then apply kernel smoothing to compute their density functions.
The dependence of the kernel estimator on bandwidth makes asymptotic bias of very high order affect the required condition on the relative magnitudes of the cross-sectional sample size ($N$) and the time-series length ($T$). 
In particular, it makes the condition on $N$ and $T$ stronger and more complicated than those typically observed in the long-panel literature without kernel smoothing.
We also consider a split-panel jackknife method to correct bias and construction of confidence intervals. 
An empirical application and Monte Carlo simulations illustrate our procedure in finite samples.

\bigskip

\noindent \textit{Keywords}: autocorrelation, density estimation, heterogeneity, incidental parameter, jackknife, kernel smoothing. 

\bigskip 

\noindent \textit{JEL Classification}: C13, C14, C23.

\end{abstract}

\newpage
\section{Introduction}
The characteristics of heterogeneity across economic units are informative for many econometric applications.
For example, there is an interest in heterogeneity in the dynamics of price deviations or changes (e.g., \citealp{klenow2010microeconomic}; \citealp{CruciniShintaniTsuruga15}).
As another example, allowing for the presence of heterogeneity may make a crucial difference in identification and estimation of production functions (e.g., \citealp{AckerbergBenkardBerryPakes07}; \citealp{Kasahara2017}).
Thus, there are many econometric studies that investigate the degree of heterogeneity using panel data (e.g., \citealp{HsiaoPesaranTahmiscioglu99}; \citealp{FernandezValLee13}; \citealp{JochmansWeidner2019}; \citealp{Okui2017}).

This paper proposes kernel-smoothing estimation for panel data to analyze heterogeneity across cross-sectional units.\footnote{An {\ttfamily R} package to implement the proposed procedure is available from the authors' websites.
}
After estimating the mean, autocovariances, and autocorrelations of each unit, we compute the kernel densities based on these estimated quantities.
This easy-to-implement procedure provides useful visual information for heterogeneity in a model-free manner. 
For example, the densities of the heterogeneous mean, variance, and first-order autocorrelation of the price deviations indicate visually the characteristics of heterogeneity in the long-run level, variance, and persistence of the price deviations across items (goods and services) that are cross-sectional units in this example.
Indeed, several empirical studies have used such estimation for various applications (e.g., \citealp[Figure 2]{Kasahara2017} and \citealp[Figure 8]{roca2017learning}), but there is no theoretical foundation for kernel-smoothing to examine heterogeneity in long-panel data.

We show consistency and asymptotic normality of the kernel density estimator based on double asymptotics under which both the cross-sectional size $N$ and the time-series length $T$ tend to infinity with the bandwidth $h$ shrinking to zero (denoted by $N, T \to \infty$ and $h \to 0$).\footnote{More precisely, 
	the double asymptotics $N, T \to \infty$ are any monotonic sequence $T=T(N) \to \infty$ as $N \to \infty$ and the bandwidth $h \to 0$ is any monotonic sequence $h = h(N, T(N)) = h(N) \to 0$ as $N, T \to \infty$ in our setting. Note that each theoretical result in this paper specifies additional conditions on the relative magnitudes of $N$, $T$, and $h$.
}
The asymptotic properties exhibit several unique features that have not been well examined in the long-panel literature.
Most importantly, asymptotic bias of even very high order affects the conditions on the relative magnitudes of $N$, $T$, and $h$ required for consistency and asymptotic normality. As a result, the different orders of asymptotic expansion we can execute have different relative magnitude conditions. 
This unique feature contrasts our analysis with the existing analyses without kernel smoothing where the required relative magnitude conditions do not depend on the order of expansions (e.g., $N / T^2 \to 0$ for asymptotic normality in \citealp{HsiaoPesaranTahmiscioglu99} and \citealp{Okui2017}).
The weakest condition (i.e., how small $T$ can be compared with $N$) can be obtained by executing an infinite order expansion. 
Even in that case, the required condition is stronger than those typically observed in the literature without kernel smoothing. 
Moreover, it requires nontrivial discussions for the expansion (e.g., the summability of the infinite-order series).
We clarify that these unique features are caused by the presence of the bandwidth $h$ and by using the estimated quantities.

Based on an infinite-order expansion, we show three asymptotic biases for the density estimation.
The first is the standard kernel-smoothing bias of order $O(h^2)$ (see, e.g., \citealp{LiRacine2007}).
The second is caused by the incidental parameter problem (\citealp{NeymanScott48} and \citealp{Nickell1981}) and is $O(1/T)$.
The third results from the nonlinearity of the kernel function and the difference between the estimated quantity and the true quantity.
We show that this is $O(1/(Th^2)) + \sum_{j=3}^{\infty} O(1/\sqrt{T^j h^{2j}})$, which is obtained only if we execute an infinite-order expansion.
By showing these asymptotic biases, we prove that the relative magnitude conditions for consistency and asymptotic normality are $N^2 / T^5 \to 0$ and $N^2 / T^3 \to 0$, respectively, when using the standard bandwidth $h \asymp N^{-1/5}$ in the density estimation with second-order kernels.

We propose to apply a split-panel jackknife method in \citet{DhaeneJochmans15} to reduce these biases.
In particular, we formally show that the half-panel jackknife (HPJ) corrects the incidental parameter bias and the second-order nonlinearity bias without inflating the asymptotic variance.
While the jackknife is useful in bias reduction especially when $T$ is small, we also show that it does not weaken the relative magnitude conditions for consistency and asymptotic normality.

We also develop confidence interval (CI) estimation and selection of bandwidth.
To construct CI, we extend the robust bias-corrected (RBC) procedure in \citet{calonico2018effect} to split-panel jackknife bias-corrected estimation. 
This method explicitly corrects all three biases above.
For the bandwidth selection, we can apply any standard procedures in the literature.
This is because, under the relative magnitude conditions, the asymptotic mean squared error (AMSE) and asymptotic distribution of the split-panel jackknife bias-corrected estimator are the same as those of the infeasible estimator based on the true quantity.

We also examine the properties of the cumulative distribution function (CDF) estimator constructed by integrating the kernel density estimator. 
This kernel CDF estimator also exhibits asymptotic bias that varies in the order of the asymptotic expansion that we execute. 
We also derive the closed form formula for the asymptotic bias. 
This is an interesting result from theoretical viewpoint because the formula for asymptotic bias for the empirical distribution is available only for Gaussian errors \citep{JochmansWeidner2019} and has not been derived in general form \citep{Okui2017}.
However, the required conditions on $N$ and $T$ for the kernel CDF estimation turn out to be stronger than those for the empirical CDF estimation derived in those studies.

We illustrate our procedures by an empirical application on heterogeneity of price deviations from the law of one price (LOP).
Our procedures reveal significant heterogeneity in the price deviations dynamics.
The split-panel jackknife bias-corrected density estimates imply much more volatile and persistent dynamics than the estimates without bias correction and the difference is visually noticeable.
This result highlights the importance of the bias correction in that the bias-corrected densities can provide distinct visual information for heterogeneity from the densities without bias correction.

\paragraph{Related literature.}
Our setting and motivation closely relate to \citet{Okui2017}, but there are several important distinctions in both theoretical and practical aspects.
First, our relative magnitude conditions are different from \citet{Okui2017} in which second-order expansions suffice to derive the conditions on estimating the moments of the quantities (e.g., the variance of the heterogeneous mean).
This feature in particular contrasts the theoretical contributions in both papers, and indeed our relative magnitude conditions are new in the literature.
Second, we show the new insight that the split-panel jackknife is applicable even to kernel estimation. 
Third, because it is well known that bootstrap inferences do not capture kernel-smoothing bias (see, e.g., \citealp{hall2013simple}), we extend the RBC inference in \citet{calonico2018effect} instead of the cross-sectional bootstrap in \citet{Okui2017}.\footnote{The failure of the cross-sectional bootstrap inference in our kernel estimation is formally shown in the previous version of this study uploaded to arXiv (arXiv:1802.08825v2).}
Finally, while \citet{Okui2017} do not clarify asymptotic biases for their empirical CDFs, we formalize those of our kernel estimators.

Our CDF estimation relates to \citet{JochmansWeidner2019} who derive the bias of the empirical distribution based on noisy measurements (e.g., estimated quantities) for the true variables of interest.
Their results are complementary to ours. 
They consider a situation where observations exhibit Gaussian errors.
We do not assume such errors. The kernel smoothing allows us to derive bias under much weaker distributional assumptions at the price of creating additional higher-order biases.

Many econometric studies examine heterogeneity in panel data (e.g., \citealp{PesaranSmith95}; \citealp{HsiaoPesaranTahmiscioglu99}; \citealp{PesaranShinSmith99}; \citealp{FernandezValLee13}).
Among them, \citet{HorowitzMarkatou96}, \citet{ArellanoBonhomme12}, and \citet{MavroeidisSasakiWelch15} propose to estimate the densities of heterogeneous quantities with short-panel data based on deconvolution techniques under some model specifications.
Compared with them, we propose model-free kernel-smoothing estimation with long-panel data.

Several studies propose model-free analyses for panel data, but do not focus on the degree of heterogeneity in the dynamics.
For example, \citet{Okui08, Okui11, Okui14} and \citet{LeeOkuiShintani13} consider homogeneous dynamics, and \citet{GalvaoKato14} study the properties of the possibly misspecified fixed effects estimator in the presence of heterogeneous dynamics.

Kernel density estimation using estimated quantities is also examined in the literature on structural estimation of auction models.
For example, \citet{mamarmershneyerov2019} and \citet{gpv2000} estimate the density of individual evaluations of auctioned goods.
In their first stage, individual evaluations of auctioned goods are estimated nonparametrically and their second stage is the kernel density estimation applied to estimated evaluations. 
They also observe that the estimation errors from the first stage affect the asymptotic behavior of the second stage estimator in a nonstandard way. 
However, their problems are different from ours. 
Their main issue is the cross-sectional correlation caused by the use of the same set of observations to estimate individual evaluations. 
As a result, their estimation errors affect the precision and the convergence rate of the second stage estimator. 
In our case, estimation errors in the first stage are cross-sectionally independent and affect the bias but not the (first-order) variance of the second stage estimator.

\paragraph{Paper organization.}
Section \ref{sec-setup} introduces our setting and density estimation.
Section \ref{sec-theory} develops the asymptotic theory, bias correction, CI estimation, bandwidth selection, and CDF estimation.
Section \ref{sec-application} presents the application.
Section \ref{sec-conclusion} concludes.
The supplementary appendix contains the proofs of the theorems, technical lemmas, other technical discussions, and Monte Carlo simulations.

\section{Kernel density estimation} \label{sec-setup}

This section describes the setting and the proposed estimation.
We explain our setting and motivation in a succinct manner because they are similar to those in \citet{Okui2017}.\footnote{Several remarks and possible extensions can be found in the previous version of this study and \citet{Okui2017}.
	For example, we can consider the presence of covariates and time effects and estimation based on other heterogeneous quantities, such as random coefficients in linear models, with minor modifications.
	In this paper, we explain our estimation briefly to save space.}

We observe panel data $\{ \{ y_{it} \}_{t=1}^T \}_{i=1}^N$ where $y_{it}$ is a scalar random variable.
We assume that $y_{it}$ is strictly stationary across time and that each individual time series $\{y_{it}\}_{t=1}^T$ is generated from some unknown probability distribution $\mathcal{L}(\{y_{it}\}_{t=1}^T; \alpha_i)$, where $\alpha_i$ is a (possibly infinite dimensional) random variable specifying the dynamics of $y_{it}$. We note that $\alpha_i$ is an abstract parameter and it does not appear in the actual implementations of our proposed procedure. Characterizing heterogenous dynamics using this abstract parameter is mathematically convenient because it allows us to keep an i.i.d. assumption. 
Existing studies without model specifications also employ this approach (e.g., \citealp{GalvaoKato14}).
We denote the conditional expectation given $\alpha_i$ by $E(\cdot | i)$.

Our goal is to examine the degree of heterogeneity of the dynamics of $y_{it}$ across units in a model-free manner.
To this end, we focus on estimating the density of the mean $\mu_i \coloneqq E(y_{it} | i)$, $k$-th autocovariance $\gamma_{k,i} \coloneqq E( (y_{it} - \mu_i)(y_{i,t-k} - \mu_i) | i )$, and $k$-th autocorrelation $\rho_{k,i} \coloneqq \gamma_{k, i} / \gamma_{0,i}$.
We first estimate $\mu_i$, $\gamma_{k,i}$, and $\rho_{k,i}$ by the sample analogues: $\hat \mu_i \coloneqq \bar y_i \coloneqq T^{-1} \sum_{t=1}^{T} y_{it}$, $\hat \gamma_{k,i} \coloneqq (T - k)^{-1} \sum_{t=k+1}^T (y_{it} - \bar y_i) (y_{i, t-k} - \bar y_i)$, and $\hat \rho_{k,i} \coloneqq \hat \gamma_{k,i} / \hat \gamma_{0,i}$.
Throughout the paper, we use the notation $\xi_i$ to represent one of $\mu_i$, $\gamma_{k,i}$, or $\rho_{k,i}$ and the notation $\hat \xi_i$ for the corresponding estimator.
The kernel estimator for the density $f_{\xi}(x)$ is given by:
\begin{align} \label{eq-dens-cdf}
	\hat f_{\hat \xi}(x)  \coloneqq \frac{1}{Nh} \sum_{i=1}^N K \left( \frac{x - \hat \xi_{i}}{h}\right),
\end{align}
where $x \in \mathbb{R}$ is a fixed point, $K:\mathbb{R} \to \mathbb{R}$ is a kernel function, and $h > 0$ is a bandwidth satisfying $h \to 0$.\footnote{We can consider estimating the joint density for $\mu_i$, $\gamma_{k, i}$, and $\rho_{k,i}$ in the same manner.}
This is a standard estimator except that we replace the true $\xi_i$ with the estimated $\hat \xi_i$.

\section{Asymptotic theory} \label{sec-theory}
This section develops our asymptotic theory, CI estimation, bandwidth selection, and CDF estimation based on the density estimator $\hat f_{\hat \xi}(x)$.
We define the notations $w_{it} \coloneqq y_{it} - \mu_i = y_{it} - E(y_{it}|i)$ and $\bar w_i \coloneqq T^{-1} \sum_{t=1}^T w_{it}$.
By construction, $y_{it} = \mu_i + w_{it}$.
Note that $\hat \mu_i = \bar y_i = \mu_i + \bar w_i$, $E(w_{it}|i) = 0$, and $\gamma_{k,i} = E(w_{it} w_{i,t-k}|i)$.

\subsection{Unique features in asymptotic investigations} \label{sec-unique}
Before formally showing the asymptotic properties, we explore the unique features of our asymptotic investigations in an informal manner.
By doing so, we clarify the mechanism behind the observation that even very high orders of asymptotic bias matter for our asymptotic analysis.

We here focus on the density estimator for $\hat \mu_i$, but similar discussions are also relevant for $\hat \gamma_{k,i}$ and $ \hat \rho_{k,i}$.
Noting that $\hat \mu_i - \mu_i = \bar w_i$, we examine the $J$-th order Taylor expansion of $\hat f_{\hat \mu}(x)$:
\begin{align}\label{eq:issue}
\begin{split}
	\hat f_{\hat \mu}(x)
	=& \frac{1}{Nh} \sum_{i=1}^N K \left( \frac{ x - \mu_i }{h}\right)\\
	& + \sum_{j = 1}^{J-1} \frac{(-1)^j}{j! Nh^{j+1}} \sum_{i=1}^N (\bar w_i)^j K^{(j)} \left( \frac{x - \mu_i }{h}\right)\\
	& + \frac{(-1)^J}{J! Nh^{J+1}} \sum_{i=1}^N (\bar w_i)^J K^{(J)} \left( \frac{x - \tilde \mu_i }{h}\right),
\end{split}
\end{align}
where $K^{(j)}$ denotes the $j$-th order derivative and $\tilde \mu_i$ is between $\mu_i$ and $\hat \mu_i$.

The first term in \eqref{eq:issue} is the infeasible density estimator based on the true $\mu_i$, and its asymptotic behavior is standard and well known in the kernel-smoothing literature.
It converges in probability to the density of interest $f_{\mu}(x)$ as $N\to \infty$ and $h\to 0$ with $Nh \to \infty$.
In addition, when $Nh^5 \to C \in [0, \infty)$ also holds, it can hold that:
\begin{align*}
	\sqrt{Nh} \left( \frac{1}{Nh} \sum_{i=1}^N K \left( \frac{ x - \mu_i }{h}\right) -f_{\mu}(x)  - h^2 \frac{\kappa_1 f_{\mu}^{''} (x)}{2}  \right) \stackrel{d}{\longrightarrow} \mathcal{N} \big( 0, \kappa_2 f_{\mu}(x) \big),
\end{align*}
where $\kappa_1 \coloneqq \int s^2 K(s) ds$ and $\kappa_2 \coloneqq \int K^2(s) ds$ and $\mathcal{N}(\mu, \sigma^2)$ is a normal distribution with mean $\mu$ and variance $\sigma^2$.

The unique features in our situation are caused from the second and third terms in \eqref{eq:issue}.
For the second term, under regularity conditions, the mean can be evaluated as:
\begin{align*}
	E\left( \frac{(-1)^j}{j!Nh^{j+1}} \sum_{i=1}^N (\bar w_i)^j K^{(j)} \left( \frac{x - \mu_i }{h}\right) \right)
	&= \frac{(-1)^j}{j!h^{j+1}} E\left( E\left( (\bar w_i)^j | \mu_i \right) K^{(j)} \left( \frac{x - \mu_i }{h}\right) \right)\\
	&= \frac{(-1)^j}{j!h^j} E( (\bar w_i)^j | \mu_i = x ) f_{\mu}(x)  \int K^{(j)}(s) ds + o\left(\frac{1}{\sqrt{T^j h^{2j}}} \right) \\
	&= O\left(\frac{1}{\sqrt{T^j h^{2j}}}\right),
\end{align*}
where we used $E( (\bar w_i)^j | \mu_i = x) = O(T^{-j/2})$ (see Assumption \ref{as-limit} below and Lemma \ref{lem-wbar} in the supplement).
Noting that $E(\bar w_i| \mu_i) = 0$, the bias caused from the second term in \eqref{eq:issue} can be written as $\sum_{j = 2}^{J-1} O(1/\sqrt{T^j h^{2j}})$.
This bias is negligible when $1/(Th^2) \to 0$, which is identical to the relative magnitude condition $N^2/T^5 \to 0$ when using the standard bandwidth $h \asymp N^{-1/5}$ in the density estimation with second-order kernels.
For the third term in \eqref{eq:issue}, the absolute mean can be evaluated as: 
\begin{align*}
	E\left| \frac{(-1)^J}{J! Nh^{J+1}} \sum_{i=1}^N (\bar w_i)^J K^{(J)} \left( \frac{x - \tilde \mu_i }{h}\right) \right|
	\le \frac{M}{h^{J+1}} E|\bar w_i|^J 
	= O\left( \frac{1}{\sqrt{T^Jh^{2J+2}}} \right),
\end{align*}
where $0 < M < \infty$ denotes a generic positive constant and we use $E|\bar w_i|^J  = O(T^{-J/2})$ (see Lemma \ref{lem-wbar}).
Hence, the third term in \eqref{eq:issue} is $O_p(1/\sqrt{T^J h^{2J+2}})$ by Markov's inequality.
Remarkably, this term does not vanish even when $1/(Th^2) \to 0$ under which the lower-order terms are negligible.
This term can be negligible only if $1/(Th^{2+2/J}) \to 0$, which implies $N^{(2 + 2/J)} / T^5 \to 0$ when $h \asymp N^{-1/5}$.
Note that $N^{(2 + 2/J)} / T^5 \to 0$ is ``stronger'' than $N^2 / T^5$.

The asymptotic investigation above exhibits several unique features.
First, it implies that the relative magnitude condition for consistency (and also that for asymptotic normality) varies in the order of the expansion.
Specifically, we need $1/(Th^{2+2/J}) \to 0$ to achieve the consistency of $\hat f_{\hat \mu}(x)$ based on the $J$-th order expansion.
Second, we can obtain the ``weakest'' relative magnitude condition $1/(Th^2) \to 0$ for consistency, only if we execute the infinite-order expansion (that is, as $J \to \infty$).
Finally, while we can derive the suitable condition $1/(Th^2) \to 0$ via the infinite-order expansion, it requires the existence of higher-order moments of $w_{it}$.
The evaluation based on the infinite-order expansion demands the existence of $E|w_{it}|^j$ for any $j$.
Hence, there is a trade-off between the relative magnitude condition and the existence of higher-order moments.

Asymptotic normality requires a further stronger condition. 
Because the rate of convergence of the kernel estimator is $\sqrt{Nh}$, it requires $N/(T^J h^{2J+1})\to 0$. 
This condition is at best $N^2/T^3 \to 0$, which is obtained under an infinite order expansion with standard bandwidth ($h \asymp N^{-1/5}$). 
Note that, as in the density estimation above, the highest order of the expansion determines the required condition for asymptotic normality. 
Such a very high order of bias cannot be corrected in practice, even though methods to correct the first few orders of bias are available in the long-panel literature (e.g., \citealp{DhaeneJochmans15}).
This result is in stark contrast to the existing studies in which bias correction improves the conditions on the relative magnitudes of $N$ and $T$.

The main reason behind these unique features is that the curvature of the summand (i.e., $K((x - \hat \mu_i)/h)$) depends on the bandwidth $h$.
Roughly speaking, as $h \to 0$, the summand function becomes steeper and more ``nonlinear.'' 
It exacerbates the bias caused by the nonlinearity and it turns out that even a very high order derivative of $K$ affects the bias. 
Alternatively, we may also interpret this problem based on the equation $K((x -\hat \mu_i)/h) = K((x - \mu_i)/h + (\mu_i -\hat \mu_i)/h))$.
The contribution of the error by using the estimated $\hat \mu_i$ is $(\mu_i - \hat \mu_i)/h$ and it increases as $h\to 0$.
Hence, the bias of the density estimator heavily depends on the magnitude of $h$ and the nonlinearity of $K$.

\subsection{Asymptotic biases for the density estimation}

We here formally show the presence of asymptotic biases of the kernel density estimator in \eqref{eq-dens-cdf}.
We conduct asymptotic investigations based on an infinite-order expansion under which the weakest possible condition on the relative magnitude of $N$ and $T$ is obtained.

We assume the following basic conditions for the data-generating process.
These are essentially the same as the assumptions in \citet{Okui2017}.

\begin{assumption}\label{as-basic}
	The sample space of $\alpha_i$ is some Polish space and $y_{it} \in \mathbb{R}$ is a scalar real random variable.
	$\{(\{y_{it}\}_{t=1}^T, \alpha_i)\}_{i=1}^N$ is i.i.d. across $i$.
\end{assumption}

\begin{assumption}\label{as-mixing-c}
	For each $i$, $\{y_{it}\}_{t=1}^{\infty}$ is strictly stationary and $\alpha$-mixing given $\alpha_i$ with mixing coefficients $\{\alpha (m|i)\}_{m=0}^\infty$.
	For any natural number $r_m \in \mathbb{N}$, there exists a sequence $\{ \alpha (m) \}_{m=0}^\infty$ such that for any $i$ and $m$, $\alpha (m|i) \le \alpha (m)$ and $\sum_{m=0}^{\infty} (m+1)^{r_m/2-1} \alpha(m) ^{\delta / (r_m+\delta)} < \infty$ for some $\delta>0$.
\end{assumption}

\begin{assumption}\label{as-w-moment-c}
	For any natural number $r_d \in \mathbb{N}$, it holds that $E|w_{it}|^{r_d+\delta} < \infty$ for some $\delta > 0$.
\end{assumption}

\begin{assumption}\label{as-rho}
	There exists a constant $\epsilon > 0$ such that $\gamma_{0,i} > \epsilon$ almost surely.
\end{assumption}

Assumptions \ref{as-basic} and \ref{as-mixing-c} require that the individual time series given $\alpha_i$ is strictly stationary across time but i.i.d. across units.
The identical distribution across $i$ is essential for our analysis.
The independence assumption across $i$ makes our asymptotic investigations tractable, while the consistency result and the same asymptotic biases could be derived even under weak cross-sectional dependence.
Note that the i.i.d. assumption does not exclude the presence of heterogeneity in panel data.
In our setting, heterogeneity is caused by differences in the realized values of $\{\alpha_i\}_{i=1}^N$ across units.
Assumption \ref{as-mixing-c} also restricts the degree of persistence of the individual time series.
The conditions for stationarity and degree of persistence require that the times series for each unit is not a unit root process and that the initial value of each time series is generated from a stationary distribution.
Assumption \ref{as-w-moment-c} requires the existence of the moments of $w_{it}$, and it allows us to derive the asymptotic biases of the estimators.
While we can develop the theoretical properties of the estimators in situations where Assumptions \ref{as-mixing-c} and \ref{as-w-moment-c} do not hold for some numbers $r_m$ and $r_d$, we cannot derive the higher-order biases based on infinite-order expansions in such situations. 
As a result, in such situations, we demand stronger conditions on the relative magnitudes as discussed in the previous section.
Assumption \ref{as-rho} allows us to derive the asymptotic properties of the kernel estimators for $\rho_{k,i}$.
All of the assumptions can be satisfied in popular panel data models.
For example, they all hold when $y_{it}$ follows a heterogeneous stationary panel autoregressive moving--average model with a Gaussian error term (e.g., $y_{it} = c_i + \phi_i y_{i,t-1} + u_{it} + \theta_i u_{i,t-1}$ with $u_{it} \sim \mathcal{N}(0, \sigma^2)$).

We also assume the following additional conditions.

\begin{assumption}\label{as-kernel}
	The kernel function $K:\mathbb{R} \to \mathbb{R}$ is bounded, symmetric, and infinitely differentiable. 
	It satisfies $\int K(s) ds = 1$, $\int |K^{(j)}(s)| ds < \infty$, $\int |s K^{(j)}(s)| ds < \infty$, $\int |s^2 K^{(j)}(s)| ds < \infty$, and $\int |s^3 K^{(j)}(s)| ds < \infty$ for any nonnegative integer $j$.
\end{assumption}

Assumption \ref{as-kernel} includes the standard conditions for the kernel function, except for infinite differentiability.
We require the differentiability in order to expand the kernel estimator for the estimated $\hat \xi_i$ at the true $\xi_i$ based on the infinite-order expansion.
Note that the symmetry of $K$ implies that $\int K^{(j)}(s) d s = 0$ for any odd $j$.

\begin{assumption}\label{as-density-kernel}
	The random variables $\mu_i \in \mathbb{R}$, $\gamma_{k,i} \in \mathbb{R}$, and $\rho_{k,i} \in (-1,1)$ are continuously distributed.
	The densities $f_{\xi}$ with $\xi=\mu$, $\gamma_k$, and $\rho_k$ are bounded away from zero near $x$ and three-times boundedly continuously differentiable near $x$.
\end{assumption}

Assumption \ref{as-density-kernel} requires that $\xi_i$ is continuously distributed without probability mass.
The continuity of the random variable is essential for implementing kernel-smoothing estimation as it rules out situations where there is no heterogeneity for $\xi_i$ (that is, the situation where $\xi_i = \xi$ for any $i$ with some constant $\xi$) and where there is finitely grouped heterogeneity (that is, $\xi_{i_1} = \xi_{i_2}$ for any $i_1, i_2 \in \mathbb{I}_g$ with some sets $\mathbb{I}_1, \mathbb{I}_2, \dots, \mathbb{I}_G$ satisfying $ \bigoplus_{g=1}^G \mathbb{I}_g = \{1, 2, \dots, N\}$).

\begin{assumption}\label{as-limit}
	The following functions are twice boundedly continuously differentiable near $x$ for any $T \in \mathbb{N}$ with finite limits at $x$ as $T \to \infty$:
	\begin{align*}
		& \sqrt{T^j} E\left( (\bar w_i)^j \middle| \mu_i =\cdot \right), 
		\qquad 
		\sqrt{T^j} E \left( (\bar w_i)^j \middle| \gamma_{k,i} = \cdot \right), 
		\qquad 
		\sqrt{T^j} E \left( (\bar w_i)^j \middle| \rho_{k,i} =\cdot \right), \\ 
		& \frac{1}{\sqrt{T^j}} E \left( \left( \sum_{t=k+1}^T (w_{it}w_{i,t-k}-\gamma_{k,i}) \right)^j \middle| \gamma_{k,i}=\cdot \right),\\
		& \frac{1}{\sqrt{T^{j_1 + j_2}}} E\left( \left( \sum_{t=k+1}^T (w_{it}w_{i,t-k}-\gamma_{k,i}) \right)^{j_1} \left( \sum_{t=1}^T (w_{it}^2-\gamma_{0,i}) \right)^{j_2} \frac{\gamma_{k,i}^{j_3}}{\gamma_{0,i}^{j_4}}  \middle| \rho_{k,i}=\cdot \right),
	\end{align*}
	for any nonnegative integers $j, j_1, j_2, j_3, j_4$.
\end{assumption}

Assumption \ref{as-limit} states the existence and smoothness of the conditional expectations.
This assumption allows us to derive the exact forms of the asymptotic biases.
The convergence rates of the terms are standard and guaranteed by Lemmas \ref{lem-wbar} and \ref{lem-wk} in Appendix \ref{sec-lemma}.
For example, the assumption requires that $T \cdot E((\bar w_i)^2 | \mu_i = \cdot) = O(1)$ and the convergence rate is consistent with the result in Lemma \ref{lem-wbar}.

The following theorem shows that the kernel density estimators are consistent and asymptotically normal but exhibit asymptotic biases. 
While the theorem assumes an infinite-order Taylor expansion and the summability of the infinite series of the asymptotic biases directly, we can show their validity under unrestrictive regularity conditions.
Because these discussions are highly technical and demand lengthy explanations, they appear in Appendices \ref{sec-expansion} and \ref{sec-series}.

\begin{theorem}\label{thm-dens}
	Let $x \in \mathbb{R}$ be an interior point in the support of $\xi_i = \mu_i$, $\gamma_{k,i}$, or $\rho_{k,i}$.
	Suppose that Assumptions \ref{as-basic}, \ref{as-mixing-c}, \ref{as-w-moment-c}, \ref{as-kernel}, \ref{as-density-kernel}, and \ref{as-limit} hold.
	In addition, if $\xi_i = \rho_{k, i}$, suppose that Assumption \ref{as-rho} also holds.
	Suppose that the infinite-order Taylor expansion of $\hat f_{\hat \xi}(x) = (Nh)^{-1} \sum_{i=1}^N K((x - \hat \xi_i) / h)$ at $\xi_i$ holds and that the infinite series of the asymptotic biases below is well defined.
	When $N,T \to \infty$ and $h \to 0$ with $Nh \to \infty$, $Nh^5 \to C \in [0, \infty)$, and $T h^2 \to \infty$, it holds that:
	\begin{align*}
		\hat f_{\hat \xi}(x) - f_{\xi} (x) = \frac{1}{Nh} \sum_{i=1}^{N} K\left( \frac{x - \xi_i }{h} \right) - f_{\xi}(x) + \frac{A_{\xi, 1}(x)}{T} +  \frac{A_{\xi, 2}(x)}{Th^2} + \sum_{j = 3}^{\infty} \frac{A_{\xi, j}(x)}{\sqrt{T^j h^{2j}}} + o_p\left( \frac{1}{Th^2} \right),
	\end{align*}
	where $A_{\xi, j}(x)$ is a nonrandom bias term that depends on $x$ and satisfies $A_{\mu, 1}(x) = 0$ for any $x$ (the formula of $A_{\xi, j}(x)$ is given in the proof).
	As a result, when $N / (T^3 h^5) \to 0$ also holds, it holds that:
	\begin{align*}
		\sqrt{Nh} \left( \hat f_{\hat \xi}(x) - f_{\xi} (x) -  h^2 \frac{\kappa_1 f_{\xi}^{''}(x)}{2} - \frac{A_{\xi, 1}(x)}{T} - \frac{A_{\xi, 2}(x)}{Th^2} \right) \stackrel{d}{\longrightarrow} \mathcal{N} \big( 0,  \kappa_2 f_{\xi}(x) \big).
	\end{align*}
\end{theorem}

The density estimator can be written as the sum of the infeasible estimator based on the true $\xi_i$, say $\hat f_{\xi}(x) \coloneqq (Nh)^{-1} \sum_{i=1}^N K((x - \xi_i)/h)$, and the asymptotic biases.
The convergence rate of the estimator is the standard order of $O_p(1/ \sqrt{Nh})$, and the asymptotic distribution is the same as that of the infeasible estimator $\hat f_{\xi}(x)$.
However, the feasible estimator exhibits asymptotic biases. 
These results also require the relative magnitude conditions of $N$, $T$, and $h$; that is, $1/(Th^2) \to 0$ and $N / (T^3 h^5) \to 0$ for consistency and asymptotic normality, respectively.

The density estimator for $\mu_i$ has two main asymptotic biases given $A_{\mu, 1}(x) = 0$, but the density estimators for $\gamma_{k,i}$ and $\rho_{k,i}$ have three main asymptotic biases, in addition to the higher-order biases.
The first bias of the form $h^2 \kappa_1 f_\xi''(x)/2$ is the standard kernel-smoothing bias.
The second bias of the form $A_{\xi, 1}(x) / T$ is the incidental parameter bias caused from estimating $\gamma_{k,i}$ and $\rho_{k,i}$ by $\hat \gamma_{k,i}$ and $\hat \rho_{k,i}$, respectively.
The estimation of $\hat \gamma_{k,i}$ and $\hat \rho_{k,i}$ involves estimating $\mu_i$ by $ \hat \mu_i = \bar y_i$ for each $i$, which becomes a source of the incidental parameter bias.
The third bias of the form $A_{\xi, 2}(x) / (Th^2)$ is the second-order nonlinearity bias caused by expanding $K((x - \hat \xi_i )/h)$ for $K(( x - \xi_i )/h)$ by Taylor expansion.
Moreover, the $j$-th order nonlinearity bias exhibits the form $A_{\xi, j}(x) / \sqrt{T^j h^{2j}}$ for $j \ge 3$.

We need the two conditions, $1/(Th^2) \to 0$ and $N / (T^3 h^5) \to 0$, to ensure the asymptotic negligibility of the higher-order nonlinearity biases.
If we use the standard bandwidth $h \asymp N^{-1/5}$ with second-order kernels, the conditions $1 / (T h^2) \to 0$ and $N / (T^3 h^5) \to 0$ imply that $N^2 / T^5 \to 0$ and that $N^2 / T^3 \to 0$, respectively, which are integrated to $N^2 / T^3 \to 0$. 
Note that while the incidental parameter bias and the second-order nonlinearity bias are also asymptotically negligible under these conditions, the practical magnitudes of these biases would be larger than those of the higher-order nonlinearity biases.

We have already discussed the source of the nonlinearity bias in Section \ref{sec-unique} so here we provide a slightly more detailed discussion of the incidental parameter bias. 
It does not appear in $\hat f_{\hat \mu}(x)$ because the estimation error in $\hat \mu_i$ (that is, $\bar w_i$) has zero mean. 
However, errors in $\gamma_{k,i}$ and $\rho_{k,i}$ are not mean-zero. 
For example, $\hat \gamma_{0,i} = \sum_{t=1}^T (y_{it} - \bar y_i)^2 /T = \gamma_{0,i} + \sum_{t=1}^T (w_{it}^2 - \gamma_{0,i}) /T - (\bar w_i)^2$ and $(\bar w_i)^2$ is not mean-zero although it converges to zero at the rate $1/T$. 
This is the source of the incidental parameter bias $A_{\gamma_0,1}/T$ and the order $1/T$ comes from the fact that $(\bar w_i)^2$ is $O_p(1/T)$.

\begin{remark} \label{rem-diff}
	Some might surmise that our kernel smoothing requires a ``weaker'' condition, such as $Nh / T^2 \to 0$, than the condition $N/T^2 \to 0$ in the existing literature because the kernel estimation is essentially taking the average number of observations in a local neighborhood that contains $Nh$ observations on average. 
	However, the above theorem clarifies that such conjecture is not true.
	The failure of the conjecture stems from the fact that, as $h\to 0$, the summands, $K ((x - \hat \xi_i)/h)$, become more nonlinear, which increases the nonlinear biases and necessitates imposing a stronger assumption to ignore higher-order nonlinear biases. 
\end{remark}

\begin{remark}
	When using higher-order kernels, the relative magnitude conditions of $N$ and $T$ for consistency and asymptotic normality are altered.
	For example, when using fourth-order kernels, the optimal bandwidth is $h \asymp N^{-1/9}$ (see, e.g.,  \citealp[Section 1.11]{LiRacine2007}).
	Then, the conditions $1/(Th^2) \to 0$ and $N/(T^3 h^5) \to 0$ are identical to $N^2 / T^9 \to 0$ and $N^{14} / T^{27} \to 0$, respectively, which are weaker than the relative magnitude conditions with second-order kernels.
	Thus, one may employ higher-order kernels especially when $T$ is much smaller than $N$.
	Nonetheless, our Monte Carlo simulations observe that the performance of the jackknife bias-corrected estimator with a second-order kernel is satisfactory even when $T$ is small.
\end{remark}

\subsection{Split-panel jackknife bias correction for density estimation}

As the incidental parameter bias and the nonlinearity biases in $\hat f_{\hat \xi}(x)$ may be severe in practice, we propose adoption of the split-panel jackknife to correct them.
Among split-panel jackknifes, here we consider half-panel jackknife (HPJ) bias correction.
For simplicity, suppose that $T$ is even.\footnote{The bias correction with odd $T$ is similar.
	See \citet[page 999]{DhaeneJochmans15} for details.
}
For $\xi_i = \mu_i$, $\gamma_{k,i}$, or $\rho_{k,i}$, we obtain the estimators $\hat f_{\hat \xi,(1)}(x)$ and $\hat f_{\hat \xi,(2)}(x)$ of $f_\xi(x)$ based on two half-panel data $\{\{y_{it}\}_{t=1}^{T/2}\}_{i=1}^N$ and $\{\{y_{it}\}_{t=T/2+1}^T \}_{i=1}^N$, respectively.
The HPJ bias-corrected estimator is $\hat f_{\hat \xi}^H(x) \coloneqq \hat f_{\hat \xi}(x) - (\bar f_{\hat \xi}(x) - \hat f_{\hat \xi}(x) )$ where $\bar f_{\hat \xi}(x) \coloneqq [\hat f_{\hat \xi,(1)}(x)+ \hat f_{\hat \xi,(2)}(x)]/2$.
The term $\bar f_{\hat \xi}(x) - \hat f_{\hat \xi}(x)$ estimates the bias in the original estimator $\hat f_{\hat \xi}(x)$.
Importantly, the bandwidths for computing $\hat f_{\hat \xi,(1)}(x)$ and $\hat f_{\hat \xi,(2)}(x)$ must be the same as that for the original estimator $\hat f_{\hat \xi}(x)$ to reduce the biases.

The next theorem formally shows that the HPJ bias-corrected estimator $\hat f_{\hat \xi}^H(x)$ does not suffer from incidental parameter bias and second-order bias, and does not alter the asymptotic variance of the estimator.

\begin{theorem}\label{thm-dens-HPJ}
	Suppose that the assumptions in Theorem \ref{thm-dens} hold.
	When $N,T \to \infty$ and $h \to 0$ with $Nh \to \infty$, $Nh^5 \to C \in [0, \infty)$, $Th^2 \to \infty$, and $N / (T^3 h^5) \to 0$, it holds that:
	\begin{align*}
		\sqrt{Nh} \left( \hat f_{\hat \xi}^H(x) - f_{\xi} (x) -  h^2 \frac{\kappa_1 f_{\xi}^{''}(x)}{2} \right) \stackrel{d}{\longrightarrow} \mathcal{N} \big( 0,  \kappa_2 f_{\xi}(x) \big).
	\end{align*}
\end{theorem}

Note that HPJ bias correction does not weaken the relative magnitude condition of $N$, $T$, and $h$ for asymptotic normality in Theorem \ref{thm-dens}; that is, $N / (T^3 h^5) \to 0$.
This is because HPJ bias correction cannot eliminate higher-order nonlinearity biases.
This result is in stark contrast to the existing literature where bias correction typically weakens the condition on the relative magnitudes of $N$ and $T$ \citep[see, e.g.,][]{DhaeneJochmans15}.

\begin{remark}
	We can also consider higher-order jackknifes to eliminate higher-order biases as in \citet{DhaeneJochmans15} and \citet{Okui2017}.
	For example, we can consider the third-order jackknife (TOJ) in the same manner as in \citet{Okui2017}, which is slightly different from the original TOJ in \citet{DhaeneJochmans15} because both studies treat different higher-order biases.
	We investigate its performance by Monte Carlo simulations in the appendix, which shows that the TOJ can work better than the HPJ, especially when the naive estimator without bias correction exhibits a large bias.
	Hence, for practical situations, we recommend the adoption of higher-order jackknifes as well as HPJ bias correction.
\end{remark}

\begin{remark}
	The half-series jackknife to correct bias of each $\hat \gamma_{k, i}$ proposed by \citet{Quenouille1949,Quenouille1956} cannot be employed for reducing the nonlinearity biases such as $A_{\gamma_k, 2}(x) / (Th^2)$.
	It corrects bias in $\hat \gamma_{k, i}$ due to estimating $\mu_i$ by $\hat \mu_i = \bar y_i$  for each $i$, so that it can reduce only the incidental parameter bias $A_{\gamma_k, 1}(x) / T$.
\end{remark}

\subsection{Confidence interval and bandwidth selection for density estimation}

This section considers CI estimation and the selection of optimal bandwidth for density estimation.

\paragraph{CI estimation.}
We propose to apply the RBC procedure in \citet{calonico2018effect} for CI estimation.
It allows us to construct a valid $1 - \alpha$ CI of $f_{\xi}(x)$ while correcting the kernel-smoothing bias $\mathcal{B}_{\xi}(x) \coloneqq h^2 \kappa_1 f_{\xi}^{''}(x) / 2$. 

The RBC procedure based on the naive estimator $\hat f_{\hat \xi}(x)$ is almost the same as the original procedure in \citet{calonico2018effect}.
We first note that the kernel-smoothing bias can be estimated by $\hat{\mathcal{B}}_{\hat \xi}(x) \coloneqq  h^2 \kappa_1 \hat f_{\hat \xi}''(x)$ where $\hat f_{\hat \xi}''(x) \coloneqq (Nb^3)^{-1} \sum_{i=1}^{N} L''((x - \hat \xi_i)/b)$ with a kernel function $L$ and bandwidth $b \to 0$.
Then, the estimator that corrects the kernel-smoothing bias is:
\begin{align*}
	\hat f_{\hat \xi}(x) - \hat{\mathcal{B}}_{\hat \xi}(x) 
	= \frac{1}{Nh} \sum_{i=1}^{N} \left( K\left(\frac{x - \hat \xi_i}{h}\right) - \kappa_1 \lambda^3 L''\left(\frac{x - \hat \xi_i}{b}\right) \right)
	\eqqcolon \frac{1}{Nh} \sum_{i=1}^{N} \mathcal{K}_i(x),
\end{align*}
where $\lambda \coloneqq h / b$.
Choosing $b$ such that $\lambda \to c$ for some $c \in (0, \infty)$ enables us to capture variance inflation caused by the bias correction while successfully removing the kernel-smoothing bias.
In practice, one can set $\lambda = 1$ by following the suggestion in \citet{calonico2018effect}.
The RBC $t$ statistic is given by:
\begin{align*}
	T_{RBC}(x) \coloneqq \frac{[\hat f_{\hat \xi}(x) - \hat{\mathcal{B}}_{\hat \xi}(x)] - f_{\xi}(x)}{\hat \sigma_{RBC}(x)},
\end{align*}
where $\hat \sigma_{RBC}^2(x)$ is the estimator of the nonasymptotic variance of $\hat f_{\hat \xi}(x) - \hat B_{\hat \xi}(x)$:
\begin{align*}
	\hat \sigma_{RBC}^2(x) \coloneqq \frac{1}{Nh^2} \left[ \frac{1}{N} \sum_{i=1}^{N} \mathcal{K}_i^2(x) - \left( \frac{1}{N} \sum_{i=1}^{N} \mathcal{K}_i(x) \right)^2  \right].
\end{align*}
It holds that $T_{RBC}(x) \stackrel{d}{\longrightarrow} \mathcal{N}(0, 1)$ under similar conditions in Theorem \ref{thm-dens}, so that we can construct the $1 - \alpha$ CI of $f_{\xi}(x)$ in the usual manner.

The RBC procedure based on the split-panel jackknife bias-corrected estimator demands some modifications.
To see this, the HPJ bias-corrected estimator that also reduces the kernel-smoothing bias can be written as follows:
\begin{align*}
	\hat f_{\hat \xi}^H(x) - \hat{\mathcal{B}}_{\hat \xi}(x)
	&= \left[ 2 \hat f_{\hat \xi}(x) - \frac{1}{2} \left( \hat f_{\hat \xi}^{(1)}(x) + \hat f_{\hat \xi}^{(2)}(x) \right) \right] - \hat{\mathcal{B}}_{\hat \xi}(x)\\
	&= \frac{1}{Nh} \sum_{i=1}^N \left[ 2 K \left( \frac{x-\hat \xi_i}{h} \right) - \frac{1}{2} \left( K \left( \frac{x - \hat \xi_i^{(1)}}{h} \right) + K \left( \frac{x - \hat \xi_i^{(2)}}{h} \right)\right) - \kappa_1 \lambda^3 L''\left(\frac{x - \hat \xi_i}{b}\right) \right]\\
	&\eqqcolon \frac{1}{Nh} \sum_{i=1}^n \mathcal{K}_i^H(x),
\end{align*}
where $\hat \xi_i^{(1)}$ and $\hat \xi_i^{(2)}$ are the estimators based on the half-series $\{y_{it}\}_{t=1}^{T/2}$ and $\{y_{it}\}_{t=T/2+1}^T$, respectively.
Then, the nonasymptotic variance of $\hat f_{\hat \xi}^H(x) - \hat{\mathcal{B}}_{\hat \xi}(x)$ can be estimated by:
\begin{align*}
	(\hat \sigma_{RBC}^H(x))^2 \coloneqq \frac{1}{Nh^2} \left[ \frac{1}{N} \sum_{i=1}^{N} (\mathcal{K}_i^H(x))^2- \left( \frac{1}{N} \sum_{i=1}^{n} \mathcal{K}_i^H(x) \right)^2 \right].
\end{align*}
As a result, the RBC $t$ statistic based on the HPJ estimator is:
\begin{align*}
	T_{RBC}^H(x) \coloneqq \frac{[\hat f_{\hat \xi}^H(x) - \hat{\mathcal{B}}_{\hat \xi}(x)] - f_{\xi}(x)}{\hat \sigma_{RBC}^H(x)}.
\end{align*}
Note that $\hat \sigma_{RBC}^H(x)$ is different from $\hat \sigma_{RBC}(x)$ above because the former also captures the finite-sample variability of HPJ bias correction.
We can construct the $1 - \alpha$ CI of $f_{\xi}(x)$ based on $T_{RBC}^H(x)$ in the usual manner.
We can also consider similar RBC procedures based on higher-order split-panel jackknife bias correction.

\begin{remark}
	Undersmoothing is often used to construct CI for the kernel density estimator. 
	However, it is not desirable in our context. 
	Undersmoothing means that we use bandwidth that converges faster than $N^{-1/5}$ so that the smoothing bias does not appear in the asymptotic distribution. 
	In our setting, the smaller is the bandwidth, the larger is the higher-order nonlinearity bias, which in turn calls for a stronger assumption on the relative magnitude of $N$ and $T$. 
	We thus prefer the method based on \citet{calonico2018effect} because we can still use the bandwidth of order $N^{-1/5}$. 
	Note also that \citet{calonico2018effect} demonstrate that their method provides better coverage than undersmoothing. 
\end{remark}

\paragraph{Bandwidth selection.}

We can select the bandwidth $h$ for the density estimation using any standard procedures based on the estimated $\hat \xi_i$.
This is because Theorem \ref{thm-dens-HPJ} shows that the AMSE and asymptotic distribution of the HPJ bias-corrected estimator $\hat f_{\hat \xi}^H(x)$ are identical to those of the infeasible estimator $\hat f_{\xi}(x) = (Nh)^{-1} \sum_{i=1}^N K((x - \xi_i)/h)$.
In our application and Monte Carlo simulations, we apply the coverage error optimal bandwidth selection procedure in \citet{calonico2018effect} because of its desirable properties as shown in the paper. Furthermore, their bandwidth tends to be larger than the bandwidth that minimizes AMSE and would be more suitable in our context because a larger bandwidth makes the nonlinearity biases smaller. 
Our Monte Carlo simulations also confirm the appropriate finite-sample properties of the procedure.

\subsection{Asymptotic biases for CDF estimation}

In this section, we consider the smoothed CDF estimator and derive its asymptotic biases.
The CDF $F_{\xi}(x) \coloneqq \Pr(\xi_i \le x)$ can be estimated by integrating the kernel density estimator:
$\hat F_{\hat \xi}(x) = \int^{x}_{-\infty} \hat f_{\hat \xi}(v) dv$. 
It is convenient to write this kernel CDF estimator as: 
\begin{align*}
	\hat F_{\hat \xi} (x) \coloneqq \frac{1}{N} \sum_{i=1}^N \mathbb{K} \left( \frac{x - \hat \xi_i}{h}\right),
\end{align*}
where $x \in \mathbb{R}$ is a fixed point, and $\mathbb{K}:\mathbb{R} \to [0,1]$ is a Borel-measurable CDF (or $\mathbb{K} (a) = \int^{x}_{-\infty} K(v) dv$).

For the CDF estimation, we need the following condition instead of Assumption \ref{as-density-kernel}.
The continuity of the random variable is essential, even for the kernel-smoothing CDF estimation.

\begin{assumption}
	\label{as-cdf}
	The random variables $\mu_i \in \mathbb{R}$, $\gamma_{k,i} \in \mathbb{R}$, and $\rho_{k,i} \in (-1,1)$ are continuously distributed.
	The CDFs $F_{\xi}$ with $\xi = \mu$, $\gamma_k$, and $\rho_k$ are three-times boundedly continuously differentiable near $x$.
\end{assumption}

The following theorem shows the presence of asymptotic biases for the kernel CDF estimator.

\begin{theorem}\label{thm-CDF}
	Let $x \in \mathbb{R}$ be an interior point in the support of $\xi_i = \mu_i$, $\gamma_{k,i}$, or $\rho_{k,i}$.	
	Suppose that Assumptions \ref{as-basic}, \ref{as-mixing-c}, \ref{as-w-moment-c}, \ref{as-kernel}, \ref{as-limit}, and \ref{as-cdf} hold.
	In addition, if $\xi_i = \rho_{k, i}$, suppose that Assumption \ref{as-rho} also holds.
	Suppose that the infinite-order Taylor expansion of $\hat F_{\hat \xi}(x) = N^{-1} \sum_{i=1}^N \mathbb{K}((x - \hat \xi_i) / h)$ at $\xi_i$ holds and that the infinite series of the asymptotic biases below is well defined.
	When $N,T \to \infty$ and $h \to 0$ with $Nh^3 \to C \in [0, \infty)$ and $T h^2 \to \infty$, it holds that:
	\begin{align*}
		\hat F_{\hat \xi}(x) - F_{\xi}(x) = \frac{1}{N} \sum_{i=1}^N \mathbb{K} \left( \frac{x - \xi_i}{h} \right) - F_{\xi}(x) + \frac{B_{\xi, 1}(x)}{T} + \frac{B_{\xi, 2}(x)}{T} + \sum_{j=3}^{\infty} \frac{B_{\xi, j}(x)}{\sqrt{T^j h^{2j - 2}}} + o_p\left( \frac{1}{\sqrt{T^3 h^4}} \right),
	\end{align*}
	where $B_{\xi, j}(x)$ is a nonrandom bias term that depends on $x$ and that satisfies $B_{\mu, 1}(x) = 0$ for any $x$ (the formula of $B_{\xi, j}(x)$ is given in the proof).
	As a result, when $N / (T^3 h^4) \to 0$ also holds, it holds that:
	\begin{align*}
		\sqrt{N} \left( \hat F_{\hat \xi}(x) - F_{\xi} (x) - \frac{B_{\xi, 1}(x)}{T} - \frac{B_{\xi, 2}(x)}{T}  \right) \stackrel{d}{\longrightarrow} \mathcal{N} \big( 0, F_{\xi}(x) [1-F_{\xi}(x)] \big).
	\end{align*}
\end{theorem}

The CDF estimator can be rearranged as the sum of the infeasible estimator based on the true $\xi_i$ and the asymptotic biases.
We present the result based on an infinite-order expansion because it yields the best possible condition of the relative magnitudes of $N$ and $T$ but it requires the validity of the infinite-order expansion, in particular the summability of the infinite series and they hold under technical regularity conditions as in the case of the density estimation in Theorem \ref{thm-dens}.
The biases of the forms $B_{\xi, 1}(x) / T$ and $B_{\xi, 2}(x) / T$ are the incidental parameter bias and the second-order nonlinearity bias, respectively.
Note that $\hat F_{\hat \mu} (x)$ does not exhibit the incidental parameter bias as in the case of the density estimation.
We also note that the standard kernel-smoothing bias of order $O(h^2)$ does not exist under asymptotic normality because it is asymptotically negligible under $Nh^3 \to C$ 
(see Lemma \ref{lem-CDF} in Appendix \ref{sec-lemma}).
Consistency and asymptotic normality require the conditions $1 / (T h^2) \to 0$ and $N / (T^3 h^4) \to 0$, respectively, which asymptotically eliminate the higher-order biases.
When using the standard bandwidth $h \asymp N^{-1/3}$ in the CDF estimation with second-order kernels, the conditions $1 / (T h^2) \to 0$ and $N / (T^3 h^4) \to 0$ are the same as $N^2 / T^3 \to 0$ and $N^7 / T^9 \to 0$, respectively, which are integrated to $N^7 / T^9 \to 0$.
Note that we can weaken the relative magnitude condition by using higher-order kernels, which leads to a larger bandwidth, as in the density estimation.

The relative magnitude conditions for the kernel CDF estimation with second-order kernels are stronger than those for the empirical CDF estimation in \citet{JochmansWeidner2019} and \citet{Okui2017}.
The empirical CDF estimation is also easier to implement in practice.
Hence, one should probably employ empirical CDF estimation in practice, and here we do not explore split-panel jackknife, CI estimation, and bandwidth selection for the kernel CDF estimation (although they are feasible).
Nonetheless, the asymptotic biases for the kernel CDF estimation in Theorem \ref{thm-CDF} are new in the literature, and they would be interesting in their own right.

\begin{remark}
	The bias of order $O(1/T)$ for $\hat{F}_{\hat \mu} (x)$ corresponds to the result in \citet{JochmansWeidner2019}. 
	They derive the asymptotic bias of the empirical distribution under Gaussian errors. 
	Suppose that $\hat \mu_i \sim \mathcal{N}( \mu_i, \sigma_i^2 /T)$ as in \citet{JochmansWeidner2019}. 
	Note that $B_{\mu, 1} (x) =0$.
	The formula for $B_{\mu ,2} (x)$ is available in the proof of Theorem \ref{thm-CDF} and becomes $0.5 \cdot \partial ( E(\sigma_i^2 | \mu_i = x) f_{\mu} (x) ) / \partial x $ in this case.\footnote{To derive this result, note that integration by parts can lead to $\int s K'(s) ds = [sK(s)]_{-\infty}^{\infty} - \int K(s) ds = -1$.}
	It is identical to the bias formula in \citet{JochmansWeidner2019}.
	Note that the bias of order $O(1/T)$ for $\hat{F}_{\hat \gamma_{k}} (x)$ and $\hat{F}_{\hat \rho_{k}} (x)$ includes $B_{\gamma_k, 1}(x)$ and $B_{\rho_k, 1}(x)$, respectively, which do not appear in \citet{JochmansWeidner2019}.
\end{remark}

\begin{remark}
	While we obtain the same bias formula of order $O(1/T)$ for the CDF estimator for $\hat \mu_i$ as that in \citet{JochmansWeidner2019}, it is still not clear whether bias formulas including higher-order terms correspond to each other in both papers. 
	The empirical CDF can be regarded as the kernel CDF by letting $h\to 0$ in the given sample (i.e., when $h \to 0$ while keeping $N$ and $T$ fixed). 
	However, the higher-order biases of our kernel CDF are derived in the joint asymptotics (i.e., $N, T \to \infty$ and $h \to 0$) and explode as $h\to 0$, so that it is not trivial how those higher-order asymptotic biases contribute as $h\to 0$, while keeping $N$ and $T$ fixed (or as $N, T \to \infty$ after $h \to 0$).
	Therefore, although we obtain the same bias formulas of order $O(1/T)$, we still hesitate to conclude definitely that our bias formula, including higher-order terms, corresponds exactly to that in \citet{JochmansWeidner2019}.  
\end{remark}

\section{Empirical application} \label{sec-application}
We apply our procedure to panel data on prices of items in US cities.
Our procedure allows us to examine the heterogeneous properties of the deviations of prices from the LOP across items and cities, and the difference in the degree of heterogeneity between goods and services.

Many empirical studies examine the heterogeneous properties of the level and variance of price deviations and the speed of price adjustment toward the long-run LOP deviation (see \citealp{AndersonVanWincoop04} for a review).
For example, \cite{EngelRogers01}, \cite{ParsleyWei01}, and \cite{CruciniShintaniTsuruga15} examine such heterogeneous properties and find that the LOP deviation dynamics are significantly heterogeneous across items and cities based on regression models.
Our investigation below complements such empirical analyses by using our model-free procedure, as it provides visual information concerning the degree of heterogeneity.

We estimate the densities of the mean $\mu_i$, variance $\gamma_{0,i}$, and first-order autocorrelation $\rho_{1,i}$.
We use the Epanechnikov kernel with the coverage error optimal bandwidth in \citet{calonico2018effect}.\footnote{We also observed similar results with different kernels and different bandwidths.}
The codes to compute the CIs and the optimal bandwidths are developed based on the {\ttfamily nprobust} package for {\ttfamily R} \citep{nprobust}.

\paragraph{Data.}
We use data from the American Chamber of Commerce Researchers Association Cost of Living Index produced by the Council of Community and Economic Research.\footnote{Mototsugu Shintani kindly provided us with the data set ready for analysis.}
The same data set is used by \cite{ParsleyWei96}, \cite{YazganYilmazkuday11}, \cite{CruciniShintaniTsuruga15}, \cite{LeeOkuiShintani13}, and \citet{Okui2017}.
The data set contains quarterly price series of 48 consumer price index items (goods and services) for 52 US cities from 1990Q1 to 2007Q4.\footnote{While the original data source contains price information for more items in additional cities, we restrict the observations to obtain a balanced panel data set, as in \cite{CruciniShintaniTsuruga15}.} 
The categorization of goods and services can be found in \citet[Table 2]{Okui2017}.

We define the LOP deviation for item $k$ in city $i$ at time $t$ as $y_{i k t}=\ln P_{i k t}-\ln P_{0kt}$, where $P_{ikt}$ is the price of item $k$ in city $i$ at time $t$ and $P_{0kt}$ is that for the benchmark city of Albuquerque, NM.
We regard each item--city pair as a cross-sectional unit, such that we focus on the degree of heterogeneity of the LOP deviations across item--city pairs. 
The number of cross-sectional units is $N = 48 \times (52 - 1) = 2448$ and the length of the time series is $T = 18 \times 4 = 72$.

\begin{figure}[!t]
	\begin{minipage}{0.33\hsize}
		\includegraphics[width=55mm, bb = 0 0 841 595]{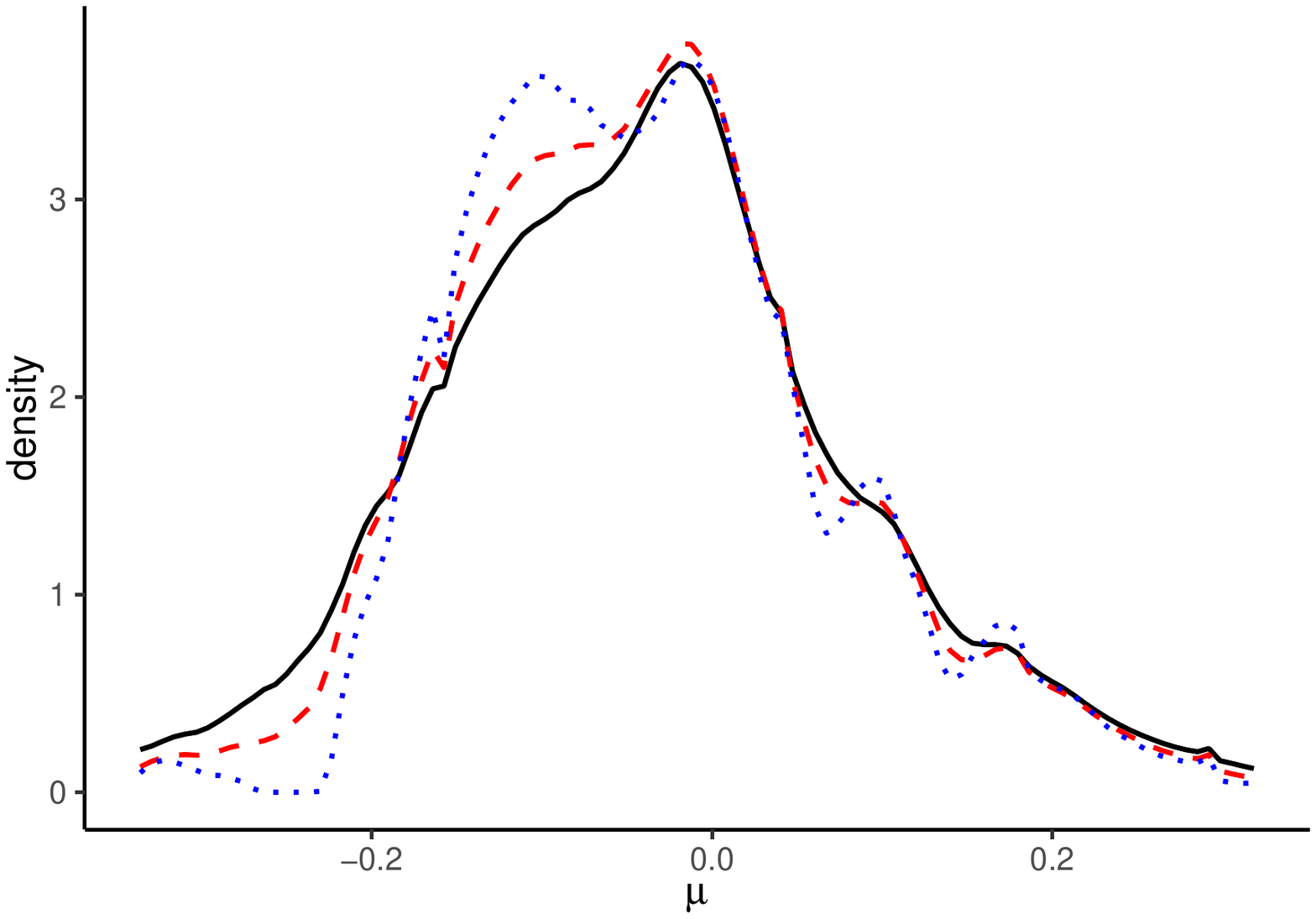}
	\end{minipage}
	\begin{minipage}{0.33\hsize}
		\includegraphics[width=55mm, bb = 0 0 841 595]{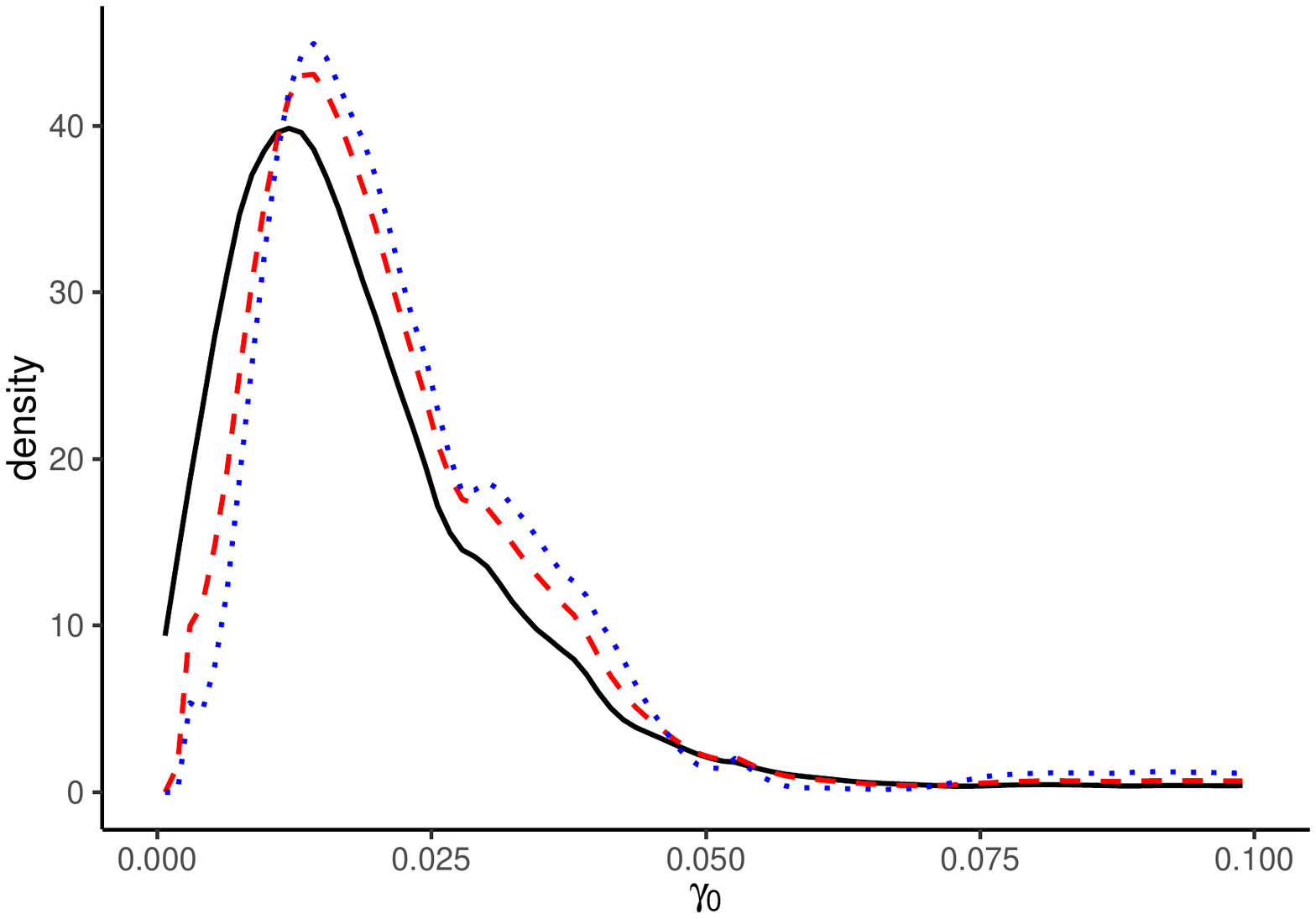}
	\end{minipage}
	\begin{minipage}{0.33\hsize}
		\includegraphics[width=55mm, bb = 0 0 841 595]{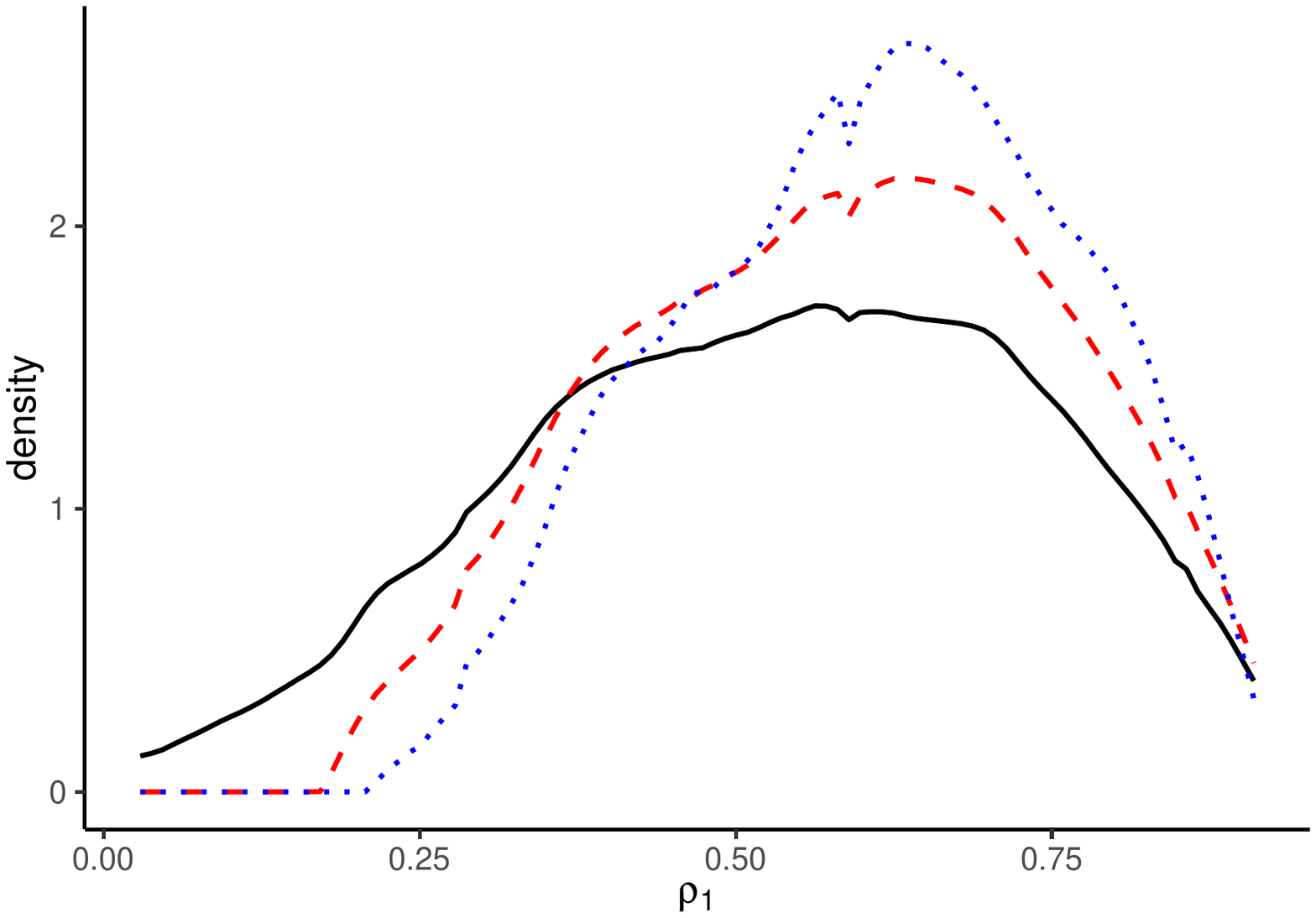}
	\end{minipage}
	\caption{The densities of $\mu$, $\gamma_0$, and $\rho_1$. 
		In each figure, the solid black line indicates the estimates without split-panel jackknife correction, the dashed red line shows the HPJ bias-corrected estimates, and the dotted blue line shows the TOJ bias-corrected estimates.}
	\label{fig:density}
	\begin{minipage}{0.33\hsize}
		\includegraphics[width=55mm, bb = 0 0 841 595]{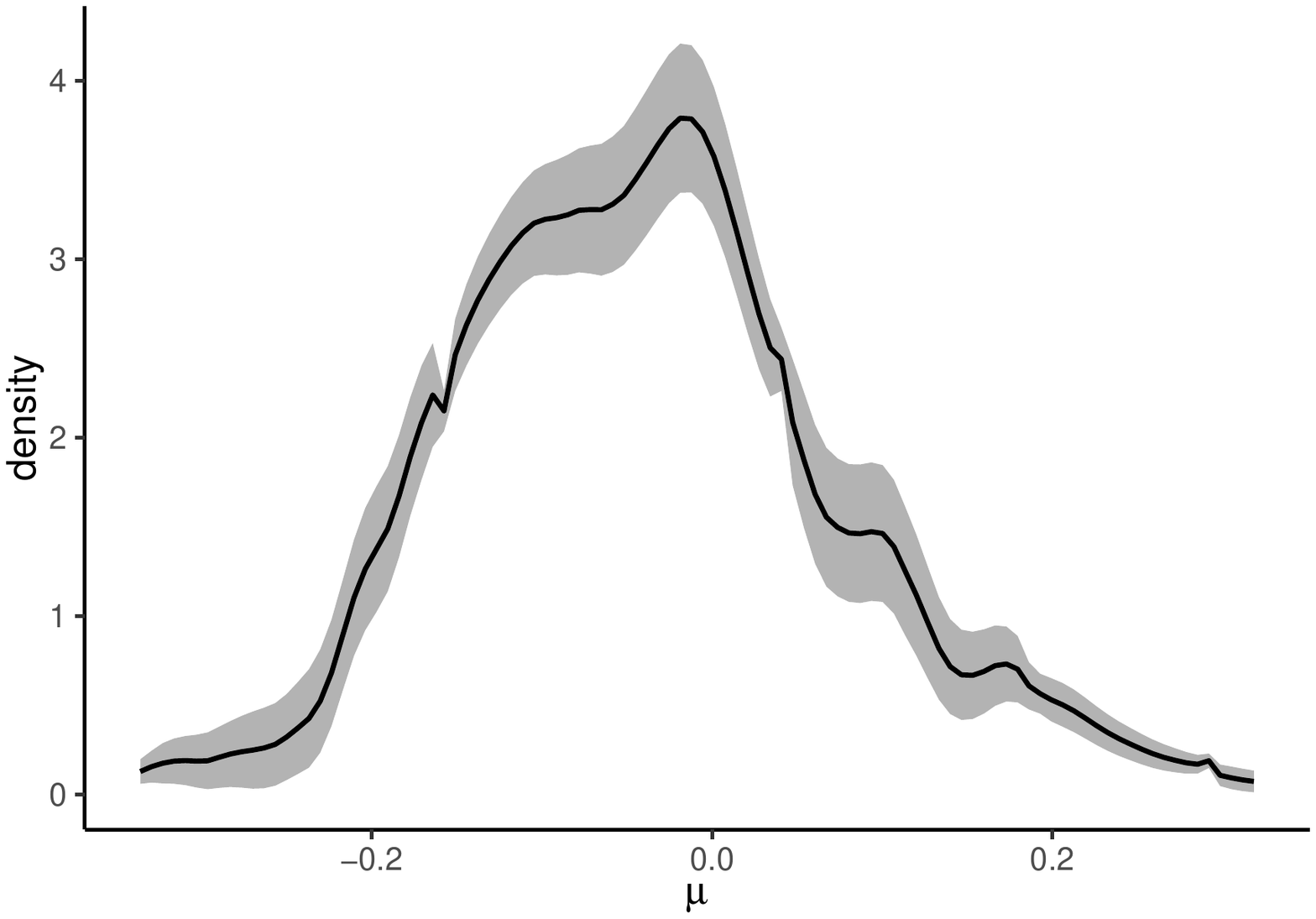}
	\end{minipage}
	\begin{minipage}{0.33\hsize}
		\includegraphics[width=55mm, bb = 0 0 841 595]{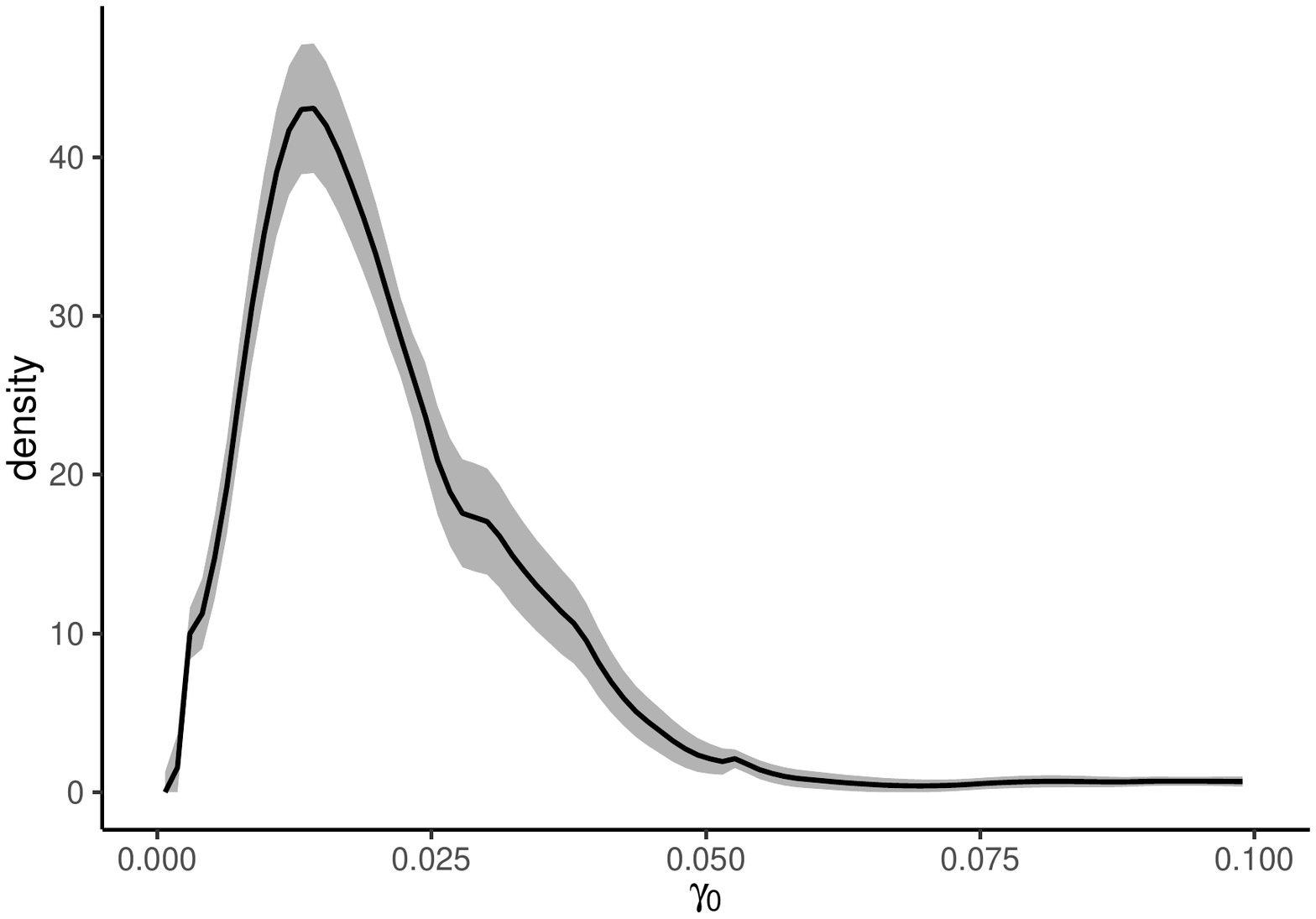}
	\end{minipage}
	\begin{minipage}{0.33\hsize}
		\includegraphics[width=55mm, bb = 0 0 841 595]{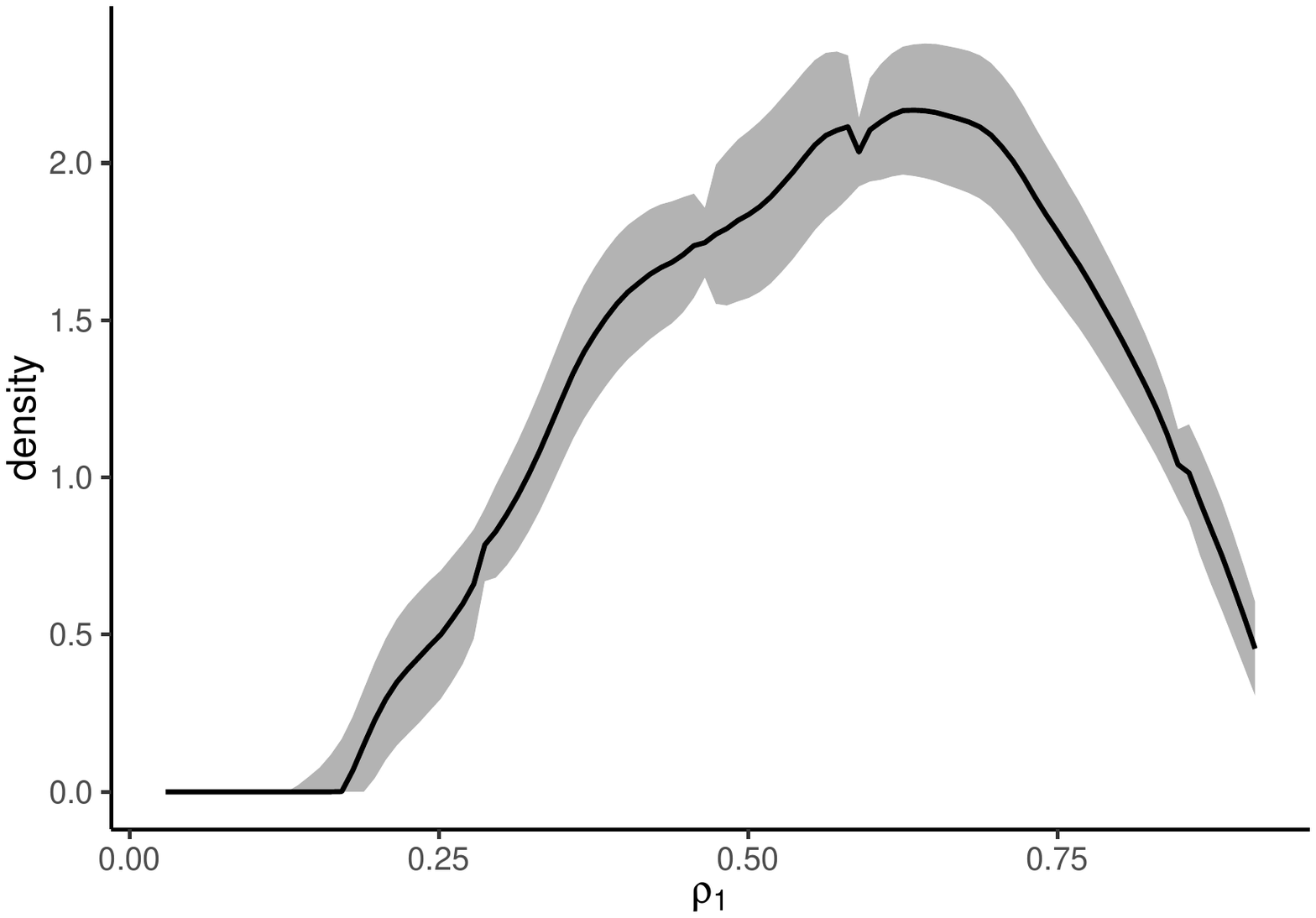}
	\end{minipage}
	\caption{The HPJ bias-corrected densities of $\mu$, $\gamma_0$, and $\rho_1$ with 95\% point-wise CIs.}
	\label{fig:ci}
\end{figure}

\paragraph{Results.}
Figure \ref{fig:density} depicts the density estimates for $\mu_i$, $\gamma_{0,i}$, and $\rho_{1,i}$.
In each panel, the solid black line indicates the density estimates without split-panel jackknife bias correction, the red dashed line shows the HPJ estimates, and the blue dotted line shows the TOJ estimates.

The estimation results with and without bias correction show that the LOP deviation dynamics are significantly heterogeneous across items.
The density estimates without bias correction for $\mu_i$ are similar to those with bias correction.
The results for $\mu_i$ also show that the mode of the heterogeneous long-run LOP deviations is close to zero, with a nearly symmetric, unimodal distribution.
In contrast, the estimates without bias correction for $\gamma_{0,i}$ and $\rho_{1,i}$ are very different from the bias-corrected estimates.
The bias-corrected estimates for $\gamma_{0,i}$ demonstrate larger variances for the LOP deviation dynamics, while the bias-corrected estimates for $\rho_{1,i}$ show more persistent dynamics with a more left-skewed distribution.
These results suggest the severe impact of the incidental parameter biases, which highlights the importance of bias correction methods.

Figure \ref{fig:ci} depicts 95\% point-wise confidence bands based on the HPJ estimates.
The confidence bands are narrow, implying that our HPJ estimates seem to be precise and reliable.

\begin{figure}[!t]
	\begin{minipage}{0.33\hsize}
		\includegraphics[width=55mm, bb = 0 0 841 595]{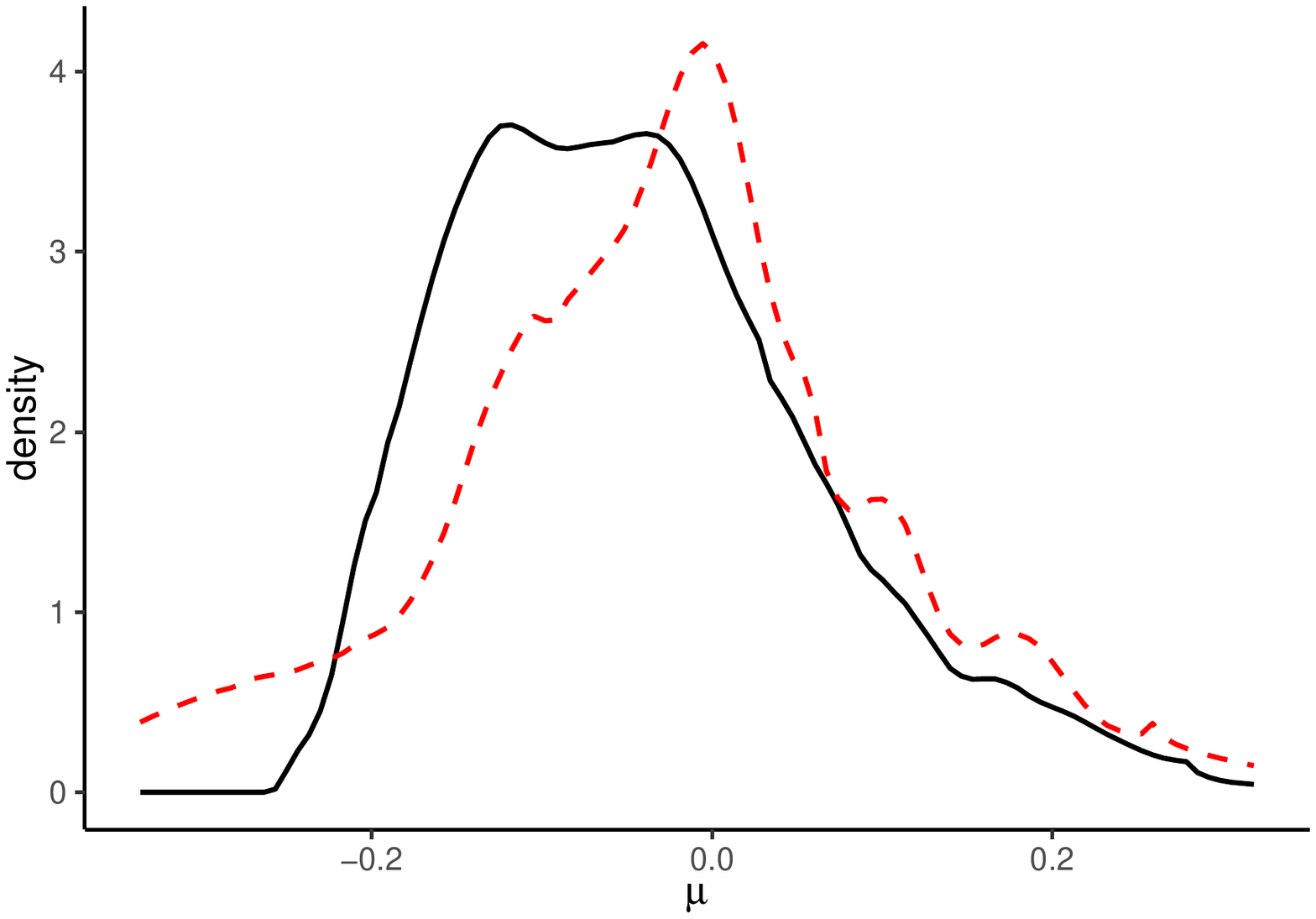}
	\end{minipage}
	\begin{minipage}{0.33\hsize}
		\includegraphics[width=55mm, bb = 0 0 841 595]{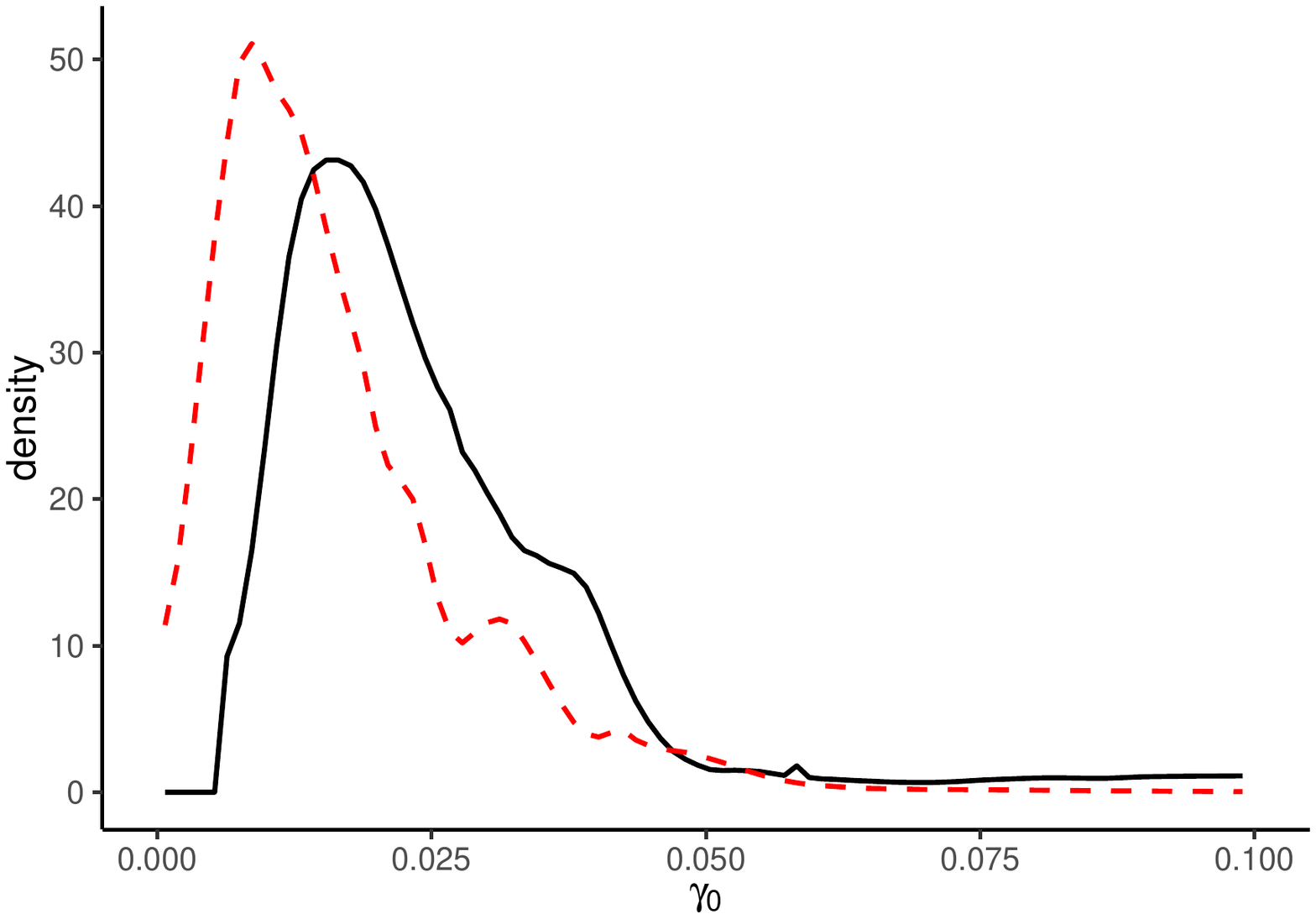}
	\end{minipage}
	\begin{minipage}{0.33\hsize}
		\includegraphics[width=55mm, bb = 0 0 841 595]{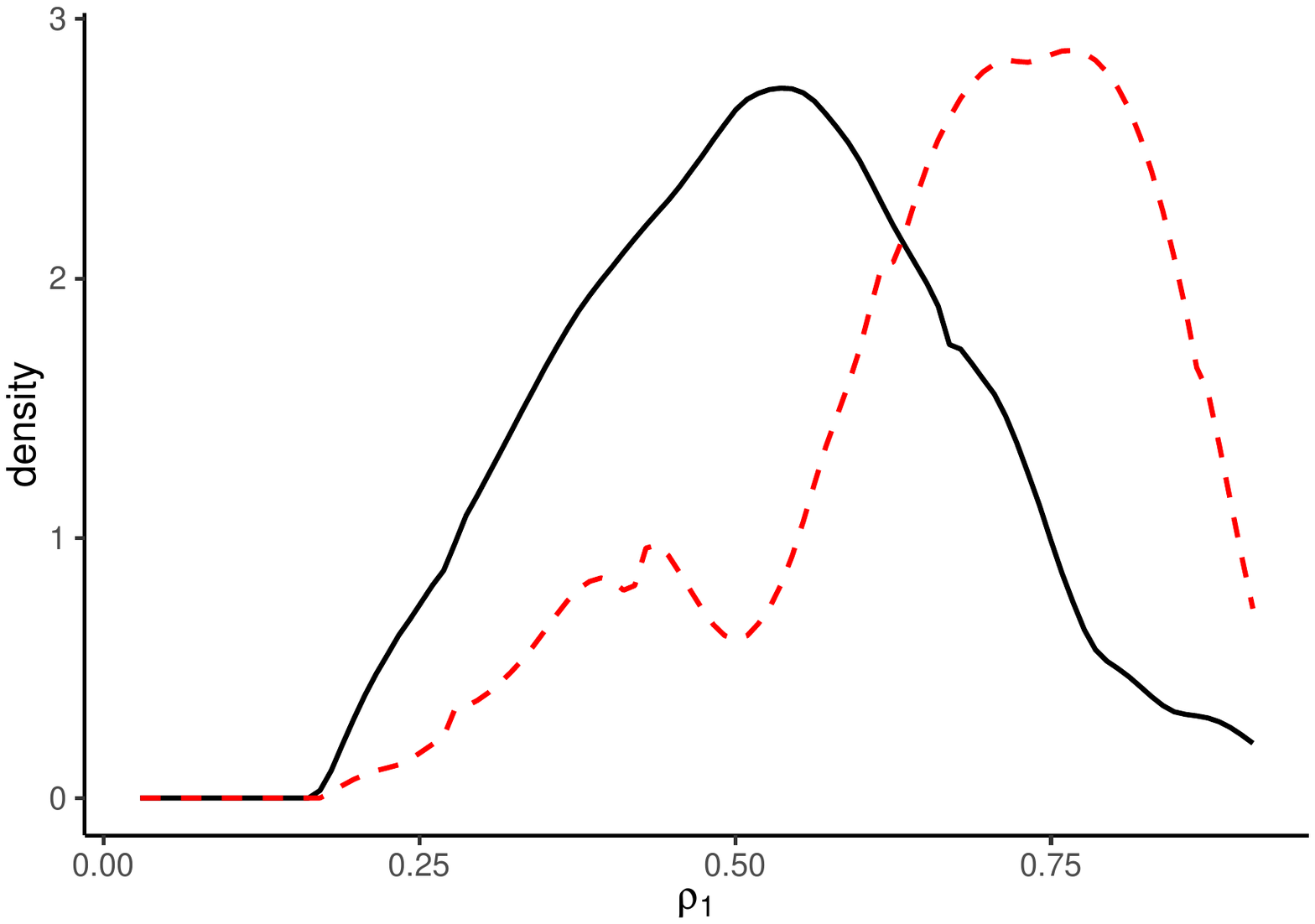}
	\end{minipage}
	\caption{The HPJ bias-corrected densities of $\mu$, $\gamma_0$, and $\rho_1$ for goods and services separately. In each figure, the solid black line indicates the HPJ bias-corrected estimates for goods, and the dashed red line shows the HPJ bias-corrected estimates for services.}
	\label{fig:gs}
\end{figure}

Figure \ref{fig:gs} illustrates the HPJ estimates of $\mu_i$, $\gamma_{0,i}$, and $\rho_{1,i}$ for goods and services separately.
The solid black lines are the HPJ estimates for goods, and the dashed red lines are those for services.
The estimated densities and CDFs show that the heterogeneous properties are significantly different between goods and services.
The densities for $\mu_i$ show that the long-run LOP deviation for goods generally tends to be larger than that for services (in an absolute sense).
The estimation results for $\gamma_{0,i}$ and $\rho_{1,i}$ show that the LOP deviation for goods tends to be more volatile but less persistent than that for services.
These results suggest that goods tend to have more volatile processes with faster adjustment speeds toward the nonnegligible long-run LOP deviation.

If we seek to examine the degree of heterogeneity of the LOP deviations across items and cities as in \citet{CruciniShintaniTsuruga15}, our model-free results are informative in their own right.
There are several possible sources of differences in the degree of heterogeneity, including the differences in trade costs across items (e.g., \citealp{AndersonVanWincoop04}) and differences in sale and nonsale prices across goods and services (e.g., \citealp{NakamuraSteinsson08}).
Furthermore, our model-free results also suggest how we should model heterogeneity when implementing structural estimation for price deviations or change.
For example, as our procedure demonstrates that the heterogeneous properties of goods and services differ, we should model unobserved heterogeneity differently for goods and services.

\section{Conclusion} \label{sec-conclusion}
This paper presented nonparametric kernel-smoothing estimation to examine the degree of heterogeneity in panel data.
The kernel density and CDF estimators are consistent and asymptotically normal under the relative magnitude conditions on the cross-sectional size $N$, time-series length $T$, and bandwidth $h$.
Because of the presence of incidental parameter bias and nonlinearity biases, the relative magnitude conditions vary in the order of the expansions.
Via infinite-order expansions, we derived the relative magnitude conditions that are suitable for microeconometric applications.
We discussed the split-panel jackknife to correct biases, the construction of CIs, and the selection of bandwidth.
We also illustrated our procedure based on an application on price deviations.

\section*{Acknowledgments}
The authors greatly appreciate the assistance of Mototsugu Shintani in providing the price panel data.
The authors would also like to thank Stephane Bonhomme, Kazuhiko Hayakawa, Koen Jochmans, Shin Kanaya, Hiroyuki Kasahara, Yoonseok Lee, Oliver Linton, Jun Ma, Yukitoshi Matsushita, Martin Weidner, Yohei Yamamoto, Yu Zhu and the participants of many conferences for helpful discussions and comments. Sebastian Calonico kindly helped us better understand the use of the {\ttfamily nprobust} package.
All remaining errors are our own. 
Part of this research was conducted while Okui was at Vrije Universiteit Amsterdam, Kyoto University, and NYU Shanghai and while Yanagi was at Hitotsubashi University. 
This work was supported by the New Faculty Startup Fund from Seoul National University, JSPS KAKENHI Grant Numbers JP25780151, JP25285067, JP15H03329, JP16K03598, JP15H06214, and JP17K13715.

\clearpage

\bibliographystyle{abbrvnat}
\bibliography{kernel.bib}

\clearpage
\renewcommand\thepage{S\arabic{page}}
\setcounter{page}{1}

\begin{center}
	{\LARGE Supplementary Appendix of ``Kernel Estimation for Panel Data with Heterogeneous Dynamics''}
	
	\bigskip
	
	{\large Ryo Okui and Takahide Yanagi}
	
	\bigskip
	
	{\large May, 2019}
	
	\bigskip
\end{center}

This supplementary appendix contains technical discussions omitted from the main text and Monte Carlo simulation results.
Appendix \ref{sec-proof} presents the proofs of the theorems in the main body of the paper.
Appendix \ref{sec-lemma} presents the technical lemmas used in the proofs of the theorems.
Appendices \ref{sec-expansion} and \ref{sec-series} present the technical discussions on the validity of the infinite-order expansions in Theorem \ref{thm-dens}.
Appendix \ref{sec-simulation} presents the simulation results.

\appendix
\section{Appendix: Proofs of the theorems} \label{sec-proof}

\setcounter{equation}{0}
\renewcommand{\theequation}{A.\arabic{equation}}

This appendix collects the proofs of the theorems.
In the following, we denote a generic positive constant by $0 < M < \infty$.

\subsection{Proof of Theorem \ref{thm-dens}}

\paragraph{The density of $\mu_i$.}
We evaluate each term in the following Taylor expansion:
\begin{align}
\hat f_{\hat \mu}(x) - f_{\mu}(x)
=& \frac{1}{Nh} \sum_{i=1}^N K \left( \frac{ x - \mu_i }{h}\right) - f_{\mu}(x)
\label{dens-mu-1}\\
& - \frac{1}{Nh^2} \sum_{i=1}^N (\hat \mu_i - \mu_i) K' \left( \frac{ x - \mu_i }{h}\right)  \label{dens-mu-2}\\
& + \sum_{j = 2}^{\infty} \frac{(-1)^j}{j! Nh^{j+1}} \sum_{i=1}^N (\hat \mu_i - \mu_i)^j K^{(j)} \left( \frac{x - \mu_i }{h}\right). \label{dens-mu-3}
\end{align}

For \eqref{dens-mu-1}, we use the standard results for the kernel density estimation.
Lemma \ref{lem-dens} under Assumptions \ref{as-basic}, \ref{as-kernel}, and \ref{as-density-kernel} shows that $(Nh)^{-1} \sum_{i=1}^N K((x - \mu_i)/h) - f_{\mu}(x) \stackrel{p}{\longrightarrow} 0$ as $N\to \infty$ and $h\to 0$ with $Nh \to \infty$.
Furthermore, Lemma \ref{lem-dens} also shows that: 
\begin{align*}
\sqrt{Nh} \left( \frac{1}{Nh} \sum_{i=1}^N K \left( \frac{ x - \mu_i }{h}\right) -f_{\mu}(x)  - h^2 \frac{\kappa_1 f_{\mu}^{''} (x)}{2}  \right) \stackrel{d}{\longrightarrow} \mathcal{N} \big( 0, \kappa_2 f_{\mu}(x) \big),
\end{align*}
as $N\to \infty$ and $h\to 0$ with $Nh \to \infty$ and $Nh^5 \to C \in [0, \infty)$.

For \eqref{dens-mu-2}, the mean is zero by the law of iterated expectations, because $\hat \mu_i - \mu_i = \bar w_i$ and $E(\bar w_i | i) = 0$.
The variance is: 
\begin{align*}	
var\left(  \frac{1}{Nh^2} \sum_{i=1}^N \bar w_i K' \left( \frac{ x - \mu_i }{h}\right) \right)
= \frac{1}{Nh^4} E\left(  (\bar w_i)^2 \left( K' \left( \frac{ x - \mu_i }{h}\right) \right)^2 \right)
= O\left( \frac{1}{NT h^{3}} \right),
\end{align*}
by Lemmas \ref{lem-wbar} and \ref{lem-kernel}.
Therefore, \eqref{dens-mu-2} is $O_p(1/\sqrt{NTh^3})$ by Markov inequality.

For the term in \eqref{dens-mu-3}, the mean is: 
\begin{align*}
& E\left( \frac{1}{j!Nh^{j+1}} \sum_{i=1}^N (\bar w_i)^j K^{(j)} \left( \frac{x - \mu_i }{h}\right) \right) \\
&= \frac{1}{j!h^{j+1}} E\left( E\left( (\bar w_i)^j | \mu_i \right) K^{(j)} \left( \frac{x - \mu_i }{h}\right) \right)\\
&= \frac{1}{j!T^{j/2} h^j} E\left( T^{j/2} (\bar w_i)^j \middle| \mu_i = x \right) f_{\mu}(x)  \int K^{(j)}(s) ds + o\left(\frac{1}{T^{j/2} h^j} \right) \\
&= (-1)^j \frac{A_{\mu, j}(x)}{\sqrt{T^j h^{2j}}} + o\left( \frac{1}{\sqrt{T^j h^{2j}}} \right),
\end{align*}
by the law of iterated expectations and Lemma \ref{lem-kernel} with the definition of 
\begin{align*}
A_{\mu, j}(x) \coloneqq \lim_{T \to \infty} \frac{(-1)^j}{j!} E\left( \sqrt{T^j} (\bar w_i)^j \middle| \mu_i = x \right) f_{\mu}(x) \int K^{(j)}(s) ds.
\end{align*}
The variance is:
\begin{align*}
var\left( \frac{1}{j! N h^{j+1}} \sum_{i=1}^N (\bar w_i)^j K^{(j)} \left( \frac{x - \mu_i}{h}\right) \right)
= O\left( \frac{1}{NT^j h^{2j+1}} \right),
\end{align*}
by Lemmas \ref{lem-wbar} and \ref{lem-kernel}.
Thus, it holds that:
\begin{align*}
\sum_{j = 2}^{\infty} \frac{(-1)^j}{j! Nh^{j+1}} \sum_{i=1}^N (\hat \mu_i - \mu_i)^j K^{(j)} \left( \frac{x - \mu_i}{h}\right)
= \sum_{j=2}^\infty \left( \frac{A_{\mu, j}(x)}{\sqrt{T^j h^{2j}}} + o_p\left( \frac{1}{\sqrt{T^j h^{2j}}} \right) \right).
\end{align*}

Consequently, we obtain the desired result for $\hat f_{\hat \mu}(x)$ by Slutsky's theorem.

\paragraph{The density of $\gamma_{k,i}$.}
We evaluate each term in the following Taylor expansion:
\begin{align}
\hat f_{\hat \gamma_k}(x) - f_{\gamma_k}(x) 
=&  \frac{1}{Nh} \sum_{i=1}^N K \left( \frac{ x - \gamma_{k,i} }{h}\right) - f_{\gamma_k}(x)
\label{dens-gam-1}\\
& - \frac{1}{Nh^2} \sum_{i=1}^N (\hat \gamma_{k,i} - \gamma_{k,i}) K' \left( \frac{ x - \gamma_{k,i} }{h}\right)  \label{dens-gam-2}\\
& + \sum_{j=2}^{\infty} \frac{(-1)^j}{j!Nh^{j+1}} \sum_{i=1}^N (\hat \gamma_{k,i} - \gamma_{k,i})^j K^{(j)} \left( \frac{ x - \gamma_{k,i} }{h}\right). \label{dens-gam-3}
\end{align}

For \eqref{dens-gam-1}, the consistency and asymptotic normality of the term are established by the same argument as for the density of $\mu_i$.

For \eqref{dens-gam-2}, we have the following equation based on the expansion for $\hat \gamma_{k,i}$:
\begin{align}
&\frac{1}{Nh^2} \sum_{i=1}^N (\hat \gamma_{k,i} - \gamma_{k,i}) K' \left( \frac{ x - \gamma_{k,i} }{h}\right) \nonumber\\
=& \frac{1}{Nh^2} \sum_{i=1}^N \left(  \frac{1}{T-k} \sum_{t=k+1}^T w_{it} w_{i,t-k} - \gamma_{k,i}\right)  K' \left( \frac{ x - \gamma_{k,i} }{h}\right) \label{dens-gam-5} \\
&- \frac{1}{Nh^2} \sum_{i=1}^N \frac{T+k}{T-k} (\bar w_i)^2 K' \left( \frac{ x - \gamma_{k,i} }{h}\right) \label{dens-gam-6} \\
&+ \frac{1}{Nh^2} \sum_{i=1}^N \frac{1}{T-k} \sum_{t=1}^k w_{it}  \bar w_{i} K' \left( \frac{ x - \gamma_{k,i} }{h}\right) \label{dens-gam-7} \\
&+ \frac{1}{Nh^2} \sum_{i=1}^N \frac{1}{T-k} \sum_{t=T-k+1}^T w_{it} \bar w_{i} K' \left( \frac{ x - \gamma_{k,i} }{h}\right). \label{dens-gam-8}
\end{align}
For \eqref{dens-gam-5}, the mean is zero by the law of iterated expectations given $E(w_{it} w_{i, t-k}|i) = \gamma_{k,i}$.
The variance is: 
\begin{align*}
var\left( \frac{1}{Nh^2} \sum_{i=1}^{N}\left( \frac{1}{T-k}\sum_{t=k+1}^{T}w_{it}w_{i,t-k} - \gamma_{k,i} \right) K' \left( \frac{ x - \gamma_{k,i} }{h}\right) \right)
= O\left( \frac{1}{NTh^3} \right),
\end{align*}
by Lemmas \ref{lem-wk} and \ref{lem-kernel}.
Thus, \eqref{dens-gam-5} is $O_{p}( 1/\sqrt{NTh^3})$.
For \eqref{dens-gam-6}, denoting $a_T(x) = E(T (\bar w_i)^2 | \gamma_{k,i} = x) f_{\gamma_k}(x)$, the mean is expanded as:
\begin{align*}
E \left( \frac{1}{Nh^2} \sum_{i=1}^N \frac{T+k}{T-k}(\bar w_{i})^2 K' \left( \frac{ x - \gamma_{k,i} }{h}\right) \right)
=& \frac{T+k}{T(T-k)h^2} E \left( E\left( T(\bar w_{i})^2 |\gamma_{k,i} \right)  K' \left( \frac{ x - \gamma_{k,i} }{h}\right) \right) \\
=& \frac{a_T(x)}{Th} \int K'(s)ds - \frac{a_T'(x)}{T} \int s K'(s) ds + o\left( \frac{1}{T} \right)\\
=& - \frac{a_T'(x) }{T} \int s K'(s) ds + o\left( \frac{1}{T} \right)\\ 
=& - \frac{A_{\gamma_k, 1}(x)}{T} + o\left( \frac{1}{T} \right),
\end{align*}
by the law of iterated expectations, Lemma \ref{lem-kernel}, and $\int K'(s) ds =0$ with the definition of: 
\begin{align*}
A_{\gamma_k, 1}(x) = \lim_{T \to \infty} a_T'(x) \int s K'(s) ds.
\end{align*}
The variance is:
\begin{align*}
var \left( \frac{1}{Nh^2} \sum_{i=1}^N \frac{T+k}{T-k} (\bar w_{i})^2 K' \left( \frac{ x - \gamma_{k,i} }{h}\right) \right)
& \le \frac{(T+k)^2}{Nh^4(T-k)^2} \sqrt{E\left( (\bar w_i)^8 \right)} \sqrt{E\left( \left( K'\left( \frac{x - \gamma_{k,i}}{h} \right) \right)^4 \right)}\\
&= O\left( \frac{1}{Nh^4} \right) \cdot O\left( \frac{1}{T^2} \right) \cdot O\left( \sqrt{h} \right)
= O \left( \frac{1}{ \sqrt{N^2 T^4 h^7}} \right),
\end{align*}
by the Cauchy--Schwarz inequality and Lemmas \ref{lem-wbar} and \ref{lem-kernel}.
Thus, \eqref{dens-gam-6} is $A_{\gamma_k, 1}(x) / T + o_p(1/T)$.
For \eqref{dens-gam-7}, the triangle inequality and the Cauchy--Schwarz inequality lead to: 
\begin{align*}
& E \left| \frac{1}{Nh^2} \sum_{i=1}^{N} \frac{1}{T-k} \sum_{t=1}^{k} w_{it}\bar w_{i} K' \left( \frac{ x - \gamma_{k,i} }{h}\right) \right|\\
&\leq \frac{1}{(T-k)h^2} E \left|\sum_{t=1}^{k} w_{it}\bar w_{i} K' \left( \frac{ x - \gamma_{k,i} }{h}\right) \right|\\
&\leq \frac{1}{(T-k)h^2} \sqrt{E\left(\left(\sum_{t=1}^{k} w_{it}\bar w_{i}\right)^{2}\right)}\sqrt{E\left(\left( K' \left( \frac{ x - \gamma_{k,i} }{h}\right)\right)^{2}\right)}\\
&= O\left( \frac{1}{Th^2} \right) \cdot O\left( \frac{1}{\sqrt{T}} \right) \cdot O\left( \sqrt{h} \right)
= O\left( \frac{1}{\sqrt{T^3 h^3}} \right),
\end{align*}
by Lemmas \ref{lem-wbar} and \ref{lem-kernel}.
Thus, \eqref{dens-gam-7} is $O_{p}( 1/\sqrt{T^3 h^3})$.
In the same manner, we can show that \eqref{dens-gam-8} is also $O_{p}( 1/\sqrt{T^3 h^3})$.
These results mean that \eqref{dens-gam-2} is $A_{\gamma_k, 1}(x)/T + o_p(1/T)$.

For \eqref{dens-gam-3}, it is easy to see that: 
\begin{align*}
& \frac{1}{j!Nh^{j+1}} \sum_{i=1}^N (\hat \gamma_{k,i} - \gamma_{k,i})^j K^{(j)} \left( \frac{x - \gamma_{k,i} }{h} \right)\\
=& \frac{1}{j!Nh^{j+1}} \sum_{i=1}^N \left( \frac{1}{T-k} \sum_{t=k+1}^T (w_{it} w_{i,t-k} - \gamma_{k,i}) \right)^j K^{(j)} \left( \frac{x - \gamma_{k,i}}{h} \right) + o_p\left( \frac{1}{\sqrt{T^j h^{2j}}} \right),\nonumber
\end{align*}
by the same procedures to show the order of the terms in \eqref{dens-gam-2}.
The mean of the term is:
\begin{align*}
& E\left( \frac{1}{j!Nh^{j+1}} \sum_{i=1}^N \left( \frac{1}{T-k} \sum_{t=k+1}^T (w_{it} w_{i,t-k} - \gamma_{k,i}) \right)^j K^{(j)} \left( \frac{x - \gamma_{k,i}}{h} \right) \right) \\
&= \frac{1}{j! (T-k)^{j/2} h^{j+1}} E\left( E\left( \frac{1}{(T-k)^{j/2}} \left( \sum_{t=k+1}^T (w_{it} w_{i,t-k} - \gamma_{k,i}) \right)^j \middle| \gamma_{k, i} \right( K^{(j)} \left( \frac{x - \gamma_{k,i}}{h} \right) \right) \\
&= \frac{1}{j!T^{j/2}h^j} f_{\gamma_k}(x)  E\left( \frac{1}{T^{j/2}} \left( \sum_{t=k+1}^T (w_{it} w_{i,t-k} - \gamma_{k,i}) \right)^j \middle| \gamma_{k,i}=x \right) \int K^{(j)}(s)ds + o\left(\frac{1}{T^{j/2}h^{j}}\right)\\
&= (-1)^j \frac{A_{\gamma_k, j}(x)}{\sqrt{T^j h^{2j}}} + o\left(\frac{1}{\sqrt{T^j h^{2j}}}\right),
\end{align*}
by the law of iterated expectations and Lemma \ref{lem-kernel} with the definition of
\begin{align*}
A_{\gamma_k, j}(x) \coloneqq \lim_{T \to \infty} (-1)^j \frac{f_{\gamma_k}(x)}{j!} E\left( \frac{1}{\sqrt{T^j}} \left( \sum_{t=k+1}^T (w_{it} w_{i,t-k} - \gamma_{k,i}) \right)^j \middle| \gamma_{k,i} = x \right) \int K^{(j)}(s)ds.
\end{align*}
The variance of the term is:
\begin{align*}
var \left( \frac{1}{Nh^{j+1}} \sum_{i=1}^N \left( \frac{1}{T-k} \sum_{t=k+1}^T (w_{it} w_{i,t-k} - \gamma_{k,i}) \right)^j K^{(j)} \left( \frac{x - \gamma_{k,i} }{h} \right) \right)
= O\left( \frac{1}{NT^j h^{2j+1}} \right), 
\end{align*}
by Lemmas \ref{lem-wk} and \ref{lem-kernel}. 
Thus, it holds that: 
\begin{align*}
\sum_{j=2}^{\infty} \frac{(-1)^j}{j!Nh^{j+1}} \sum_{i=1}^N (\hat \gamma_{k,i} - \gamma_{k,i})^j K^{(j)} \left( \frac{ x - \gamma_{k,i} }{h}\right)  
= \sum_{j=2}^{\infty} \left( \frac{A_{\gamma_k, j}(x)}{\sqrt{T^j h^{2j}}} + o_p\left(\frac{1}{\sqrt{T^j h^{2j}}}\right) \right).
\end{align*}	

Consequently, we obtain the desired result for $\hat f_{\hat \gamma_k}(x)$ by Slutsky's theorem.

\paragraph{The density of $\rho_{k,i}$.}
We regard $K( (x - \hat \rho_{k,i})/h ) = K( (x - \hat \gamma_{k,i}/\hat \gamma_{0,i} )/h)$ as a function of two variables $(\hat \gamma_{k,i}, \hat \gamma_{0,i} )$.
Taylor's theorem for multivariate functions leads to:
\begin{align}
& \hat f_{\hat \rho_k}(x) - f_{\rho_k}(x) \nonumber \\
=& \frac{1}{Nh} \sum_{i=1}^N K \left( \frac{ x - \rho_{k,i} }{h}\right) - f_{\rho_k}(x) \label{dens-rho-1} \\
& - \frac{1}{Nh} \sum_{i=1}^N \sum_{j_1+j_2=1} (\hat \gamma_{k,i} - \gamma_{k,i})^{j_1} (\hat \gamma_{0,i} - \gamma_{0,i})^{j_2}  \frac{\partial^{j_1 + j_2}}{\partial a^{j_1} \partial b^{j_2}} K \left( \frac{ x- a/b}{h}\right) \Big|_{a = \gamma_{k,i}, b = \gamma_{0,i}} \label{dens-rho-2} \\
& + \sum_{j=2}^{\infty} \frac{(-1)^j}{Nh} \sum_{i=1}^N \sum_{j_1+j_2=j} \frac{1}{j_1!j_2!} (\hat \gamma_{k,i} - \gamma_{k,i})^{j_1} (\hat \gamma_{0,i} - \gamma_{0,i})^{j_2}  \frac{\partial^{j_1 + j_2}}{\partial a^{j_1} \partial b^{j_2}} K \left( \frac{ x- a/b}{h}\right) \Big|_{a = \gamma_{k,i}, b = \gamma_{0,i}}. \label{dens-rho-3}
\end{align}
We evaluate each term below. 

For \eqref{dens-rho-1}, the consistency and asymptotic normality of the term are established by the same argument as for the density of $\mu_i$.

\eqref{dens-rho-2} contains two terms.
Of these, we consider only $(Nh^2)^{-1} \sum_{i=1}^N \gamma_{0,i}^{-1} (\hat \gamma_{k,i} - \gamma_{k,i}) K'( (x - \rho_{k,i}) / h)$, because the other term can be evaluated by the same argument.
However, this term is analogous to that in \eqref{dens-gam-2}, so it can be evaluated by the same argument.
This means that \eqref{dens-rho-2} can be written as $A_{\rho_k, 1}(x)/T + o_p(1/T)$ for a nonrandom $A_{\rho_k, 1}(x)$.

For \eqref{dens-rho-3}, we evaluate the mean of the term:
\begin{align*}
\sum_{j_1+j_2=j} \frac{1}{j_1!j_2!} (\hat \gamma_{k,i} - \gamma_{k,i})^{j_1} (\hat \gamma_{0,i} - \gamma_{0,i})^{j_2}  \frac{\partial^{j_1 + j_2}}{\partial a^{j_1} \partial b^{j_2}} K \left( \frac{ x- a/b}{h}\right) \Big|_{a = \gamma_{k,i}, b = \gamma_{0,i}},
\end{align*}
which contains $j+1$ terms.
Among these, we consider only $(j!Nh^{j+1})^{-1} \sum_{i=1}^N \gamma_{0,i}^{-j} (\hat \gamma_{k,i} - \gamma_{k,i})^j  K^{(j)}((x - \rho_{k,i})/h)$, as the other terms can be evaluated in the same manner.
However, this term is analogous to that in \eqref{dens-gam-3}, so it can be evaluated by the same argument.
This means that \eqref{dens-rho-3} can be written as: 
\begin{align*}
& \sum_{j=2}^{\infty} \frac{(-1)^j}{Nh} \sum_{i=1}^N \sum_{j_1+j_2=j} \frac{1}{j_1!j_2!} (\hat \gamma_{k,i} - \gamma_{k,i})^{j_1} (\hat \gamma_{0,i} - \gamma_{0,i})^{j_2}  \frac{\partial^{j_1 + j_2}}{\partial a^{j_1} \partial b^{j_2}} K \left( \frac{ x- a/b}{h}\right) \Big|_{a = \gamma_{k,i}, b = \gamma_{0,i}} \\
&= \sum_{j=2}^{\infty} \left( \frac{A_{\rho_k, j}(x)}{\sqrt{T^j h^{2j}}} + o_p\left( \frac{1}{\sqrt{T^j h^{2j}}} \right) \right),
\end{align*}
for a nonrandom $A_{\rho_k, j}(x)$.

Consequently, we have the desired result for $\hat f_{\hat \rho_k}(x)$ by Slutsky's theorem.	
\qed

\subsection{Proof of Theorem \ref{thm-dens-HPJ}}
We show the proof for the density estimator of $\gamma_{k,i}$ only. Those of $\mu_i$ and $\rho_{k,i}$ are the same.
The proof of Theorem \ref{thm-dens} has shown that: 
\begin{align*}
\hat f_{\hat \gamma_k}(x) - f_{\gamma_k}(x) 
= \frac{1}{Nh} \sum_{i=1}^N K\left( \frac{x - \gamma_{k,i}}{h} \right) - f_{\gamma_k}(x) + \frac{A_{\gamma_k, 1}(x)}{T} + \frac{A_{\gamma_k, 2}(x)}{Th^2} +  O_p\left( \frac{1}{\sqrt{T^3 h^6}} \right).
\end{align*}
This result implies that the estimators based on the half-panel data are: 
\begin{align*}
\hat f_{\hat \gamma_k,(l)}(x) - f_{\gamma_k}(x) 
= \frac{1}{Nh} \sum_{i=1}^N K\left( \frac{x - \gamma_{k,i}}{h} \right) - f_{\gamma_k}(x) + \frac{2A_{\gamma_k, 1}(x)}{T} + \frac{2A_{\gamma_k, 2}(x)}{Th^2} + O_p\left( \frac{1}{\sqrt{T^3 h^6}} \right),
\end{align*}
for $l=1,2$.
As a result, the HPJ bias-corrected estimator satisfies:
\begin{align*}
\hat f_{\hat \gamma_k}^H(x) -  f_{\gamma_k}(x) =  \frac{1}{Nh} \sum_{i=1}^N K\left( \frac{x - \gamma_{k,i}}{h} \right) - f_{\gamma_k}(x) + O_p\left( \frac{1}{\sqrt{T^3 h^6}} \right).
\end{align*}
Therefore, the same argument as for the term in \eqref{dens-gam-1} leads to the desired result.
\qed

\subsection{Proof of Theorem \ref{thm-CDF}}

\paragraph{The CDF of $\mu_i$.}
We evaluate each term in the following Taylor expansion:
\begin{align}
\hat F_{\hat \mu}(x) - F_{\mu}(x) 
=&  \frac{1}{N} \sum_{i=1}^N \mathbb{K} \left( \frac{x-\mu_i}{h}\right) - F_{\mu} (x) 
\label{cdf-mu-1}\\
& - \frac{1}{Nh} \sum_{i=1}^N (\hat \mu_i - \mu_i) K \left( \frac{x - \mu_i}{h}\right)  \label{cdf-mu-2}\\
& + \frac{1}{2Nh^2} \sum_{i=1}^N (\hat \mu_i - \mu_i)^2 K' \left( \frac{x -  \mu_i}{h}\right) \label{cdf-mu-3} \\
& + \sum_{j=3}^{\infty}  \frac{(-1)^j}{j! N h^j} \sum_{i=1}^N (\hat \mu_i - \mu_i)^j K^{(j-1)} \left( \frac{x -  \mu_i}{h}\right). \label{cdf-mu-4}
\end{align}

For the term in \eqref{cdf-mu-1}, Lemma \ref{lem-CDF} under Assumptions \ref{as-basic}, \ref{as-kernel}, and \ref{as-cdf} shows that: 
\begin{align*}
\frac{1}{N} \sum_{i=1}^N \mathbb{K} \left( \frac{x-\mu_i}{h}\right) - F_{\mu}(x)
\stackrel{p}{\longrightarrow} 0,
\end{align*}
as $N\to \infty$ and $h\to 0$.
Moreover, Lemma \ref{lem-CDF} also shows that:
\begin{align*}
\sqrt{N} \left( \frac{1}{N} \sum_{i=1}^N \mathbb{K} \left( \frac{x-\mu_i}{h}\right) -F_{\mu} (x) \right) 
\stackrel{d}{\longrightarrow} \mathcal{N} \big( 0,  F_{\mu} (x) [1- F_{\mu}(x)] \big),
\end{align*}
as $N\to \infty$ and $h\to 0$ with $Nh^4  \to 0$.

For \eqref{cdf-mu-2}, the mean is zero given $\hat \mu_i - \mu_i = \bar w_i$ and $E(\bar w_i | i) = 0$.
The variance is: 
\begin{align*}
var\left( \frac{1}{Nh} \sum_{i=1}^N (\hat \mu_i - \mu_i) K \left( \frac{ x - \mu_i}{h}\right) \right)
= O\left( \frac{1}{NT h} \right),
\end{align*}
by Lemmas \ref{lem-mu} and \ref{lem-kernel}.
Thus, \eqref{cdf-mu-2} is $O_p(1 / \sqrt{NTh})$ by Markov inequality.

For \eqref{cdf-mu-3}, we define $c_T(x) \coloneqq E(T (\bar w_i)^2|\mu_i=x) f_{\mu}(x)$.
The mean is: 
\begin{align*}
E\left( \frac{1}{2Nh^2} \sum_{i=1}^N (\hat \mu_i - \mu_i)^2 K' \left( \frac{x - \mu_i}{h}\right) \right)
&= \frac{1}{2Th^2} E\left( E[T(\bar w_i)^2|\mu_i] K' \left( \frac{x - \mu_i}{h}\right) \right) \\
&= \frac{1}{2Th} c_T(x) \int K'(s)ds - \frac{1}{2T} c_T'(x) \int sK'(s)ds + o\left( \frac{1}{T} \right) \\
&= -\frac{1}{2T} c_T'(x) \int sK'(s)ds + o\left( \frac{1}{T} \right) \\
&= \frac{B_{\mu, 2}(x)}{T} + o\left( \frac{1}{T} \right),
\end{align*}
by the law of iterated expectations, Lemma \ref{lem-kernel}, and $\int K'(s)=0$ with the definition of
\begin{align*}
B_{\mu, 2}(x) \coloneqq - \lim_{T \to \infty} \frac{c_T'(x)}{2} \int sK'(s)ds.
\end{align*}
The variance is: 
\begin{align*}
var\left( \frac{1}{2Nh^2} \sum_{i=1}^N (\hat \mu_i - \mu_i)^2 K' \left( \frac{x - \mu_i}{h}\right) \right) 
= O\left( \frac{1}{NT^2h^3} \right),
\end{align*}
by Lemmas \ref{lem-wbar} and \ref{lem-kernel}.
Thus, \eqref{cdf-mu-3} can be written as $B_{\mu, 2}(x)/T + o_p(1/T)$.

For the term in \eqref{cdf-mu-4}, the mean is:
\begin{align*}
& E\left( \frac{1}{j! N h^j} \sum_{i=1}^N (\hat \mu_i - \mu_i)^j K^{(j-1)} \left( \frac{x - \mu_i}{h}\right) \right) \\
&= \frac{1}{j! T^{j/2} h^j} E\left( E\left( T^{j/2} (\bar w_i)^j \middle| \mu_i \right) K^{(j-1)} \left( \frac{x - \mu_i}{h}\right) \right) \\
&= \frac{1}{j! \sqrt{T^j h^{2j-2}}} E\left( \sqrt{T^j} (\bar w_i)^j \middle| \mu_i = x \right) f_{\mu}(x) \int K^{(j-1)}(s) ds  + o\left( \frac{1}{\sqrt{T^j h^{2j-2}}} \right) \\
&= (-1)^j \frac{B_{\mu, j}(x)}{T^j h^{2j-2}} + o\left( \frac{1}{\sqrt{T^j h^{2j-2}}} \right),
\end{align*}
by the law of iterated expectations and Lemmas \ref{lem-wbar} and \ref{lem-kernel} with the definition of 
\begin{align*}
B_{\mu, j}(x) \coloneqq \lim_{T \to \infty} \frac{(-1)^j}{j!} E\left( \sqrt{T^j} (\bar w_i)^j \middle| \mu_i = x \right) f_{\mu}(x) \int K^{(j-1)}(s) ds.
\end{align*}
The variance is: 
\begin{align*}
var\left( \frac{1}{j! N h^j} \sum_{i=1}^N (\hat \mu_i - \mu_i)^j K^{(j-1)} \left( \frac{x - \mu_i}{h}\right) \right)
= O\left( \frac{1}{NT^j h^{2j-1}} \right),
\end{align*}
by Lemmas \ref{lem-mu} and \ref{lem-kernel}.
Thus, \eqref{cdf-mu-4} can be written as: 
\begin{align*}
\sum_{j=3}^{\infty} \frac{(-1)^j}{j! N h^j} \sum_{i=1}^N (\hat \mu_i - \mu_i)^j K^{(j-1)} \left( \frac{x -  \mu_i}{h}\right) 
= \sum_{j=3}^{\infty} \left( \frac{B_{\mu, j}(x)}{T^j h^{2j-2}} + o_p\left( \frac{1}{\sqrt{T^j h^{2j-2}}} \right) \right).
\end{align*}

Consequently, we obtain the desired result for $\hat F_{\hat \mu}(x)$ by Slutsky's theorem.

\paragraph{The CDF of $\gamma_{k,i}$.}
We evaluate each term in the following Taylor expansion:
\begin{align}
\hat F_{\hat \gamma_k}(x) - F_{\gamma_k}(x) 
=&  \frac{1}{N} \sum_{i=1}^N \mathbb{K} \left( \frac{x- \gamma_{k,i}}{h}\right) 
- F_{\gamma_k} (x) \label{cdf-gam-1}\\
& - \frac{1}{Nh} \sum_{i=1}^N (\hat \gamma_{k,i} - \gamma_{k,i}) K \left( \frac{x-\gamma_{k,i}}{h}\right)  \label{cdf-gam-2}\\
& + \frac{1}{2Nh^2} \sum_{i=1}^N (\hat \gamma_{k,i} - \gamma_{k,i})^2 K' \left( \frac{x- \gamma_{k,i}}{h}\right) \label{cdf-gam-3}\\
& + \sum_{j = 3}^{\infty}  \frac{(-1)^j}{j! N h^j} \sum_{i=1}^N (\hat \gamma_{k,i} - \gamma_{k,i})^j K^{(j-1)} \left( \frac{x- \gamma_{k,i}}{h}\right). \label{cdf-gam-4}
\end{align}

For \eqref{cdf-gam-1}, consistency and asymptotic normality are established by the same arguments as for the CDF of $\mu_i$.

For \eqref{cdf-gam-2}, we have the following equation by the expansion for $\hat \gamma_{k,i}$:
\begin{align}
\frac{1}{Nh} \sum_{i=1}^N (\hat \gamma_{k,i} - \gamma_{k,i}) K \left( \frac{ x- \gamma_{k,i}}{h}\right)
=& \frac{1}{Nh} \sum_{i=1}^N \left(  \frac{1}{T-k} \sum_{t=k+1}^T w_{it} w_{i,t-k} - \gamma_{k,i}\right) K \left( \frac{x - \gamma_{k,i}}{h}\right) \label{cdf-gam-5} \\
&- \frac{1}{Nh} \sum_{i=1}^N \frac{T+k}{T-k} (\bar w_i)^2 K \left( \frac{ x-\gamma_{k,i} }{h}\right) \label{cdf-gam-6} \\
&+ \frac{1}{Nh} \sum_{i=1}^N \frac{1}{T-k} \sum_{t=1}^k w_{it}  \bar w_{i} K \left( \frac{x- \gamma_{k,i}}{h}\right) \label{cdf-gam-7} \\
&+ \frac{1}{Nh} \sum_{i=1}^N \frac{1}{T-k} \sum_{t=T-k+1}^T w_{it} \bar w_{i} K \left( \frac{ x-\gamma_{k,i} }{h}\right). \label{cdf-gam-8}
\end{align}
For the term in \eqref{cdf-gam-5}, the mean is zero and the variance is: 
\begin{align*}
var\left( \frac{1}{Nh} \sum_{i=1}^{N}\left( \frac{1}{T-k}\sum_{t=k+1}^{T}w_{it}w_{i,t-k} - \gamma_{k,i} \right) K \left( \frac{ x-\gamma_{k,i}}{h}\right) \right)
= O\left( \frac{1}{NTh} \right),
\end{align*}
by Lemmas \ref{lem-wk} and \ref{lem-kernel}.
Thus, \eqref{cdf-gam-5} is $O_{p}( 1/\sqrt{NTh})$.
For \eqref{cdf-gam-6}, the mean is:
\begin{align*}
E \left( \frac{1}{Nh} \frac{T+k}{T-k} \sum_{i=1}^N (\bar w_{i})^2 K \left( \frac{x- \gamma_{k,i}}{h}\right) \right)
&= \frac{T+k}{hT(T-k)} E \left( E[T(\bar w_i)^2|\gamma_{k,i}] K \left( \frac{x-\gamma_{k,i}}{h}\right) \right) \\
&= \frac{1}{T} E(T (\bar w_i)^2|\gamma_{k,i}=x) f_{\gamma_k}(x) \int K(s) ds + o\left( \frac{1}{T} \right)\\
&= - \frac{B_{\gamma_k, 1}(x)}{T} + o\left( \frac{1}{T} \right),
\end{align*}
by the law of iterated expectations and Lemma \ref{lem-kernel} with the definition of
\begin{align*}
B_{\gamma_k, 1}(x) \coloneqq - \lim_{T \to \infty} T \cdot E((\bar w_i)^2|\gamma_{k,i}=x) f_{\gamma_k}(x) \int K(s) ds.
\end{align*}
The variance is: 
\begin{align*}
var \left( \frac{1}{Nh} \frac{T+k}{T-k} \sum_{i=1}^N (\bar w_{i})^2 K \left( \frac{ x-\gamma_{k,i}}{h}\right) \right)
= O\left( \frac{1}{NT^2h} \right),
\end{align*}
by Lemmas \ref{lem-wbar} and \ref{lem-kernel}.
Thus, \eqref{cdf-gam-6} is $B_{\gamma_k, 1}(x)/T + o_p(1/T)$.
For \eqref{cdf-gam-7}, the absolute mean is:
\begin{align*}
E \left| \frac{1}{Nh} \sum_{i=1}^{N} \frac{1}{T-k} \sum_{t=1}^{k} w_{it}\bar w_{i} K \left( \frac{ x-\gamma_{k,i} }{h}\right) \right|
&\le \frac{1}{(T-k)h} \sqrt{E\left( \left(\sum_{t=1}^{k} w_{it}\bar w_{i}\right)^{2}\right)}\sqrt{E\left( K^2 \left( \frac{ x-\gamma_{k,i} }{h}\right)\right)}\\
&= O\left( \frac{1}{Th} \right) \cdot O\left( \frac{1}{\sqrt{T}} \right) \cdot O\left( \sqrt{h} \right)\\
&= O\left( \frac{1}{\sqrt{T^3h}} \right),
\end{align*}
by Lemmas \ref{lem-wbar} and \ref{lem-kernel}.
Thus, \eqref{cdf-gam-7} is $O_{p}(1/\sqrt{T^3h})$.
For \eqref{cdf-gam-8}, we can show that it is $O_{p}(1/\sqrt{T^3h})$ by the same argument.
Thus, \eqref{cdf-gam-2} is $B_{\gamma_k, 1}(x)/T + o_p(1/T)$. 

For \eqref{cdf-gam-3}, it is easy to see that: 
\begin{align*}
& \frac{1}{2Nh^2} \sum_{i=1}^N (\hat \gamma_{k,i} - \gamma_{k,i})^2 K^{'} \left( \frac{x-\gamma_{k,i}}{h} \right)\\
&= \frac{1}{2Nh^2} \sum_{i=1}^N \left( \frac{1}{T-k} \sum_{t=k+1}^T w_{it}w_{i,t-k} - \gamma_{k,i} \right)^2 K' \left( \frac{x-\gamma_{k,i}}{h} \right) + o_p\left( \frac{1}{T} \right),
\end{align*}
by similar procedures, to show the orders of \eqref{cdf-gam-5}, \eqref{cdf-gam-6}, \eqref{cdf-gam-7}, and \eqref{cdf-gam-8}.
Introducing the shorthand notation $d_T(x) \coloneqq E[(\sum_{t=k+1}^T [w_{it}w_{i,t-k} - \gamma_{k,i}])^2|\gamma_{k,i}=x] f_{\gamma_k}(x) / (T-k)$, the mean of the term is: 
\begin{align*}
& E\left( \frac{1}{2Nh^2} \sum_{i=1}^N \left( \frac{1}{T-k} \sum_{t=k+1}^T w_{it}w_{i,t-k} - \gamma_{k,i} \right)^2 K' \left( \frac{x-\gamma_{k,i}}{h} \right) \right) \\
=& \frac{1}{2(T-k)h^2} E\left( E\left( \frac{1}{T-k} \left(\sum_{t=k+1}^T w_{it}w_{i,t-k} - \gamma_{k,i} \right)^2 \middle| \gamma_{k,i} \right)  K' \left( \frac{x-\gamma_{k,i}}{h} \right) \right)\\
=& \frac{1}{2(T-k)h} d_T(x) \int K'(s) ds -  \frac{1}{2(T-k)} d_T'(x) \int s K'(s) ds + o\left( \frac{1}{T} \right)\\
=& - \frac{1}{2T} d_T'(x) \int sK'(s) ds + o\left( \frac{1}{T} \right)\\
=& \frac{B_{\gamma_k, 2}(x)}{T} + o\left( \frac{1}{T} \right),
\end{align*}
by Lemma \ref{lem-kernel} and $\int K'(s)ds = 0$ with the definition of 
\begin{align*}
B_{\gamma_k, 2}(x) \coloneqq - \lim_{T \to \infty} \frac{d_T'(x)}{2} \int sK'(s) ds.
\end{align*}
The variance of the term is: 
\begin{align*}
var\left( \frac{1}{2Nh^2} \sum_{i=1}^N \left( \frac{1}{T-k} \sum_{t=k+1}^T w_{it}w_{i,t-k} - \gamma_{k,i} \right)^2 K' \left( \frac{x-\gamma_{k,i}}{h} \right) \right)
= O\left( \frac{1}{NT^2h^3} \right),
\end{align*}
by Lemmas \ref{lem-wk} and \ref{lem-kernel}.
Thus, \eqref{cdf-gam-3} is $B_{\gamma_k, 2}(x)/T + o_p(1/T)$.

For \eqref{cdf-gam-4}, it is easy to see that:
\begin{align*}
& \frac{1}{j! N h^j} \sum_{i=1}^N (\hat \gamma_{k,i} - \gamma_{k,i})^j K^{(j-1)} \left( \frac{x- \gamma_{k,i}}{h}\right) \\
&= \frac{1}{j! N h^j} \sum_{i=1}^N \left( \frac{1}{T-k} \sum_{t=k+1}^T w_{it}w_{i,t-k} - \gamma_{k,i} \right)^j K^{(j-1)} \left( \frac{x- \gamma_{k,i}}{h}\right) + o_p\left( \frac{1}{\sqrt{T^j h^{2j-2}}} \right),
\end{align*}
by the same argument as for \eqref{cdf-gam-3}.
The mean of the term is: 
\begin{align*}
& E \left( \frac{1}{j!Nh^j} \sum_{i=1}^N \left( \frac{1}{T-k} \sum_{t=k+1}^T w_{it}w_{i,t-k} - \gamma_{k,i} \right)^j K^{(j-1)} \left( \frac{x- \gamma_{k,i}}{h}\right) \right) \\
&= \frac{1}{j!(T-k)^{j/2} h^j} E \left( \frac{1}{(T-k)^{j/2}} E\left( \left( \sum_{t=k+1}^T (w_{it}w_{i,t-k} - \gamma_{k,i}) \right)^j \middle| \gamma_{k, i} \right)  K^{(j-1)} \left( \frac{x- \gamma_{k,i}}{h}\right) \right) \\
&= \frac{1}{ j!\sqrt{T^j h^{2j-2}}} \left( \frac{1}{\sqrt{T^j}} E\left( \left(\sum_{t=k+1}^T (w_{it}w_{i,t-k} - \gamma_{k,i}) \right)^j \middle| \gamma_{k,i} = x \right) f_{\gamma_k}(x) \right) \int K^{(j-1)}(s) ds + o_p\left( \frac{1}{ \sqrt{T^j h^{2j-2}}} \right)\\
&= (-1)^j \frac{B_{\gamma_k, j}(x)}{ \sqrt{T^j h^{2j-2}}} + o_p\left( \frac{1}{ \sqrt{T^j h^{2j-2}}} \right),
\end{align*}
by Lemma \ref{lem-kernel} with the definition of 
\begin{align*}
B_{\gamma_k, j}(x) \coloneqq \lim_{T \to \infty} \frac{(-1)^j}{j!} E\left( \frac{1}{\sqrt{T^j}} \left(\sum_{t=k+1}^T (w_{it}w_{i,t-k} - \gamma_{k,i}) \right)^j \middle| \gamma_{k,i} = x \right) f_{\gamma_k}(x) \int K^{(j-1)}(s) ds.
\end{align*}
The variance is: 
\begin{align*}
var \left( \frac{1}{j! N h^j} \sum_{i=1}^N \left( \frac{1}{T-k} \sum_{t=k+1}^T w_{it}w_{i,t-k} - \gamma_{k,i} \right)^j K^{(j-1)} \left( \frac{x- \gamma_{k,i}}{h}\right) \right)
= O\left( \frac{1}{N T^j h^{2j-1}} \right),
\end{align*}
by Lemmas \ref{lem-gamma} and \ref{lem-kernel}.
Thus, \eqref{cdf-gam-4} can be written as: 
\begin{align*}
\sum_{j = 3}^{\infty} \frac{(-1)^j}{j! N h^j} \sum_{i=1}^N (\hat \gamma_{k,i} - \gamma_{k,i})^j K^{(j-1)} \left( \frac{x- \gamma_{k,i}}{h}\right)
= \sum_{j=3}^{\infty} \left( \frac{B_{\gamma_k, j}(x)}{ \sqrt{T^j h^{2j-2}}} + o_p\left( \frac{1}{ \sqrt{T^j h^{2j-2}}} \right) \right).
\end{align*}

Consequently, we obtain the desired result for $\hat F_{\hat \gamma_k}(x)$ by Slutsky's theorem. 

\paragraph{The CDF of $\rho_{k,i}$.}
We regard $\mathbb{K}( (x-\hat \rho_{k,i})/h) = \mathbb{K}((x-\hat \gamma_{k,i}/\hat \gamma_{0,i})/h)$ as a function of two variables $(\hat \gamma_{k,i}, \hat \gamma_{0,i} )$.
Taylor's theorem for multivariate functions leads to:
\begin{align}
&\hat F_{\hat \rho_k}(x) - F_{\rho_k}(x) \nonumber \\
=& \frac{1}{N} \sum_{i=1}^N \mathbb{K} \left( \frac{ x- \rho_{k,i}}{h}\right) - F_{\rho_k}(x) \label{cdf-rho-1} \\
& + \frac{1}{N} \sum_{i=1}^N \sum_{j_1+j_2=1} (\hat \gamma_{k,i} - \gamma_{k,i})^{j_1} (\hat \gamma_{0,i} - \gamma_{0,i})^{j_2}  \frac{\partial^{j_1 + j_2}}{\partial a^{j_1} \partial b^{j_2}} \mathbb{K} \left( \frac{ x- a/b}{h}\right) \Big|_{a = \gamma_{k,i}, b = \gamma_{0,i}} \label{cdf-rho-2} \\
& + \frac{1}{N} \sum_{i=1}^N \sum_{j_1+j_2=2} \frac{1}{j_1!j_2!} (\hat \gamma_{k,i} - \gamma_{k,i})^{j_1} (\hat \gamma_{0,i} - \gamma_{0,i})^{j_2}  \frac{\partial^{j_1 + j_2}}{\partial a^{j_1} \partial b^{j_2}} \mathbb{K} \left( \frac{ x- a/b}{h}\right) \Big|_{a = \gamma_{k,i}, b = \gamma_{0,i}} \label{cdf-rho-3} \\
& + \sum_{j=3}^{\infty} \frac{1}{N} \sum_{i=1}^N \sum_{j_1+j_2=j} \frac{1}{j_1!j_2!} (\hat \gamma_{k,i} - \gamma_{k,i})^{j_1} (\hat \gamma_{0,i} - \gamma_{0,i})^{j_2}  \frac{\partial^{j_1 + j_2}}{\partial a^{j_1} \partial b^{j_2}} \mathbb{K} \left( \frac{ x- a/b}{h}\right) \Big|_{a = \gamma_{k,i}, b = \gamma_{0,i}}. \label{cdf-rho-4}
\end{align}

For \eqref{cdf-rho-1}, the consistency and asymptotic normality of the term are established by the same argument as for the CDF of $\mu_i$.

\eqref{cdf-rho-2} contains two terms.
Of these, we focus only on $(Nh)^{-1} \sum_{i=1}^N \gamma_{0,i}^{-1} (\hat \gamma_{k,i} - \gamma_{k,i}) K ( (x-\rho_{k,i})/h)$, as the other term can be evaluated in the same manner.
However, this term is analogous to that in \eqref{cdf-gam-2}, so it can be evaluated by the same argument.
This means that \eqref{cdf-rho-2} can be written as $B_{\rho_k, 1}(x) + o_p(1/T)$ for a nonrandom $B_{\rho_k, 1}(x)$.

\eqref{cdf-rho-3} contains three terms.
Of these, we focus only on $(2Nh^2)^{-1} \sum_{i=1}^N \gamma_{0,i}^{-2} (\hat \gamma_{k,i} - \gamma_{k,i})^2 K'((x - \rho_{k,i})/h)$, as the other terms can be evaluated in the same manner.
However, this term is analogous to that in \eqref{cdf-gam-3}, so it can be evaluated by the same argument.
This means that \eqref{cdf-rho-3} is also $B_{\rho_k, 2}(x) / T + o_p(1/T)$ for a nonrandom $B_{\rho_k, 2}(x)$.

For \eqref{cdf-rho-4}, we evaluate the mean of the term
\begin{align*}
\sum_{j_1+j_2=j} \frac{1}{j_1!j_2!} (\hat \gamma_{k,i} - \gamma_{k,i})^{j_1} (\hat \gamma_{0,i} - \gamma_{0,i})^{j_2}  \frac{\partial^{j_1 + j_2}}{\partial a^{j_1} \partial b^{j_2}} \mathbb{K} \left( \frac{ x- a/b}{h}\right) \Big|_{a = \gamma_{k,i}, b = \gamma_{0,i}},
\end{align*}
which contains $j+1$ terms.
Of these terms, we consider only: 
\begin{align*}
\frac{1}{j! N h^j} \sum_{i=1}^N \frac{1}{\gamma_{0,i}^j} (\hat \gamma_{k,i} - \gamma_{k,i})^j K^{(j-1)}\left( \frac{ x - \rho_{k,i} }{h} \right),
\end{align*}
as the other terms can be evaluated in the same manner.
However, this term is analogous to that in \eqref{cdf-gam-4}, so we can evaluate it using the same argument.
This means that \eqref{cdf-rho-4} can be written as: 
\begin{align*}
& \sum_{j=3}^{\infty} \frac{1}{N} \sum_{i=1}^N \sum_{j_1+j_2=j} \frac{1}{j_1!j_2!} (\hat \gamma_{k,i} - \gamma_{k,i})^{j_1} (\hat \gamma_{0,i} - \gamma_{0,i})^{j_2}  \frac{\partial^{j_1 + j_2}}{\partial a^{j_1} \partial b^{j_2}} \mathbb{K} \left( \frac{ x- a/b}{h}\right) \Big|_{a = \gamma_{k,i}, b = \gamma_{0,i}} \\
&= \sum_{j=3}^{\infty} \left( \frac{B_{\rho_k, j}(x)}{\sqrt{T^j h^{2j-2}}} + o\left( \frac{1}{\sqrt{T^j h^{2j - 2}}} \right) \right),
\end{align*}
for a nonrandom $B_{\rho_k, j}(x)$.

Consequently, we have the desired result for $\hat F_{\hat \rho_k}(x)$ by Slutsky's theorem.	
\qed

\section{Appendix: Lemmas} \label{sec-lemma}

This appendix contains the technical lemmas used to demonstrate the theorems in the main body.

We first present the lemmas for which the proofs are given in \citet{Okui2017}.

\begin{lemma}\label{lem-wbar}
	Suppose that Assumptions \ref{as-basic}, \ref{as-mixing-c}, and \ref{as-w-moment-c} hold for 
	$r_m = r$ and $r_d =r$ with a natural number $r$.
	Then, it holds that $E((\bar w_i)^r ) = O(T^{-r/2})$.
\end{lemma}

\begin{lemma}\label{lem-mu}
	Suppose that Assumptions \ref{as-basic}, \ref{as-mixing-c}, and \ref{as-w-moment-c} hold for $r_m = r$ and $r_d =r$ with a natural number $r$.
	Then, it holds that $E\left( (\hat \mu_i -\mu_i)^{r}\right) = O(T^{-r/2})$.
\end{lemma}

\begin{lemma}\label{lem-wk}
	Suppose that Assumptions \ref{as-basic}, \ref{as-mixing-c}, and \ref{as-w-moment-c} hold for $r_m = r$ and $r_d = 2r$ with a natural number $r$.
	Then, it holds that $E((\sum_{t=k+1}^T (w_{it}w_{i,t-k} - \gamma_{k,i}))^r) = O(T^{r/2})$.
\end{lemma}

\begin{lemma}\label{lem-gamma}
	Suppose that Assumptions \ref{as-basic}, \ref{as-mixing-c}, and \ref{as-w-moment-c} hold  for $r_m = 2r$ and $r_d=2r$ with a natural number $r$.
	Then, it holds that $E((\hat\gamma_{k,i}-\gamma_{k,i})^{r}) = O(T^{-r/2})$.
\end{lemma}

\begin{lemma}\label{lem-rho}
	Suppose that Assumptions \ref{as-basic}, \ref{as-mixing-c}, \ref{as-w-moment-c}, and \ref{as-rho} hold for $r_m = 2r$ and $r_d=2r$ with a natural number $r$.
	Then, it holds that $E((\hat\rho_{k,i}-\rho_{k,i})^{r}) = O(T^{-r/2})$.
\end{lemma}

We repeatedly use the following lemmas to prove our theorems.
The proofs are similar to those in \citet{pagan1999nonparametric} and \citet{LiRacine2007}, and are omitted.

\begin{lemma}\label{lem-kernel}
	Consider a continuous random variable $X \in \mathbb{R}$, a random vector $Y = (Y_1, Y_2, \dots, Y_d)^\top \in \mathbb{R}^d$, and an interior point $x \in \mathbb{R}$.
	Suppose that a function $g_1: \mathbb{R} \to \mathbb{R}$ satisfies $\int |g_1(s)| ds < \infty$, $\int |s g_1(s)| ds < \infty$, and $\int |s^2 g_1(s)| ds < \infty$, and that $E[g_2(X,Y)|X=\cdot] : \mathbb{R} \to \mathbb{R}$ and the density $f_X:\mathbb{R} \to \mathbb{R}$ are twice boundedly continuously differentiable at $x$.
	It holds that 
	\begin{align*}
	E\left( g_1\left( \frac{x-X}{h}\right) g_2(X,Y) \right) 
	&= h A(x) \int g_1(s) ds - h^2 A'(x) \int s g_1(s) ds + o(h^2)\\
	&= O(h) + O(h^2) + o(h^2),
	\end{align*}
	where $A(x) \coloneqq E[g_2(X,Y)|X=x] f_X(x)$.
\end{lemma}

Note that the above result implies that, if we set $g_2(x,y)=1$ (constant):
\begin{align*}
E\left( g_1\left( \frac{x-X}{h}\right) \right) 
&= h f_X(x) \int g_1(s) ds - h^2 f'_X(x) \int s g_1(s) ds + o(h^2)\\
&= O(h) + O(h^2) + o(h^2).
\end{align*}

Suppose that $\{X_i\}_{i=1}^N$ is a random sample of a continuous random variable $X \in \mathbb{R}$.
We denote the density and CDF of $X$ by $f_X(\cdot)$ and $F_X(\cdot) = \Pr(X \le \cdot)$, respectively.

\begin{lemma}\label{lem-dens}
	Let $\hat f_X(x) \coloneqq (Nh)^{-1} \sum_{i=1}^N K((x - X_i)/h)$ be the kernel density estimator.
	Let $x \in \mathbb{R}$ be a fixed interior point in the support of $X$.
	Suppose that the kernel function $K: \mathbb{R} \to \mathbb{R}$ is symmetric and satisfies $\int K(s)ds=1$, $\kappa_1 = \int s^2 K(s) ds < \infty$, $\kappa_2 = \int K^2(s) ds < \infty$, and $\int | s^3 K(s)| ds < \infty$, and that $f_X$ is bounded away from zero and three-times boundedly continuously differentiable near $x$.
	When $N \to \infty$ and $h \to 0$ with $Nh \to \infty$, it holds that $E( \hat f_X(x) ) = f_X(x) + h^2 \kappa_1 f''_X(x) / 2 + o(h^2)$ and $var( \hat f_X(x) ) = \kappa_2 f_X(x) / (Nh) + o( (Nh)^{-1} )$.
	Moreover, when $\int |K(s)|^3 ds < \infty$ and $Nh^5 \to C \in [0, \infty)$, it holds that $\sqrt{Nh} ( \hat f_X(x) - f_X(x) - h^2 \kappa_1 f''_X(x) / 2 ) \stackrel{d}{\longrightarrow} \mathcal{N} (0, \kappa_2 f_X(x) )$.
\end{lemma}

\begin{lemma}\label{lem-CDF}
	Let $\hat F_X(x) \coloneqq N^{-1} \sum_{i=1}^N \mathbb{K}((x-X_i)/h)$ be the kernel CDF estimator.
	Let $x \in \mathbb{R}$ be a fixed interior point in the support of $X$.
	Let $K(s) = d \mathbb{K}(s)/ds$ be the derivative.
	Suppose that $K: \mathbb{R} \to \mathbb{R}$ is symmetric and satisfies $\int K(s) ds =1$, $\kappa_1 = \int s^2 K(s) ds < \infty$, $\int |s^3K(s)| ds<\infty$, and $\int |s K(s) | \mathbb{K}(s) ds < \infty$, and that $F_X$ is three-times boundedly continuously differentiable near $x$.
	When $N \to \infty$ and $h \to 0$, then $E( \hat F_X(x) ) = F_X(x) + h^2 \kappa_1 f_X'(x) /2 + o(h^2)$ and $var( \hat F_X(x) ) = F_X(x) [ 1- F_X(x) ] / N + o( N^{-1} )$.
	Moreover, when $Nh^4 \to 0$ also holds, it holds that $\sqrt{N} ( \hat F_X(x) - F_X(x) ) \stackrel{d}{\longrightarrow} \mathcal{N} (0, F_X(x) [ 1 - F_X(x) ] )$.
\end{lemma}

\section{Appendix: The validity of the infinite-order Taylor expansion}\label{sec-expansion}

This appendix discusses the validity of the infinite-order Taylor expansion for the density estimation in Theorem \ref{thm-dens}.
The discussion for the expansion of the CDF estimation in Theorem \ref{thm-CDF} is similar.

The infinite-order Taylor expansion of $\hat f_{\hat \xi}(x) = (Nh)^{-1} \sum_{i=1}^N K((x - \hat \xi_i) / h)$ is:
\begin{align*}
\hat f_{\hat \xi}(x) = \frac{1}{Nh} \sum_{i=1}^N \sum_{j=0}^\infty \frac{(-1)^j }{j!} \frac{(\hat \xi_i - \xi_i)^j}{h^j} K^{(j)} \left( \frac{x - \xi_i}{h} \right).
\end{align*}
It holds if the remainder term of the finite-order Taylor expansion converges to zero as the order of the expansion increases. We show that it is the case with probability approaching one.
The remainder term is given by:
\begin{align*}
\frac{1}{N h^{j+1}} \frac{(-1)^j}{j!} \sum_{i=1}^N (\hat \xi_i - \xi_i )^j K^{(j)} \left( \frac{ x- \tilde \xi_i}{h} \right),
\end{align*}
where $\tilde \xi_i$ is between $\hat \xi_i$ and $\xi_i$. 
It is sufficient to argue that it converges to zero, as $j \to \infty$, with probability approaching one.
We observe that: 
\begin{equation}\label{eq:remainder1}
\begin{split}
& \left| \frac{1}{N h^{j+1}} \frac{(-1)^j}{j!} \sum_{i=1}^N (\hat \xi_i - \xi_i )^j K^{(j)} \left( \frac{ x- \tilde \xi_i}{h} \right) \right| \\
&\le \left( \frac{1}{h^j} \max_{1 \le i \le N} |\hat \xi_i - \xi_i|^j \right) \left( \frac{1}{ j ! h }\max_{1 \le i \le N}  \left| K^{(j)} \left( \frac{ x- \tilde \xi_i}{h} \right) \right| \right).
\end{split}
\end{equation}

We argue that the term in the first parenthesis of \eqref{eq:remainder1} converges to zero, as $j \to \infty$, with probability approaching one. Note that the convergence holds when $\max_{1 \le i \le N} |\hat \xi_i - \xi_i| / h < 1$. For this, we observe that for any fixed $\varepsilon > 0$ and positive integer $r \ge 1$:
\begin{equation}\label{eq:remainder2}
\begin{split}
\Pr\left( \frac{1}{h} \max_{1 \le i \le N} |\hat \xi_i - \xi_i| \le \varepsilon \right)
=	\Pr\left( \frac{1}{h^r} \max_{1 \le i \le N} |\hat \xi_i - \xi_i|^r \le \varepsilon^r \right)
& = \left( \Pr\left( |\hat \xi_i - \xi_i|^r \le \varepsilon^r h^r \right) \right)^N\\
& \ge \left( 1 - \frac{E|\hat \xi_i - \xi_i|^r}{\varepsilon^r h^r} \right)^N \\
& \ge \left( 1 - \frac{M}{\sqrt{T^r h^{2r}}} \right)^N,
\end{split}
\end{equation}
by Assumption \ref{as-basic}, Markov's inequality, and Lemma \ref{lem-mu}, \ref{lem-gamma}, or \ref{lem-rho} with fixed $M > 0$.
The probability on the left-hand side of \eqref{eq:remainder2} thus converges to one if $(1 - 1 /\sqrt{T^r h^{2r}})^N \to 1$.
Based on the binomial theorem, we observe that:
\begin{align*}
\left( 1 - \frac{1}{\sqrt{T^r h^{2r}}} \right)^N 
&= \sum_{l = 0}^N \binom{N}{l} \left( - \frac{1}{\sqrt{T^r h^{2 r}}} \right)^l \\
&= 1 - \frac{N}{\sqrt{T^r h^{2 r}}} + \frac{N(N-1)}{2! (\sqrt{T^r h^{2 r}})^2} - \frac{N(N-1)(N-2)}{3! (\sqrt{T^r h^{2 r}})^3} + \cdots + \left(- \frac{1}{\sqrt{T^r h^{2 r}}} \right)^N.
\end{align*}
As a result, the probability on the left-hand side of \eqref{eq:remainder2} converges to one if $N / \sqrt{T^r h^{2r}} \to 0$ for sufficiently large $r$ as $N \to \infty$ and $Th^2 \to \infty$.
By taking $\varepsilon <1 $, we obtain the desired result.
We note that the condition is significantly weaker than the relative magnitudes condition in Theorem \ref{thm-dens}. 
Hence, the term in the first parenthesis of \eqref{eq:remainder1} converges to zero, as $j \to \infty$,  with probability approaching one.

In a similar manner, we can observe that the term in the second parenthesis of \eqref{eq:remainder1} converges to zero with probability approaching one under regularity conditions.

Therefore, the infinite-order Taylor expansion in Theorem \ref{thm-dens} holds under regularity conditions.

\section{Appendix: The infinite series of the asymptotic biases}\label{sec-series}

This appendix discusses the conditions under which the infinite series of the asymptotic biases is well defined (i.e., summable and convergent).
We focus on the density estimator for $\mu_i$ only, because the discussions for the other estimators are similar.

To examine the series of the asymptotic biases, we focus on the nonlinearity biases of the density estimator $\hat f_{\hat \mu}(x)$.
Let $e_{T, j}(x) \coloneqq E( T^{j/2} (\bar w_i)^j | \mu_i = x ) f_{\mu}(x)$.
For the nonlinearity bias in \eqref{dens-mu-3} of the proof of the density estimation, we observe that:
\begin{align*}
& E\left( \frac{1}{j!Nh^{j+1}} \sum_{i=1}^N (\bar w_i)^j K^{(j)} \left( \frac{x - \mu_i }{h}\right) \right) \\
&= \frac{1}{j!h^{j+1}} E\left( E\left( (\bar w_i)^j | \mu_i \right) K^{(j)} \left( \frac{x - \mu_i }{h}\right) \right)\\
&= \frac{1}{j!T^{j/2} h^j} e_{T,j}(x) \int K^{(j)}(s) ds - \frac{1}{j!T^{j/2} h^{j-1}} e'_{T,j}(x) \int s K^{(j)}(s) ds + \frac{1}{j!T^{j/2} h^{j-2}} \int e''_{T,j}(\tilde x) s^2 K^{(j)}(s) ds,
\end{align*}
by the law of iterated expectations, the change of variables, and Taylor's theorem with $\tilde x$ located between $x - sh$ and $x$.
The equation for any odd $j$ is equal to:
\begin{align*}
-\frac{1}{j!T^{j/2} h^{j-1}} e'_{T,j}(x) \int s K^{(j)}(s) ds + \frac{1}{j!T^{j/2} h^{j-2}} \int e''_{T,j}(\tilde x) s^2 K^{(j)}(s) ds,
\end{align*}
because of the symmetry of $K$.
On the contrary, the equation for any even $j$ is equal to: 
\begin{align*}
\frac{1}{j!T^{j/2} h^j} e_{T,j}(x) \int K^{(j)}(s) ds + \frac{1}{j!T^{j/2} h^{j-2}} \int e''_{T,j}(\tilde x) s^2 K^{(j)}(s) ds.
\end{align*}

We focus on the summability of the series of the biases for odd $j$ only, because the discussion for even $j$ is the same.
The partial sum of the series of the biases for odd $j$ can be written as: 
\begin{align*}
& \sum_{j = 3}^{2n + 1} \left( - \frac{1}{j!T^{j/2} h^{j-1}} e'_{T,j}(x) \int s K^{(j)}(s) ds + \frac{1}{j!T^{j/2} h^{j-2}} \int e''_{T,j}(\tilde x) s^2 K^{(j)}(s) ds \right) \\
& = \sum_{l = 1}^n \left( - \frac{1}{(2l + 1)!T^{(2l + 1)/2} h^{2l}} e'_{T, 2l+1}(x) \int s K^{(2l + 1)}(s) ds \right. \\
& \quad \left. + \frac{1}{(2l + 1)!T^{(2l + 1)/2} h^{2l - 1}} \int e''_{T,2l+1}(\tilde x) s^2 K^{(2l + 1)}(s) ds \right) \\
&= - \left(\sum_{l =1}^n S_{1l} \right) +  \left(\sum_{l =1}^n S_{2l} \right),
\end{align*}
where we define the variables $S_{1l} \coloneqq e'_{T,2l+1}(x) \int s K^{(2l + 1)}(s) ds / [(2l + 1)!T^{(2l + 1)/2} h^{2l}]$ and $S_{2l} \coloneqq \int e''_{T,2l+1}(\tilde x) s^2 K^{(2l + 1)}(s) ds / [(2l + 1)!T^{(2l + 1)/2} h^{2l-1}]$.

We examine the series of $S_{1l}$ only. The discussion for $S_{2l}$ is the same.
The ratio test means that the series $\sum_{l = 1}^{\infty} S_{1l}$ is summable and convergent if $\lim_{l \to \infty} |S_{1, l + 1}/ S_{1l}| < 1$.
We observe that:
\begin{align*}
\left| \frac{S_{1,l+1}}{S_{1,l}} \right| = \left| \frac{1}{Th^2} \frac{1}{(2l+3)(2l+2)} \frac{e_{T,2l+3}'(x)}{e_{T,2l+1}'(x)} \frac{\int sK^{(2l + 3)}(s) ds}{\int sK^{(2l + 1)}(s) ds} \right|,
\end{align*}
for any $l \ge 1$.
It converges to zero as $l \to \infty$ if $e_{T,2l+3}'(x) / e_{T,2l+1}'(x) = O(1)$ over $l$ and if $\int sK^{(2l + 3)}(s) ds / \int sK^{(2l + 1)}(s) ds = O(l)$.
The former is a regularity condition, and we can simply assume it.
The latter is also a regularity condition, and we can easily show its validity by assuming that $K$ is the Gaussian kernel function.
Hence, the condition for the ratio test is satisfied, implying that the series $\sum_{l = 1}^{\infty} S_{1l}$ is summable and convergent.

The above discussions imply that the infinite series of the asymptotic biases is well defined under regularity conditions.

\section{Appendix: Monte Carlo simulations} \label{sec-simulation}
This section presents the results of the Monte Carlo simulations.
We here focus on the density estimation only.
The number of simulation replications is 5,000.

\paragraph{Design.}
We generate the data using the AR(1) process $y_{it}= (1-\phi_i)\varsigma_i + \phi_i y_{i,t-1} + \sqrt{(1-\phi_i^2)\sigma_i^2} u_{it}$ where $u_{it} \sim \mathcal{N}(0,1)$, $y_{i0} \sim \mathcal{N}(\varsigma_i , \sigma_i^2)$, and $u_{i0} \sim \mathcal{N}(0, 1)$.
Note that this design satisfies $\mu_i = \varsigma_i$, $\gamma_{0,i} = \sigma_i^2$, and $\rho_{1,i} = \phi_i$.
We generate the unit-specific random variables $\varsigma_i \sim \mathcal{N}(-1, 1)$, $\phi_i \sim 2 \cdot Beta(2, 4) - 1$, and $\sigma_i^2 \sim 3 \cdot Beta(3, 2)$.
We consider $N = 250, 500, 1000$ and $T = 12, 24, 48, 96$.

\paragraph{Estimators.}
We estimate the densities of $\mu_i$, $\gamma_{0,i}$, and $\rho_{1,i}$ at their 20\%, 40\%, 60\%, and 80\% quantiles based on four estimators.
The first is the naive estimator (NE) without split-panel jackknife bias correction.
The second and third are the HPJ and TOJ estimators.
The fourth is the infeasible estimator (IE) based on the true $\mu_i$, $\gamma_{0,i}$, and $\rho_{1,i}$.
For all estimators, we use the Epanechnikov kernel and the coverage error optimal bandwidth in \citet{calonico2018effect}.

\paragraph{Results.}
Tables \ref{table:mean}, \ref{table:acov}, and \ref{table:acor} present the simulation results for the densities of $\mu_i$, $\gamma_{0,i}$, and $\rho_{1,i}$, respectively.
The tables report the true values of the parameters and the bias and standard deviation (std) of each estimator.
They also report the coverage probability (cp) of the 95\% CI computed by the RBC procedure based on each estimator.
Table \ref{table:band} also describes the mean and the standard deviation of the selected bandwidths for NE and IE.
Note that we use the same bandwidth for HPJ and TOJ as NE as discussed in the main body.

The NE exhibits large biases, especially with small $T$.
In particular, the biases of the density for $\gamma_{0,i}$ and $\rho_{1,i}$ are crucial because of the incidental parameter biases.
As a result, the coverage probabilities of the NE are much smaller than 0.95.
The performance of the NE improves as $T$ grows, but it is unsatisfactory for several parameters even when $T = 96$.
These results highlight the importance of bias correction even with relatively large $T$.

The performances of the HPJ and TOJ are significantly better than the NE.
The HPJ and TOJ operate well especially when the NE exhibits large biases.
The TOJ outperforms the HPJ when the HPJ exhibits relatively large biases, as a result of relatively large higher-order nonlinearity biases.
Furthermore, for several parameters, the TOJ operates as well as the IE in terms of bias reduction and coverage probability.
Note that the TOJ may inflate the estimation variability, especially when $T$ is small, but such cost is inevitable when our goal is to conduct unbiased inferences.

These simulation results demonstrate the severity of the incidental parameter biases and the nonlinearity biases and the success of the split-panel jackknife and the RBC inference.
We thus recommend the RBC inference based on the split-panel jackknife bias-corrected estimation.

\begin{landscape}
	\begin{table}[!t]
		{\scriptsize
			\caption{Monte Carlo simulation results for $\mu$}
			\label{table:mean} 
			\begin{center}
				\begin{tabular}{lrrrcrrrcrrrcrrrcrrr}
					\hline\hline
					\multicolumn{1}{l}{\bfseries }&\multicolumn{3}{c}{\bfseries }&\multicolumn{1}{c}{\bfseries }&\multicolumn{3}{c}{\bfseries NE}&\multicolumn{1}{c}{\bfseries }&\multicolumn{3}{c}{\bfseries HPJ}&\multicolumn{1}{c}{\bfseries }&\multicolumn{3}{c}{\bfseries TOJ}&\multicolumn{1}{c}{\bfseries }&\multicolumn{3}{c}{\bfseries IE}\tabularnewline
					\cline{6-8} \cline{10-12} \cline{14-16} \cline{18-20}
					\multicolumn{1}{l}{}&\multicolumn{1}{c}{true}&\multicolumn{1}{c}{$N$}&\multicolumn{1}{c}{$T$}&\multicolumn{1}{c}{}&\multicolumn{1}{c}{bias}&\multicolumn{1}{c}{std}&\multicolumn{1}{c}{cp}&\multicolumn{1}{c}{}&\multicolumn{1}{c}{bias}&\multicolumn{1}{c}{std}&\multicolumn{1}{c}{cp}&\multicolumn{1}{c}{}&\multicolumn{1}{c}{bias}&\multicolumn{1}{c}{std}&\multicolumn{1}{c}{cp}&\multicolumn{1}{c}{}&\multicolumn{1}{c}{bias}&\multicolumn{1}{c}{std}&\multicolumn{1}{c}{cp}\tabularnewline
					\hline
					at $\mu$'s 20\%Q&$0.280$&$ 250$&$12$&&$-0.013$&$0.032$&$0.914$&&$-0.008$&$0.039$&$0.938$&&$-0.007$&$0.051$&$0.945$&&$-0.005$&$0.034$&$0.941$\tabularnewline
					&$0.280$&$ 250$&$24$&&$-0.009$&$0.032$&$0.940$&&$-0.006$&$0.038$&$0.950$&&$-0.005$&$0.047$&$0.951$&&$-0.004$&$0.034$&$0.946$\tabularnewline
					&$0.280$&$ 250$&$48$&&$-0.007$&$0.033$&$0.936$&&$-0.006$&$0.037$&$0.942$&&$-0.006$&$0.043$&$0.947$&&$-0.004$&$0.034$&$0.946$\tabularnewline
					&$0.280$&$ 250$&$96$&&$-0.007$&$0.034$&$0.938$&&$-0.006$&$0.036$&$0.938$&&$-0.006$&$0.040$&$0.942$&&$-0.005$&$0.034$&$0.943$\tabularnewline
					&$0.280$&$ 500$&$12$&&$-0.011$&$0.024$&$0.916$&&$-0.007$&$0.030$&$0.940$&&$-0.006$&$0.042$&$0.948$&&$-0.004$&$0.026$&$0.949$\tabularnewline
					&$0.280$&$ 500$&$24$&&$-0.008$&$0.024$&$0.938$&&$-0.005$&$0.029$&$0.947$&&$-0.005$&$0.037$&$0.951$&&$-0.003$&$0.026$&$0.950$\tabularnewline
					&$0.280$&$ 500$&$48$&&$-0.006$&$0.026$&$0.931$&&$-0.005$&$0.030$&$0.936$&&$-0.005$&$0.036$&$0.938$&&$-0.004$&$0.026$&$0.940$\tabularnewline
					&$0.280$&$ 500$&$96$&&$-0.005$&$0.025$&$0.944$&&$-0.004$&$0.028$&$0.946$&&$-0.004$&$0.031$&$0.946$&&$-0.004$&$0.026$&$0.944$\tabularnewline
					&$0.280$&$1000$&$12$&&$-0.011$&$0.019$&$0.900$&&$-0.006$&$0.024$&$0.936$&&$-0.005$&$0.034$&$0.945$&&$-0.002$&$0.019$&$0.952$\tabularnewline
					&$0.280$&$1000$&$24$&&$-0.007$&$0.019$&$0.933$&&$-0.004$&$0.023$&$0.948$&&$-0.003$&$0.030$&$0.953$&&$-0.003$&$0.020$&$0.948$\tabularnewline
					&$0.280$&$1000$&$48$&&$-0.005$&$0.019$&$0.945$&&$-0.003$&$0.022$&$0.953$&&$-0.003$&$0.027$&$0.953$&&$-0.003$&$0.019$&$0.950$\tabularnewline
					&$0.280$&$1000$&$96$&&$-0.004$&$0.019$&$0.947$&&$-0.003$&$0.021$&$0.950$&&$-0.003$&$0.024$&$0.952$&&$-0.003$&$0.019$&$0.949$\tabularnewline
					\hline
					at $\mu$'s 40\%Q&$0.386$&$ 250$&$12$&&$-0.033$&$0.035$&$0.743$&&$-0.020$&$0.042$&$0.848$&&$-0.014$&$0.054$&$0.890$&&$-0.008$&$0.038$&$0.899$\tabularnewline
					&$0.386$&$ 250$&$24$&&$-0.022$&$0.036$&$0.830$&&$-0.013$&$0.041$&$0.883$&&$-0.009$&$0.050$&$0.902$&&$-0.008$&$0.038$&$0.906$\tabularnewline
					&$0.386$&$ 250$&$48$&&$-0.016$&$0.036$&$0.865$&&$-0.010$&$0.040$&$0.897$&&$-0.008$&$0.046$&$0.910$&&$-0.007$&$0.037$&$0.909$\tabularnewline
					&$0.386$&$ 250$&$96$&&$-0.013$&$0.037$&$0.888$&&$-0.009$&$0.039$&$0.901$&&$-0.008$&$0.043$&$0.908$&&$-0.008$&$0.037$&$0.905$\tabularnewline
					&$0.386$&$ 500$&$12$&&$-0.031$&$0.027$&$0.693$&&$-0.018$&$0.034$&$0.847$&&$-0.011$&$0.045$&$0.899$&&$-0.006$&$0.029$&$0.910$\tabularnewline
					&$0.386$&$ 500$&$24$&&$-0.021$&$0.027$&$0.811$&&$-0.012$&$0.031$&$0.892$&&$-0.008$&$0.040$&$0.913$&&$-0.006$&$0.028$&$0.916$\tabularnewline
					&$0.386$&$ 500$&$48$&&$-0.015$&$0.028$&$0.856$&&$-0.009$&$0.031$&$0.895$&&$-0.007$&$0.036$&$0.912$&&$-0.006$&$0.029$&$0.907$\tabularnewline
					&$0.386$&$ 500$&$96$&&$-0.011$&$0.028$&$0.884$&&$-0.008$&$0.030$&$0.904$&&$-0.007$&$0.033$&$0.911$&&$-0.007$&$0.028$&$0.914$\tabularnewline
					&$0.386$&$1000$&$12$&&$-0.030$&$0.020$&$0.620$&&$-0.016$&$0.026$&$0.850$&&$-0.010$&$0.036$&$0.911$&&$-0.005$&$0.022$&$0.917$\tabularnewline
					&$0.386$&$1000$&$24$&&$-0.020$&$0.021$&$0.781$&&$-0.010$&$0.025$&$0.888$&&$-0.005$&$0.033$&$0.921$&&$-0.004$&$0.022$&$0.911$\tabularnewline
					&$0.386$&$1000$&$48$&&$-0.013$&$0.022$&$0.847$&&$-0.007$&$0.025$&$0.905$&&$-0.005$&$0.030$&$0.921$&&$-0.004$&$0.022$&$0.912$\tabularnewline
					&$0.386$&$1000$&$96$&&$-0.010$&$0.022$&$0.879$&&$-0.006$&$0.023$&$0.910$&&$-0.005$&$0.026$&$0.918$&&$-0.005$&$0.022$&$0.911$\tabularnewline
					\hline
					at $\mu$'s 60\%Q&$0.386$&$ 250$&$12$&&$-0.033$&$0.035$&$0.752$&&$-0.019$&$0.043$&$0.854$&&$-0.014$&$0.055$&$0.888$&&$-0.007$&$0.037$&$0.909$\tabularnewline
					&$0.386$&$ 250$&$24$&&$-0.023$&$0.036$&$0.820$&&$-0.014$&$0.042$&$0.878$&&$-0.010$&$0.051$&$0.904$&&$-0.008$&$0.038$&$0.907$\tabularnewline
					&$0.386$&$ 250$&$48$&&$-0.016$&$0.036$&$0.870$&&$-0.010$&$0.039$&$0.900$&&$-0.008$&$0.045$&$0.912$&&$-0.008$&$0.037$&$0.907$\tabularnewline
					&$0.386$&$ 250$&$96$&&$-0.012$&$0.037$&$0.887$&&$-0.009$&$0.040$&$0.900$&&$-0.007$&$0.043$&$0.907$&&$-0.008$&$0.037$&$0.912$\tabularnewline
					&$0.386$&$ 500$&$12$&&$-0.031$&$0.027$&$0.691$&&$-0.018$&$0.033$&$0.849$&&$-0.012$&$0.044$&$0.898$&&$-0.006$&$0.029$&$0.911$\tabularnewline
					&$0.386$&$ 500$&$24$&&$-0.022$&$0.027$&$0.809$&&$-0.012$&$0.032$&$0.883$&&$-0.008$&$0.040$&$0.905$&&$-0.006$&$0.029$&$0.904$\tabularnewline
					&$0.386$&$ 500$&$48$&&$-0.015$&$0.028$&$0.858$&&$-0.008$&$0.032$&$0.893$&&$-0.006$&$0.037$&$0.908$&&$-0.006$&$0.029$&$0.902$\tabularnewline
					&$0.386$&$ 500$&$96$&&$-0.010$&$0.029$&$0.888$&&$-0.006$&$0.031$&$0.904$&&$-0.005$&$0.034$&$0.914$&&$-0.005$&$0.029$&$0.914$\tabularnewline
					&$0.386$&$1000$&$12$&&$-0.030$&$0.020$&$0.628$&&$-0.016$&$0.026$&$0.851$&&$-0.009$&$0.036$&$0.912$&&$-0.005$&$0.022$&$0.918$\tabularnewline
					&$0.386$&$1000$&$24$&&$-0.020$&$0.021$&$0.777$&&$-0.010$&$0.025$&$0.887$&&$-0.006$&$0.033$&$0.919$&&$-0.004$&$0.022$&$0.916$\tabularnewline
					&$0.386$&$1000$&$48$&&$-0.013$&$0.022$&$0.855$&&$-0.007$&$0.024$&$0.907$&&$-0.005$&$0.030$&$0.920$&&$-0.004$&$0.022$&$0.919$\tabularnewline
					&$0.386$&$1000$&$96$&&$-0.009$&$0.022$&$0.881$&&$-0.005$&$0.024$&$0.898$&&$-0.004$&$0.027$&$0.911$&&$-0.004$&$0.022$&$0.914$\tabularnewline
					\hline
					at $\mu$'s 80\%Q&$0.280$&$ 250$&$12$&&$-0.012$&$0.032$&$0.924$&&$-0.008$&$0.039$&$0.939$&&$-0.007$&$0.051$&$0.946$&&$-0.004$&$0.033$&$0.951$\tabularnewline
					&$0.280$&$ 250$&$24$&&$-0.009$&$0.033$&$0.932$&&$-0.006$&$0.039$&$0.942$&&$-0.006$&$0.049$&$0.947$&&$-0.004$&$0.034$&$0.942$\tabularnewline
					&$0.280$&$ 250$&$48$&&$-0.007$&$0.033$&$0.945$&&$-0.006$&$0.036$&$0.945$&&$-0.006$&$0.043$&$0.943$&&$-0.005$&$0.033$&$0.947$\tabularnewline
					&$0.280$&$ 250$&$96$&&$-0.005$&$0.033$&$0.943$&&$-0.005$&$0.036$&$0.946$&&$-0.005$&$0.040$&$0.945$&&$-0.004$&$0.034$&$0.944$\tabularnewline
					&$0.280$&$ 500$&$12$&&$-0.011$&$0.024$&$0.915$&&$-0.007$&$0.031$&$0.938$&&$-0.005$&$0.041$&$0.948$&&$-0.003$&$0.026$&$0.946$\tabularnewline
					&$0.280$&$ 500$&$24$&&$-0.008$&$0.025$&$0.932$&&$-0.005$&$0.030$&$0.944$&&$-0.005$&$0.038$&$0.947$&&$-0.003$&$0.025$&$0.946$\tabularnewline
					&$0.280$&$ 500$&$48$&&$-0.006$&$0.025$&$0.940$&&$-0.004$&$0.028$&$0.946$&&$-0.004$&$0.034$&$0.952$&&$-0.003$&$0.025$&$0.946$\tabularnewline
					&$0.280$&$ 500$&$96$&&$-0.004$&$0.026$&$0.946$&&$-0.003$&$0.028$&$0.949$&&$-0.003$&$0.031$&$0.945$&&$-0.003$&$0.026$&$0.948$\tabularnewline
					&$0.280$&$1000$&$12$&&$-0.010$&$0.018$&$0.912$&&$-0.006$&$0.023$&$0.948$&&$-0.004$&$0.033$&$0.949$&&$-0.002$&$0.019$&$0.948$\tabularnewline
					&$0.280$&$1000$&$24$&&$-0.007$&$0.018$&$0.934$&&$-0.004$&$0.022$&$0.949$&&$-0.004$&$0.030$&$0.949$&&$-0.003$&$0.019$&$0.951$\tabularnewline
					&$0.280$&$1000$&$48$&&$-0.005$&$0.019$&$0.942$&&$-0.004$&$0.022$&$0.946$&&$-0.003$&$0.027$&$0.950$&&$-0.003$&$0.020$&$0.944$\tabularnewline
					&$0.280$&$1000$&$96$&&$-0.004$&$0.019$&$0.944$&&$-0.003$&$0.021$&$0.950$&&$-0.002$&$0.024$&$0.955$&&$-0.003$&$0.019$&$0.950$\tabularnewline
					\hline
		\end{tabular}\end{center}}
	\end{table}
	
	\begin{table}[!t]
		{\scriptsize
			\caption{Monte Carlo simulation results for $\gamma_0$}
			\label{table:acov} 
			\begin{center}
				\begin{tabular}{lrrrcrrrcrrrcrrrcrrr}
					\hline\hline
					\multicolumn{1}{l}{\bfseries }&\multicolumn{3}{c}{\bfseries }&\multicolumn{1}{c}{\bfseries }&\multicolumn{3}{c}{\bfseries NE}&\multicolumn{1}{c}{\bfseries }&\multicolumn{3}{c}{\bfseries HPJ}&\multicolumn{1}{c}{\bfseries }&\multicolumn{3}{c}{\bfseries TOJ}&\multicolumn{1}{c}{\bfseries }&\multicolumn{3}{c}{\bfseries IE}\tabularnewline
					\cline{6-8} \cline{10-12} \cline{14-16} \cline{18-20}
					\multicolumn{1}{l}{}&\multicolumn{1}{c}{true}&\multicolumn{1}{c}{$N$}&\multicolumn{1}{c}{$T$}&\multicolumn{1}{c}{}&\multicolumn{1}{c}{bias}&\multicolumn{1}{c}{std}&\multicolumn{1}{c}{cp}&\multicolumn{1}{c}{}&\multicolumn{1}{c}{bias}&\multicolumn{1}{c}{std}&\multicolumn{1}{c}{cp}&\multicolumn{1}{c}{}&\multicolumn{1}{c}{bias}&\multicolumn{1}{c}{std}&\multicolumn{1}{c}{cp}&\multicolumn{1}{c}{}&\multicolumn{1}{c}{bias}&\multicolumn{1}{c}{std}&\multicolumn{1}{c}{cp}\tabularnewline
					\hline
					at $\gamma_0$'s 20\%Q&$0.646$&$ 250$&$12$&&$ 0.173$&$0.072$&$0.302$&&$ 0.197$&$0.106$&$0.524$&&$ 0.109$&$0.165$&$0.905$&&$-0.013$&$0.068$&$0.921$\tabularnewline
					&$0.646$&$ 250$&$24$&&$ 0.131$&$0.074$&$0.564$&&$ 0.083$&$0.097$&$0.879$&&$ 0.011$&$0.137$&$0.944$&&$-0.014$&$0.067$&$0.920$\tabularnewline
					&$0.646$&$ 250$&$48$&&$ 0.078$&$0.071$&$0.830$&&$ 0.025$&$0.084$&$0.943$&&$-0.007$&$0.108$&$0.947$&&$-0.013$&$0.068$&$0.918$\tabularnewline
					&$0.646$&$ 250$&$96$&&$ 0.038$&$0.070$&$0.926$&&$ 0.001$&$0.077$&$0.942$&&$-0.010$&$0.090$&$0.942$&&$-0.012$&$0.068$&$0.919$\tabularnewline
					&$0.646$&$ 500$&$12$&&$ 0.174$&$0.058$&$0.114$&&$ 0.211$&$0.087$&$0.284$&&$ 0.127$&$0.142$&$0.858$&&$-0.009$&$0.053$&$0.928$\tabularnewline
					&$0.646$&$ 500$&$24$&&$ 0.136$&$0.058$&$0.331$&&$ 0.093$&$0.080$&$0.789$&&$ 0.017$&$0.119$&$0.948$&&$-0.010$&$0.053$&$0.928$\tabularnewline
					&$0.646$&$ 500$&$48$&&$ 0.083$&$0.057$&$0.702$&&$ 0.030$&$0.070$&$0.933$&&$-0.005$&$0.095$&$0.944$&&$-0.009$&$0.054$&$0.926$\tabularnewline
					&$0.646$&$ 500$&$96$&&$ 0.042$&$0.055$&$0.897$&&$ 0.004$&$0.062$&$0.948$&&$-0.009$&$0.075$&$0.939$&&$-0.009$&$0.053$&$0.928$\tabularnewline
					&$0.646$&$1000$&$12$&&$ 0.176$&$0.046$&$0.014$&&$ 0.223$&$0.071$&$0.091$&&$ 0.145$&$0.120$&$0.777$&&$-0.007$&$0.041$&$0.932$\tabularnewline
					&$0.646$&$1000$&$24$&&$ 0.140$&$0.045$&$0.102$&&$ 0.099$&$0.064$&$0.669$&&$ 0.019$&$0.102$&$0.949$&&$-0.007$&$0.042$&$0.928$\tabularnewline
					&$0.646$&$1000$&$48$&&$ 0.085$&$0.045$&$0.512$&&$ 0.032$&$0.057$&$0.922$&&$-0.004$&$0.081$&$0.946$&&$-0.007$&$0.042$&$0.933$\tabularnewline
					&$0.646$&$1000$&$96$&&$ 0.046$&$0.043$&$0.834$&&$ 0.007$&$0.050$&$0.950$&&$-0.007$&$0.063$&$0.945$&&$-0.007$&$0.040$&$0.939$\tabularnewline
					\hline
					at $\gamma_0$'s 40\%Q&$0.701$&$ 250$&$12$&&$-0.094$&$0.057$&$0.558$&&$ 0.008$&$0.087$&$0.917$&&$ 0.030$&$0.142$&$0.915$&&$-0.010$&$0.068$&$0.929$\tabularnewline
					&$0.701$&$ 250$&$24$&&$-0.026$&$0.061$&$0.888$&&$ 0.027$&$0.089$&$0.926$&&$ 0.021$&$0.134$&$0.927$&&$-0.011$&$0.070$&$0.918$\tabularnewline
					&$0.701$&$ 250$&$48$&&$-0.003$&$0.064$&$0.933$&&$ 0.015$&$0.085$&$0.935$&&$ 0.002$&$0.118$&$0.934$&&$-0.010$&$0.068$&$0.933$\tabularnewline
					&$0.701$&$ 250$&$96$&&$ 0.000$&$0.067$&$0.929$&&$ 0.002$&$0.081$&$0.926$&&$-0.005$&$0.101$&$0.927$&&$-0.012$&$0.070$&$0.918$\tabularnewline
					&$0.701$&$ 500$&$12$&&$-0.097$&$0.047$&$0.382$&&$ 0.014$&$0.071$&$0.918$&&$ 0.043$&$0.118$&$0.903$&&$-0.008$&$0.052$&$0.931$\tabularnewline
					&$0.701$&$ 500$&$24$&&$-0.024$&$0.047$&$0.877$&&$ 0.034$&$0.070$&$0.907$&&$ 0.029$&$0.108$&$0.929$&&$-0.008$&$0.054$&$0.924$\tabularnewline
					&$0.701$&$ 500$&$48$&&$ 0.000$&$0.050$&$0.941$&&$ 0.021$&$0.066$&$0.936$&&$ 0.007$&$0.094$&$0.942$&&$-0.008$&$0.053$&$0.928$\tabularnewline
					&$0.701$&$ 500$&$96$&&$ 0.001$&$0.052$&$0.930$&&$ 0.003$&$0.064$&$0.931$&&$-0.005$&$0.081$&$0.937$&&$-0.010$&$0.053$&$0.926$\tabularnewline
					&$0.701$&$1000$&$12$&&$-0.098$&$0.037$&$0.224$&&$ 0.021$&$0.058$&$0.913$&&$ 0.059$&$0.102$&$0.885$&&$-0.006$&$0.040$&$0.935$\tabularnewline
					&$0.701$&$1000$&$24$&&$-0.025$&$0.036$&$0.847$&&$ 0.035$&$0.055$&$0.883$&&$ 0.032$&$0.086$&$0.929$&&$-0.007$&$0.040$&$0.927$\tabularnewline
					&$0.701$&$1000$&$48$&&$ 0.000$&$0.038$&$0.936$&&$ 0.020$&$0.053$&$0.930$&&$ 0.006$&$0.076$&$0.940$&&$-0.008$&$0.041$&$0.929$\tabularnewline
					&$0.701$&$1000$&$96$&&$ 0.003$&$0.039$&$0.941$&&$ 0.006$&$0.049$&$0.943$&&$-0.004$&$0.066$&$0.941$&&$-0.006$&$0.040$&$0.928$\tabularnewline
					\hline
					at $\gamma_0$'s 60\%Q&$0.623$&$ 250$&$12$&&$-0.221$&$0.062$&$0.064$&&$-0.114$&$0.102$&$0.795$&&$-0.054$&$0.187$&$0.938$&&$-0.011$&$0.068$&$0.950$\tabularnewline
					&$0.623$&$ 250$&$24$&&$-0.134$&$0.062$&$0.471$&&$-0.054$&$0.094$&$0.919$&&$-0.026$&$0.164$&$0.950$&&$-0.011$&$0.068$&$0.946$\tabularnewline
					&$0.623$&$ 250$&$48$&&$-0.072$&$0.063$&$0.814$&&$-0.025$&$0.088$&$0.943$&&$-0.019$&$0.141$&$0.952$&&$-0.010$&$0.070$&$0.942$\tabularnewline
					&$0.623$&$ 250$&$96$&&$-0.042$&$0.066$&$0.904$&&$-0.018$&$0.085$&$0.945$&&$-0.021$&$0.123$&$0.951$&&$-0.012$&$0.068$&$0.947$\tabularnewline
					&$0.623$&$ 500$&$12$&&$-0.220$&$0.050$&$0.005$&&$-0.110$&$0.085$&$0.733$&&$-0.044$&$0.159$&$0.933$&&$-0.009$&$0.052$&$0.947$\tabularnewline
					&$0.623$&$ 500$&$24$&&$-0.131$&$0.050$&$0.280$&&$-0.046$&$0.079$&$0.912$&&$-0.011$&$0.143$&$0.952$&&$-0.008$&$0.052$&$0.950$\tabularnewline
					&$0.623$&$ 500$&$48$&&$-0.072$&$0.049$&$0.740$&&$-0.020$&$0.070$&$0.949$&&$-0.011$&$0.120$&$0.952$&&$-0.007$&$0.051$&$0.950$\tabularnewline
					&$0.623$&$ 500$&$96$&&$-0.040$&$0.050$&$0.891$&&$-0.014$&$0.067$&$0.948$&&$-0.015$&$0.102$&$0.949$&&$-0.010$&$0.051$&$0.952$\tabularnewline
					&$0.623$&$1000$&$12$&&$-0.222$&$0.039$&$0.000$&&$-0.111$&$0.067$&$0.614$&&$-0.042$&$0.125$&$0.934$&&$-0.006$&$0.039$&$0.953$\tabularnewline
					&$0.623$&$1000$&$24$&&$-0.130$&$0.040$&$0.092$&&$-0.040$&$0.065$&$0.914$&&$-0.001$&$0.120$&$0.949$&&$-0.006$&$0.039$&$0.951$\tabularnewline
					&$0.623$&$1000$&$48$&&$-0.071$&$0.038$&$0.598$&&$-0.015$&$0.057$&$0.952$&&$-0.004$&$0.100$&$0.955$&&$-0.006$&$0.039$&$0.950$\tabularnewline
					&$0.623$&$1000$&$96$&&$-0.037$&$0.038$&$0.861$&&$-0.009$&$0.054$&$0.947$&&$-0.010$&$0.087$&$0.944$&&$-0.006$&$0.040$&$0.951$\tabularnewline
					\hline
					at $\gamma_0$'s 80\%Q&$0.433$&$ 250$&$12$&&$-0.205$&$0.045$&$0.004$&&$-0.142$&$0.074$&$0.546$&&$-0.103$&$0.135$&$0.885$&&$-0.008$&$0.060$&$0.934$\tabularnewline
					&$0.433$&$ 250$&$24$&&$-0.144$&$0.050$&$0.183$&&$-0.086$&$0.079$&$0.817$&&$-0.056$&$0.142$&$0.932$&&$-0.006$&$0.060$&$0.938$\tabularnewline
					&$0.433$&$ 250$&$48$&&$-0.094$&$0.055$&$0.603$&&$-0.050$&$0.080$&$0.905$&&$-0.036$&$0.136$&$0.943$&&$-0.006$&$0.059$&$0.943$\tabularnewline
					&$0.433$&$ 250$&$96$&&$-0.056$&$0.057$&$0.832$&&$-0.026$&$0.078$&$0.936$&&$-0.021$&$0.122$&$0.946$&&$-0.004$&$0.059$&$0.945$\tabularnewline
					&$0.433$&$ 500$&$12$&&$-0.206$&$0.034$&$0.000$&&$-0.143$&$0.058$&$0.322$&&$-0.104$&$0.108$&$0.846$&&$-0.004$&$0.043$&$0.958$\tabularnewline
					&$0.433$&$ 500$&$24$&&$-0.143$&$0.038$&$0.026$&&$-0.084$&$0.062$&$0.761$&&$-0.052$&$0.113$&$0.933$&&$-0.004$&$0.045$&$0.949$\tabularnewline
					&$0.433$&$ 500$&$48$&&$-0.093$&$0.042$&$0.403$&&$-0.046$&$0.063$&$0.897$&&$-0.028$&$0.111$&$0.947$&&$-0.004$&$0.046$&$0.936$\tabularnewline
					&$0.433$&$ 500$&$96$&&$-0.055$&$0.044$&$0.764$&&$-0.023$&$0.062$&$0.933$&&$-0.018$&$0.100$&$0.952$&&$-0.003$&$0.045$&$0.945$\tabularnewline
					&$0.433$&$1000$&$12$&&$-0.205$&$0.026$&$0.000$&&$-0.142$&$0.044$&$0.097$&&$-0.101$&$0.083$&$0.786$&&$-0.003$&$0.034$&$0.950$\tabularnewline
					&$0.433$&$1000$&$24$&&$-0.143$&$0.030$&$0.000$&&$-0.082$&$0.049$&$0.649$&&$-0.049$&$0.091$&$0.920$&&$-0.002$&$0.034$&$0.951$\tabularnewline
					&$0.433$&$1000$&$48$&&$-0.091$&$0.032$&$0.155$&&$-0.041$&$0.049$&$0.887$&&$-0.020$&$0.088$&$0.946$&&$-0.002$&$0.034$&$0.946$\tabularnewline
					&$0.433$&$1000$&$96$&&$-0.054$&$0.033$&$0.642$&&$-0.021$&$0.048$&$0.940$&&$-0.012$&$0.081$&$0.947$&&$-0.003$&$0.034$&$0.949$\tabularnewline
					\hline
		\end{tabular}\end{center}}
	\end{table}
	
	\begin{table}[!t]
		{\scriptsize
			\caption{Monte Carlo simulation results for $\rho_1$}
			\label{table:acor} 
			\begin{center}
				\begin{tabular}{lrrrcrrrcrrrcrrrcrrr}
					\hline\hline
					\multicolumn{1}{l}{\bfseries }&\multicolumn{3}{c}{\bfseries }&\multicolumn{1}{c}{\bfseries }&\multicolumn{3}{c}{\bfseries NE}&\multicolumn{1}{c}{\bfseries }&\multicolumn{3}{c}{\bfseries HPJ}&\multicolumn{1}{c}{\bfseries }&\multicolumn{3}{c}{\bfseries TOJ}&\multicolumn{1}{c}{\bfseries }&\multicolumn{3}{c}{\bfseries IE}\tabularnewline
					\cline{6-8} \cline{10-12} \cline{14-16} \cline{18-20}
					\multicolumn{1}{l}{}&\multicolumn{1}{c}{true}&\multicolumn{1}{c}{$N$}&\multicolumn{1}{c}{$T$}&\multicolumn{1}{c}{}&\multicolumn{1}{c}{bias}&\multicolumn{1}{c}{std}&\multicolumn{1}{c}{cp}&\multicolumn{1}{c}{}&\multicolumn{1}{c}{bias}&\multicolumn{1}{c}{std}&\multicolumn{1}{c}{cp}&\multicolumn{1}{c}{}&\multicolumn{1}{c}{bias}&\multicolumn{1}{c}{std}&\multicolumn{1}{c}{cp}&\multicolumn{1}{c}{}&\multicolumn{1}{c}{bias}&\multicolumn{1}{c}{std}&\multicolumn{1}{c}{cp}\tabularnewline
					\hline
					at $\rho_1$'s 20\%Q&$0.609$&$ 250$&$12$&&$ 0.148$&$0.089$&$0.457$&&$ 0.089$&$0.152$&$0.862$&&$-0.004$&$0.288$&$0.931$&&$-0.007$&$0.080$&$0.945$\tabularnewline
					&$0.609$&$ 250$&$24$&&$ 0.066$&$0.085$&$0.825$&&$-0.018$&$0.131$&$0.947$&&$-0.076$&$0.222$&$0.938$&&$-0.011$&$0.080$&$0.952$\tabularnewline
					&$0.609$&$ 250$&$48$&&$ 0.027$&$0.083$&$0.930$&&$-0.017$&$0.110$&$0.946$&&$-0.019$&$0.160$&$0.954$&&$-0.009$&$0.080$&$0.943$\tabularnewline
					&$0.609$&$ 250$&$96$&&$ 0.008$&$0.082$&$0.944$&&$-0.014$&$0.098$&$0.947$&&$-0.015$&$0.124$&$0.947$&&$-0.010$&$0.081$&$0.943$\tabularnewline
					&$0.609$&$ 500$&$12$&&$ 0.151$&$0.067$&$0.302$&&$ 0.094$&$0.115$&$0.805$&&$-0.002$&$0.228$&$0.930$&&$-0.007$&$0.060$&$0.944$\tabularnewline
					&$0.609$&$ 500$&$24$&&$ 0.070$&$0.065$&$0.717$&&$-0.013$&$0.102$&$0.947$&&$-0.068$&$0.176$&$0.941$&&$-0.006$&$0.061$&$0.949$\tabularnewline
					&$0.609$&$ 500$&$48$&&$ 0.029$&$0.063$&$0.909$&&$-0.014$&$0.087$&$0.951$&&$-0.016$&$0.134$&$0.950$&&$-0.006$&$0.062$&$0.944$\tabularnewline
					&$0.609$&$ 500$&$96$&&$ 0.010$&$0.062$&$0.942$&&$-0.010$&$0.076$&$0.947$&&$-0.009$&$0.098$&$0.954$&&$-0.007$&$0.060$&$0.947$\tabularnewline
					&$0.609$&$1000$&$12$&&$ 0.154$&$0.050$&$0.149$&&$ 0.099$&$0.088$&$0.665$&&$-0.004$&$0.184$&$0.933$&&$-0.004$&$0.046$&$0.950$\tabularnewline
					&$0.609$&$1000$&$24$&&$ 0.073$&$0.048$&$0.539$&&$-0.008$&$0.078$&$0.950$&&$-0.060$&$0.138$&$0.933$&&$-0.005$&$0.046$&$0.949$\tabularnewline
					&$0.609$&$1000$&$48$&&$ 0.031$&$0.048$&$0.868$&&$-0.012$&$0.069$&$0.944$&&$-0.014$&$0.108$&$0.949$&&$-0.005$&$0.046$&$0.946$\tabularnewline
					&$0.609$&$1000$&$96$&&$ 0.014$&$0.047$&$0.937$&&$-0.006$&$0.058$&$0.950$&&$-0.005$&$0.079$&$0.958$&&$-0.005$&$0.047$&$0.943$\tabularnewline
					\hline
					at $\rho_1$'s 40\%Q&$0.823$&$ 250$&$12$&&$ 0.054$&$0.093$&$0.903$&&$ 0.139$&$0.156$&$0.801$&&$ 0.105$&$0.283$&$0.915$&&$-0.015$&$0.091$&$0.942$\tabularnewline
					&$0.823$&$ 250$&$24$&&$ 0.041$&$0.094$&$0.919$&&$ 0.027$&$0.146$&$0.935$&&$-0.051$&$0.244$&$0.940$&&$-0.016$&$0.089$&$0.949$\tabularnewline
					&$0.823$&$ 250$&$48$&&$ 0.012$&$0.090$&$0.949$&&$-0.017$&$0.124$&$0.952$&&$-0.039$&$0.183$&$0.951$&&$-0.015$&$0.091$&$0.944$\tabularnewline
					&$0.823$&$ 250$&$96$&&$-0.004$&$0.094$&$0.942$&&$-0.021$&$0.114$&$0.944$&&$-0.024$&$0.144$&$0.951$&&$-0.015$&$0.090$&$0.946$\tabularnewline
					&$0.823$&$ 500$&$12$&&$ 0.056$&$0.070$&$0.869$&&$ 0.142$&$0.119$&$0.673$&&$ 0.104$&$0.219$&$0.903$&&$-0.011$&$0.067$&$0.950$\tabularnewline
					&$0.823$&$ 500$&$24$&&$ 0.045$&$0.070$&$0.895$&&$ 0.031$&$0.112$&$0.934$&&$-0.048$&$0.192$&$0.947$&&$-0.010$&$0.069$&$0.947$\tabularnewline
					&$0.823$&$ 500$&$48$&&$ 0.016$&$0.069$&$0.941$&&$-0.012$&$0.099$&$0.952$&&$-0.034$&$0.153$&$0.944$&&$-0.012$&$0.070$&$0.944$\tabularnewline
					&$0.823$&$ 500$&$96$&&$ 0.003$&$0.069$&$0.953$&&$-0.013$&$0.087$&$0.953$&&$-0.015$&$0.115$&$0.953$&&$-0.011$&$0.068$&$0.951$\tabularnewline
					&$0.823$&$1000$&$12$&&$ 0.059$&$0.053$&$0.770$&&$ 0.148$&$0.091$&$0.459$&&$ 0.105$&$0.167$&$0.876$&&$-0.008$&$0.052$&$0.951$\tabularnewline
					&$0.823$&$1000$&$24$&&$ 0.045$&$0.054$&$0.834$&&$ 0.030$&$0.088$&$0.931$&&$-0.049$&$0.153$&$0.940$&&$-0.009$&$0.052$&$0.950$\tabularnewline
					&$0.823$&$1000$&$48$&&$ 0.019$&$0.053$&$0.932$&&$-0.008$&$0.077$&$0.953$&&$-0.029$&$0.123$&$0.950$&&$-0.009$&$0.052$&$0.947$\tabularnewline
					&$0.823$&$1000$&$96$&&$ 0.003$&$0.052$&$0.952$&&$-0.013$&$0.067$&$0.951$&&$-0.016$&$0.095$&$0.957$&&$-0.009$&$0.051$&$0.952$\tabularnewline
					\hline
					at $\rho_1$'s 60\%Q&$0.889$&$ 250$&$12$&&$-0.083$&$0.089$&$0.744$&&$ 0.087$&$0.143$&$0.910$&&$ 0.103$&$0.248$&$0.924$&&$-0.016$&$0.092$&$0.931$\tabularnewline
					&$0.889$&$ 250$&$24$&&$-0.014$&$0.091$&$0.928$&&$ 0.056$&$0.135$&$0.933$&&$ 0.028$&$0.216$&$0.944$&&$-0.013$&$0.095$&$0.930$\tabularnewline
					&$0.889$&$ 250$&$48$&&$-0.005$&$0.091$&$0.943$&&$ 0.005$&$0.121$&$0.946$&&$-0.025$&$0.172$&$0.945$&&$-0.013$&$0.092$&$0.935$\tabularnewline
					&$0.889$&$ 250$&$96$&&$-0.009$&$0.093$&$0.940$&&$-0.011$&$0.110$&$0.942$&&$-0.016$&$0.132$&$0.943$&&$-0.014$&$0.094$&$0.931$\tabularnewline
					&$0.889$&$ 500$&$12$&&$-0.078$&$0.067$&$0.701$&&$ 0.095$&$0.110$&$0.867$&&$ 0.132$&$0.196$&$0.889$&&$-0.011$&$0.069$&$0.946$\tabularnewline
					&$0.889$&$ 500$&$24$&&$-0.011$&$0.070$&$0.928$&&$ 0.060$&$0.106$&$0.918$&&$ 0.029$&$0.173$&$0.949$&&$-0.012$&$0.069$&$0.944$\tabularnewline
					&$0.889$&$ 500$&$48$&&$-0.003$&$0.069$&$0.947$&&$ 0.006$&$0.094$&$0.953$&&$-0.023$&$0.139$&$0.936$&&$-0.011$&$0.070$&$0.943$\tabularnewline
					&$0.889$&$ 500$&$96$&&$-0.007$&$0.071$&$0.938$&&$-0.009$&$0.086$&$0.943$&&$-0.014$&$0.108$&$0.945$&&$-0.011$&$0.069$&$0.944$\tabularnewline
					&$0.889$&$1000$&$12$&&$-0.077$&$0.051$&$0.587$&&$ 0.097$&$0.084$&$0.793$&&$ 0.154$&$0.162$&$0.804$&&$-0.010$&$0.052$&$0.944$\tabularnewline
					&$0.889$&$1000$&$24$&&$-0.008$&$0.053$&$0.929$&&$ 0.063$&$0.082$&$0.886$&&$ 0.030$&$0.138$&$0.944$&&$-0.010$&$0.053$&$0.942$\tabularnewline
					&$0.889$&$1000$&$48$&&$ 0.000$&$0.052$&$0.947$&&$ 0.009$&$0.073$&$0.950$&&$-0.021$&$0.111$&$0.947$&&$-0.008$&$0.053$&$0.944$\tabularnewline
					&$0.889$&$1000$&$96$&&$-0.004$&$0.053$&$0.946$&&$-0.005$&$0.065$&$0.948$&&$-0.010$&$0.086$&$0.942$&&$-0.009$&$0.053$&$0.941$\tabularnewline
					\hline
					at $\rho_1$'s 80\%Q&$0.790$&$ 250$&$12$&&$-0.247$&$0.075$&$0.099$&&$-0.048$&$0.120$&$0.886$&&$-0.069$&$0.194$&$0.911$&&$-0.015$&$0.086$&$0.919$\tabularnewline
					&$0.790$&$ 250$&$24$&&$-0.105$&$0.084$&$0.656$&&$ 0.021$&$0.119$&$0.930$&&$ 0.052$&$0.179$&$0.938$&&$-0.011$&$0.087$&$0.921$\tabularnewline
					&$0.790$&$ 250$&$48$&&$-0.045$&$0.086$&$0.863$&&$ 0.014$&$0.108$&$0.934$&&$ 0.007$&$0.145$&$0.938$&&$-0.012$&$0.087$&$0.920$\tabularnewline
					&$0.790$&$ 250$&$96$&&$-0.025$&$0.086$&$0.901$&&$-0.003$&$0.097$&$0.934$&&$-0.010$&$0.113$&$0.935$&&$-0.012$&$0.086$&$0.921$\tabularnewline
					&$0.790$&$ 500$&$12$&&$-0.240$&$0.058$&$0.026$&&$-0.031$&$0.094$&$0.888$&&$-0.082$&$0.156$&$0.887$&&$-0.009$&$0.065$&$0.928$\tabularnewline
					&$0.790$&$ 500$&$24$&&$-0.103$&$0.064$&$0.555$&&$ 0.027$&$0.092$&$0.929$&&$ 0.063$&$0.144$&$0.930$&&$-0.010$&$0.067$&$0.922$\tabularnewline
					&$0.790$&$ 500$&$48$&&$-0.043$&$0.066$&$0.848$&&$ 0.018$&$0.086$&$0.935$&&$ 0.008$&$0.121$&$0.938$&&$-0.009$&$0.066$&$0.926$\tabularnewline
					&$0.790$&$ 500$&$96$&&$-0.022$&$0.066$&$0.900$&&$-0.001$&$0.076$&$0.938$&&$-0.010$&$0.094$&$0.934$&&$-0.009$&$0.065$&$0.931$\tabularnewline
					&$0.790$&$1000$&$12$&&$-0.237$&$0.043$&$0.001$&&$-0.024$&$0.071$&$0.891$&&$-0.110$&$0.109$&$0.836$&&$-0.007$&$0.050$&$0.936$\tabularnewline
					&$0.790$&$1000$&$24$&&$-0.097$&$0.048$&$0.455$&&$ 0.037$&$0.071$&$0.931$&&$ 0.073$&$0.116$&$0.921$&&$-0.007$&$0.051$&$0.932$\tabularnewline
					&$0.790$&$1000$&$48$&&$-0.039$&$0.050$&$0.839$&&$ 0.021$&$0.066$&$0.948$&&$ 0.009$&$0.099$&$0.950$&&$-0.008$&$0.050$&$0.937$\tabularnewline
					&$0.790$&$1000$&$96$&&$-0.021$&$0.050$&$0.905$&&$-0.001$&$0.059$&$0.946$&&$-0.013$&$0.077$&$0.936$&&$-0.008$&$0.051$&$0.924$\tabularnewline
					\hline
		\end{tabular}\end{center}}
	\end{table}
	
	\begin{table}[!t]
		{\scriptsize
			\caption{Monte Carlo simulation results for bandwidths}
			\label{table:band}
			\begin{center}
				\begin{tabular}{lrrcrrcrrcrrcrrcrrcrr}
					\hline\hline
					\multicolumn{1}{l}{\bfseries }&\multicolumn{2}{c}{\bfseries }&\multicolumn{1}{c}{\bfseries }&\multicolumn{2}{c}{\bfseries $\mu$ NE}&\multicolumn{1}{c}{\bfseries }&\multicolumn{2}{c}{\bfseries $\mu$ IE}&\multicolumn{1}{c}{\bfseries }&\multicolumn{2}{c}{\bfseries $\gamma_0$ NE}&\multicolumn{1}{c}{\bfseries }&\multicolumn{2}{c}{\bfseries $\gamma_0$ IE}&\multicolumn{1}{c}{\bfseries }&\multicolumn{2}{c}{\bfseries $\rho_1$ NE}&\multicolumn{1}{c}{\bfseries }&\multicolumn{2}{c}{\bfseries $\rho_1$ IE}\tabularnewline
					\cline{5-6} \cline{8-9} \cline{11-12} \cline{14-15} \cline{17-18} \cline{20-21}
					\multicolumn{1}{l}{}&\multicolumn{1}{c}{$N$}&\multicolumn{1}{c}{$T$}&\multicolumn{1}{c}{}&\multicolumn{1}{c}{mean}&\multicolumn{1}{c}{std}&\multicolumn{1}{c}{}&\multicolumn{1}{c}{mean}&\multicolumn{1}{c}{std}&\multicolumn{1}{c}{}&\multicolumn{1}{c}{mean}&\multicolumn{1}{c}{std}&\multicolumn{1}{c}{}&\multicolumn{1}{c}{mean}&\multicolumn{1}{c}{std}&\multicolumn{1}{c}{}&\multicolumn{1}{c}{mean}&\multicolumn{1}{c}{std}&\multicolumn{1}{c}{}&\multicolumn{1}{c}{mean}&\multicolumn{1}{c}{std}\tabularnewline
					\hline
					at 20\% Q&$ 250$&$12$&&$0.889$&$0.277$&&$0.807$&$0.255$&&$0.374$&$0.025$&&$0.387$&$0.053$&&$0.316$&$0.092$&&$0.320$&$0.098$\tabularnewline
					&$ 250$&$24$&&$0.850$&$0.259$&&$0.807$&$0.254$&&$0.370$&$0.024$&&$0.388$&$0.052$&&$0.311$&$0.096$&&$0.317$&$0.097$\tabularnewline
					&$ 250$&$48$&&$0.831$&$0.261$&&$0.805$&$0.254$&&$0.373$&$0.030$&&$0.386$&$0.052$&&$0.314$&$0.096$&&$0.318$&$0.097$\tabularnewline
					&$ 250$&$96$&&$0.819$&$0.260$&&$0.810$&$0.260$&&$0.379$&$0.039$&&$0.388$&$0.055$&&$0.317$&$0.098$&&$0.318$&$0.098$\tabularnewline
					&$ 500$&$12$&&$0.786$&$0.237$&&$0.714$&$0.232$&&$0.316$&$0.015$&&$0.324$&$0.032$&&$0.282$&$0.082$&&$0.285$&$0.086$\tabularnewline
					&$ 500$&$24$&&$0.759$&$0.235$&&$0.716$&$0.229$&&$0.311$&$0.014$&&$0.325$&$0.034$&&$0.276$&$0.085$&&$0.283$&$0.087$\tabularnewline
					&$ 500$&$48$&&$0.734$&$0.230$&&$0.716$&$0.232$&&$0.313$&$0.017$&&$0.324$&$0.032$&&$0.278$&$0.085$&&$0.283$&$0.086$\tabularnewline
					&$ 500$&$96$&&$0.727$&$0.227$&&$0.717$&$0.235$&&$0.317$&$0.022$&&$0.324$&$0.032$&&$0.282$&$0.087$&&$0.284$&$0.087$\tabularnewline
					&$1000$&$12$&&$0.699$&$0.219$&&$0.636$&$0.207$&&$0.267$&$0.009$&&$0.273$&$0.019$&&$0.252$&$0.072$&&$0.254$&$0.078$\tabularnewline
					&$1000$&$24$&&$0.671$&$0.209$&&$0.630$&$0.202$&&$0.263$&$0.009$&&$0.273$&$0.019$&&$0.247$&$0.076$&&$0.253$&$0.077$\tabularnewline
					&$1000$&$48$&&$0.656$&$0.206$&&$0.637$&$0.208$&&$0.265$&$0.011$&&$0.273$&$0.019$&&$0.247$&$0.076$&&$0.253$&$0.078$\tabularnewline
					&$1000$&$96$&&$0.650$&$0.210$&&$0.633$&$0.206$&&$0.267$&$0.013$&&$0.273$&$0.020$&&$0.251$&$0.076$&&$0.251$&$0.076$\tabularnewline
					\hline
					at 40\% Q&$ 250$&$12$&&$0.850$&$0.179$&&$0.792$&$0.175$&&$0.510$&$0.151$&&$0.434$&$0.097$&&$0.323$&$0.084$&&$0.314$&$0.095$\tabularnewline
					&$ 250$&$24$&&$0.824$&$0.174$&&$0.789$&$0.164$&&$0.479$&$0.103$&&$0.432$&$0.096$&&$0.318$&$0.091$&&$0.313$&$0.094$\tabularnewline
					&$ 250$&$48$&&$0.810$&$0.169$&&$0.790$&$0.171$&&$0.451$&$0.090$&&$0.435$&$0.099$&&$0.312$&$0.093$&&$0.313$&$0.093$\tabularnewline
					&$ 250$&$96$&&$0.800$&$0.170$&&$0.788$&$0.166$&&$0.443$&$0.096$&&$0.431$&$0.096$&&$0.313$&$0.094$&&$0.314$&$0.095$\tabularnewline
					&$ 500$&$12$&&$0.731$&$0.141$&&$0.679$&$0.131$&&$0.433$&$0.147$&&$0.383$&$0.086$&&$0.286$&$0.074$&&$0.280$&$0.084$\tabularnewline
					&$ 500$&$24$&&$0.710$&$0.133$&&$0.681$&$0.129$&&$0.427$&$0.096$&&$0.380$&$0.082$&&$0.280$&$0.079$&&$0.279$&$0.085$\tabularnewline
					&$ 500$&$48$&&$0.699$&$0.135$&&$0.682$&$0.131$&&$0.399$&$0.082$&&$0.381$&$0.084$&&$0.278$&$0.084$&&$0.278$&$0.085$\tabularnewline
					&$ 500$&$96$&&$0.692$&$0.134$&&$0.684$&$0.131$&&$0.390$&$0.077$&&$0.382$&$0.085$&&$0.276$&$0.082$&&$0.279$&$0.084$\tabularnewline
					&$1000$&$12$&&$0.625$&$0.098$&&$0.586$&$0.105$&&$0.340$&$0.139$&&$0.339$&$0.074$&&$0.256$&$0.067$&&$0.247$&$0.075$\tabularnewline
					&$1000$&$24$&&$0.610$&$0.102$&&$0.588$&$0.107$&&$0.387$&$0.094$&&$0.339$&$0.074$&&$0.247$&$0.071$&&$0.247$&$0.075$\tabularnewline
					&$1000$&$48$&&$0.598$&$0.102$&&$0.584$&$0.099$&&$0.356$&$0.072$&&$0.341$&$0.077$&&$0.244$&$0.073$&&$0.246$&$0.075$\tabularnewline
					&$1000$&$96$&&$0.595$&$0.104$&&$0.587$&$0.099$&&$0.345$&$0.070$&&$0.341$&$0.076$&&$0.246$&$0.075$&&$0.247$&$0.075$\tabularnewline
					\hline
					at 60\% Q&$ 250$&$12$&&$0.842$&$0.165$&&$0.788$&$0.165$&&$0.308$&$0.061$&&$0.415$&$0.131$&&$0.323$&$0.070$&&$0.327$&$0.082$\tabularnewline
					&$ 250$&$24$&&$0.823$&$0.173$&&$0.786$&$0.165$&&$0.364$&$0.132$&&$0.411$&$0.129$&&$0.322$&$0.071$&&$0.324$&$0.081$\tabularnewline
					&$ 250$&$48$&&$0.811$&$0.174$&&$0.787$&$0.164$&&$0.410$&$0.149$&&$0.413$&$0.129$&&$0.322$&$0.074$&&$0.327$&$0.080$\tabularnewline
					&$ 250$&$96$&&$0.801$&$0.171$&&$0.784$&$0.156$&&$0.420$&$0.144$&&$0.414$&$0.131$&&$0.324$&$0.078$&&$0.325$&$0.080$\tabularnewline
					&$ 500$&$12$&&$0.731$&$0.139$&&$0.682$&$0.136$&&$0.253$&$0.028$&&$0.363$&$0.117$&&$0.282$&$0.057$&&$0.290$&$0.072$\tabularnewline
					&$ 500$&$24$&&$0.710$&$0.128$&&$0.680$&$0.129$&&$0.290$&$0.098$&&$0.363$&$0.117$&&$0.281$&$0.060$&&$0.291$&$0.072$\tabularnewline
					&$ 500$&$48$&&$0.698$&$0.138$&&$0.683$&$0.135$&&$0.344$&$0.131$&&$0.362$&$0.118$&&$0.287$&$0.066$&&$0.290$&$0.073$\tabularnewline
					&$ 500$&$96$&&$0.689$&$0.134$&&$0.678$&$0.129$&&$0.363$&$0.130$&&$0.364$&$0.119$&&$0.288$&$0.070$&&$0.290$&$0.072$\tabularnewline
					&$1000$&$12$&&$0.623$&$0.098$&&$0.586$&$0.099$&&$0.212$&$0.012$&&$0.317$&$0.105$&&$0.249$&$0.049$&&$0.258$&$0.065$\tabularnewline
					&$1000$&$24$&&$0.609$&$0.108$&&$0.585$&$0.103$&&$0.233$&$0.049$&&$0.320$&$0.107$&&$0.250$&$0.053$&&$0.258$&$0.065$\tabularnewline
					&$1000$&$48$&&$0.597$&$0.097$&&$0.584$&$0.106$&&$0.281$&$0.105$&&$0.320$&$0.106$&&$0.255$&$0.057$&&$0.258$&$0.063$\tabularnewline
					&$1000$&$96$&&$0.593$&$0.102$&&$0.584$&$0.099$&&$0.312$&$0.116$&&$0.318$&$0.105$&&$0.256$&$0.061$&&$0.258$&$0.063$\tabularnewline
					\hline
					at 80\% Q&$ 250$&$12$&&$0.889$&$0.271$&&$0.809$&$0.257$&&$0.373$&$0.116$&&$0.404$&$0.137$&&$0.338$&$0.082$&&$0.312$&$0.057$\tabularnewline
					&$ 250$&$24$&&$0.851$&$0.260$&&$0.806$&$0.250$&&$0.370$&$0.131$&&$0.407$&$0.139$&&$0.317$&$0.060$&&$0.310$&$0.054$\tabularnewline
					&$ 250$&$48$&&$0.832$&$0.261$&&$0.809$&$0.258$&&$0.384$&$0.145$&&$0.404$&$0.135$&&$0.309$&$0.054$&&$0.311$&$0.056$\tabularnewline
					&$ 250$&$96$&&$0.817$&$0.251$&&$0.808$&$0.256$&&$0.397$&$0.144$&&$0.409$&$0.139$&&$0.308$&$0.053$&&$0.310$&$0.054$\tabularnewline
					&$ 500$&$12$&&$0.787$&$0.243$&&$0.715$&$0.233$&&$0.323$&$0.097$&&$0.358$&$0.121$&&$0.297$&$0.064$&&$0.265$&$0.040$\tabularnewline
					&$ 500$&$24$&&$0.761$&$0.239$&&$0.719$&$0.230$&&$0.319$&$0.109$&&$0.358$&$0.121$&&$0.272$&$0.046$&&$0.265$&$0.040$\tabularnewline
					&$ 500$&$48$&&$0.748$&$0.237$&&$0.718$&$0.228$&&$0.331$&$0.125$&&$0.358$&$0.123$&&$0.262$&$0.038$&&$0.264$&$0.039$\tabularnewline
					&$ 500$&$96$&&$0.730$&$0.233$&&$0.717$&$0.223$&&$0.348$&$0.129$&&$0.362$&$0.122$&&$0.262$&$0.039$&&$0.264$&$0.039$\tabularnewline
					&$1000$&$12$&&$0.704$&$0.210$&&$0.637$&$0.205$&&$0.287$&$0.091$&&$0.317$&$0.111$&&$0.258$&$0.050$&&$0.223$&$0.026$\tabularnewline
					&$1000$&$24$&&$0.677$&$0.214$&&$0.635$&$0.205$&&$0.274$&$0.089$&&$0.315$&$0.108$&&$0.227$&$0.031$&&$0.225$&$0.029$\tabularnewline
					&$1000$&$48$&&$0.658$&$0.210$&&$0.633$&$0.208$&&$0.287$&$0.105$&&$0.317$&$0.112$&&$0.220$&$0.024$&&$0.224$&$0.027$\tabularnewline
					&$1000$&$96$&&$0.655$&$0.215$&&$0.639$&$0.208$&&$0.300$&$0.112$&&$0.314$&$0.107$&&$0.221$&$0.024$&&$0.225$&$0.028$\tabularnewline
					\hline
		\end{tabular}\end{center}}
	\end{table}
\end{landscape}

\end{document}